\documentclass{jfm}

\usepackage{graphicx}
\usepackage{array}
\usepackage{booktabs}

\usepackage{xkeyval,xcolor}
\makeatletter
\newlength{\sfp@hseplen}\newlength{\sfp@vseplen}
\define@cmdkey{subfigpos}[sfp@]{pos}[ul]{}
\define@cmdkey{subfigpos}[sfp@]{font}[\small]{}
\define@cmdkey{subfigpos}[sfp@]{vsep}[2\baselineskip]{\setlength{\sfp@vseplen}{\sfp@vsep}}
\define@cmdkey{subfigpos}[sfp@]{hsep}[10pt]{\setlength{\sfp@hseplen}{\sfp@hsep}}
\newcommand{\subfigimg}[3][,]{%
  \setkeys{Gin,subfigpos}{pos,font,vsep,hsep,#1}
  \setbox1=\hbox{\includegraphics{#3}}
  \ifnum\pdfstrcmp{\sfp@pos}{ul}=0
    \leavevmode\rlap{\usebox1}
    \rlap{\hspace*{\sfp@hsep}\raisebox{\dimexpr\ht1-\sfp@vsep}{\sfp@font{#2}}}
    \phantom{\usebox1}
  \else\ifnum\pdfstrcmp{\sfp@pos}{ur}=0
    \leavevmode\usebox1
    \llap{\raisebox{\dimexpr\ht1-\sfp@vsep}{\sfp@font{#2}}\hspace*{\sfp@hsep}}
  \else\ifnum\pdfstrcmp{\sfp@pos}{lr}=0
    \leavevmode\usebox1
    \llap{\raisebox{\sfp@vsep}{\sfp@font{#2}}\hspace*{\sfp@hsep}}
  \else
    \leavevmode\rlap{\usebox1}
    \rlap{\hspace*{\sfp@hseplen}\raisebox{\sfp@vsep}{\sfp@font{#2}}}
    \phantom{\usebox1}
  \fi\fi\fi
}
\makeatother

\usepackage{amsmath,amsfonts,amssymb}
\usepackage{subfigure}
\usepackage{float}
\usepackage{color}
\usepackage{multirow}
\usepackage{algorithm}
\usepackage{algpseudocode}

\begin{document}

\newtheorem{lemma}{Lemma}
\newtheorem{corollary}{Corollary}

\shorttitle{A neural network approach for the blind deconvolution of turbulent flows} 
\shortauthor{R.Maulik and O.San} 

\title{A neural network approach for the blind deconvolution of turbulent flows}

\author
 {
 R. Maulik
  \and
  O. San
  \corresp{\email{osan@okstate.edu}}
  }

\affiliation
{
School of Mechanical \& Aerospace Engineering, Oklahoma State University, Stillwater, OK 74078, USA
}
\maketitle

\begin{abstract}
We present a single-layer feed-forward artificial neural network architecture trained through a supervised learning approach for the deconvolution of flow variables from their coarse-grained computations such as those encountered in large eddy simulations. We stress that the deconvolution procedure proposed in this investigation is blind, i.e. the deconvolved field is computed without any pre-existing information about the filtering procedure or kernel. This may be conceptually contrasted to the celebrated approximate deconvolution approaches where a filter shape is predefined for an iterative deconvolution process. We demonstrate that the proposed blind deconvolution network performs exceptionally well in the a-priori testing of two-dimensional Kraichnan, three-dimensional Kolmogorov and compressible stratified turbulence test cases and shows promise in forming the backbone of a physics-augmented data-driven closure for the Navier-Stokes equations.
\end{abstract}

\section{Introduction}\label{Intro}

In the past few decades, an exponential increase in computational power, algorithmic advances and experimental data collection strategies have seen an explosion in modeling efforts which leverage information obtained from physical data. The ability to \emph{learn} from the physics of an experiment represents an attractive augmentation to the laws that have been painstakingly developed from the first principles of fluid mechanics \citep{schmidt2009distilling}. As highlighted by \cite{kutz2017deep}, data-driven learning combined with well established governing laws represent an extraordinary opportunity for the mitigation of the challenges faced by many modeling mechanisms that are purely developed from first principles.

The use of machine learning techniques have traditionally been limited to the flow control community \citep{lee1997application,milano2002neural,gautier2015closed} but there has been recent interest in their utilization for subgrid scale modeling for high Reynolds ($Re$) number flows. For instance, artificial neural networks (ANNs) have recently been utilized for the purpose of subgrid scale model classification \citep{gamahara2017searching} and Reynolds stress anisotropy tensor prediction \citep{ling2016reynolds,wang2017physics}. Together with other machine learning based techniques such as Gaussian process regression \citep{raissi2017hidden,raissi2016deep,zhang2015machine,wang2016physics}, field inversion \citep{duraisamy2015new,parish2016paradigm}, symbolic regression \citep{weatheritt2016novel,brunton2016discovering} and several different algorithms \citep{ling2015evaluation,ling2016machine,tracey2013application,bright2013compressive}, data-driven approaches such as ANNs are poised to form the backbone of the next great leap in closure modeling for nonlinear conservation laws.

A chief motivation for this research lies in the recent advances in the image processing community which exploit the general body of machine learning techniques used for the reconstruction of noisy or blurred images \citep{cichocki2002adaptive}. In particular we endeavour to justify the implementation of ANN based machine learning strategies for the purpose of subfilter scale recovery in turbulence closure modeling. In this work, we have developed a single-layer feed-forward ANN to identify a nonlinear relationship between the low-pass spatially filtered and coarse-grained (but unfiltered) field variables for settings in two-dimensional (2D) and three-dimensional (3D) homogeneous isotropic turbulence as well as a stratified turbulence case exhibiting moderate compressibility in the limit of infinite Reynolds numbers. A fundamental basis for the selection of a single hidden layer is to reduce training durations and to constrain the degrees of freedom of our trained network (by reducing the number of weights). We emphasize that the performance demanded from our ANN is \emph{statistical} in nature and generalization is vital. Our trained ANN, once deployed, performs a data recovery procedure utilizing the statistical relationship between filtered and true data. Our argument for the choices presented above are given by the fact that a successful implementation of the proposed idea would connect potential applications across many different time and length scales.

The homogeneous and isotropic turbulence framework is chosen from the point of view of a well established understanding of the energy cascade mechanism in the inertial range. In essence, the statistical recovery of scaling behavior in the inertial range shall represent our contribution from the first principles of fluid mechanics. The compressible stratified turbulence test case serves a dual purpose. Not only is it utilized to evaluate the performance of our proposed architecture in a perfectly a-priori fashion, but it is also used for testing the suitability of the trained ANN for deconvolution across different flow physics. To that end, we generate training data sets from one of the 3D homogeneous isotropic or compressible stratified turbulence simulations and test it on the other. This may be considered to be more challenging than a purely a-priori analysis and would expose the universal nature of the training in our data-driven framework. These tests reveal further information related to the suitability of the data-driven closure across different physical regimes.

In addition to the various subfilter recovery analyses outlined above, the ability of our proposed framework to stabilize aliasing error is also studied for the different test cases. For this purpose, we investigate the predictive performance of the proposed procedure by solving a noisy data recovery problem in each setting. This noisy data recovery problem requires the estimation of a relationship between inputs which are perturbed by noise and their true counterparts to examine its suitability for coarse-grained large eddy simulations which often exhibit severe aliasing error. Besides the cross-validation within the framework of the different test cases for this noisy data problem, we also utilize training and testing data from different flow physics (for instance training between homogeneous isotropic and stratified turbulence test cases) to test the universality of this regularization behavior.

In particular, the approach outlined in our study may be considered to be analogous to the \emph{approximate deconvolution} methodology \citep{stolz1999approximate} to recover subfilter contributions of low-pass spatially filtered flow fields. The chief difference is the lack of assumption of any filtering kernel (Gaussian or otherwise); a fact that necessitates the utilization of additional regularization in case of flows which may exhibit distinctly non-Gaussian distributions \citep{stolz2001approximate}. A potential advantage to emphasize here is that a data-driven closure could be incorporated to \emph{learn} the nonlinear deconvolution procedure for challenging flows. This could, conceptually, aid in a significant increase of the accuracy of reconstructed subfilter contributions to the turbulence shear stresses. We propose this hypothesis since it has been rigorously established that single and multilayered neural networks are capable of acting as universal function approximators \citep{hornik1989multilayer} and could represent a continuous underlying `natural' low-pass spatial filter shape effectively. We validate our hypothesis by testing the ability of the trained ANN to capture inertial length scales accurately (i.e., while respecting the \cite{kolmogorov1941local} and \cite{kraichnan1967inertial} scaling laws) and by examining the probability density functions of the true, distorted (through low-pass spatial filtering or noise) and reconstructed field variables. Apart from this, our approach is also cross-examined with state of the art structural closures which require an explicitly specified low-pass spatial filter \citep{layton2012approximate}. A cross-validation strategy is also detailed and performed to ensure that the performance of the proposed framework is physics-based and not an artifact of data-localization.

\section{Artificial Neural Networks}

\subsection{Architecture}
\label{sec:arch_ann}

The basic structure of the simple feed-forward ANN consists of layers possessing a predefined number of unit cells called neurons. Each of these layers has an associated transfer function and each unit cell has an associated bias. Any input to the neuron has a bias added to it followed by activation through the transfer function. To describe this process using matrix operations, we have for a single neuron in the $l^{th}$ layer receiving a vector of inputs $\mathbf{S}^l$ from the $(l-1)^{th}$ layer given by \citep{demuth2014neural}
\begin{align}\label{ANN_1}
  \mathbf{S}^l = \mathbf{W}^l \mathbf{X}^{l-1},
\end{align}
where $\mathbf{W}^l$ stands for a matrix of weights linking the $l-1$ and $l$ layers with $\mathbf{X}^{l-1}$ being the output of the $(l-1)^{th}$ layer. The output of the $l^{th}$ layer is now given by
\begin{align}\label{ANN_2}
  \mathbf{X}^l = f(\mathbf{S}^l + \mathbf{B}^l),
\end{align}
where $\mathbf{B}^l$ is the vector of biasing parameters for the $l^{th}$ layer. Every node (or unit cell) has an associated transfer function which acts on its input and bias to produce an output which is \emph{fed forward} through the network. The nodes which take the raw input value of our training data set (i.e., the nodes of the first layer in the network) perform no computation as they do not have any biasing or activation through a transfer function. The output layers generally have a linear activation function with a bias which implies a simple summation of inputs incident to a unit cell with its associated bias. In this investigation, we have used one hidden layer of neurons between the set of inputs and targets with a Tan-Sigmoid activation function which can be expressed as
\begin{align}\label{ANN_3}
  f(a) = \frac{1-e^{-2a}}{1+e^{-2a}} ,
\end{align}
where the transfer function $f$ calculates the neuron's output given its net input. In theory, any differentiable function can qualify as an activation function \citep{zhang1998forecasting}, however, only a small number of functions which are bounded, monotonically increasing and differentiable are used for this purpose.


\subsection{Extreme Learning Machine}

In this section we detail the extreme learning machine (ELM) approach to generalized single-layer feed-forward ANN training. This methodology was proposed in \cite{huang2004extreme} for extremely fast training of a single-layer feed-forward ANN based on the principles of the least-squares approximation. We would like to note that the ELM training approach has previously been utilized for image deblurring but using the principles of classification in a classical deep convolutional network as against the principles of regression utilized here \citep{wang2011image}. For the ease of description, let us define a few matrices for the single-layer feed-forward network. This is a generalization of the architecture introduced in Section~\ref{sec:arch_ann}. Our input matrix is
\begin{align}
    \mathbf{X}^{0} =
    \begin{bmatrix}
    \mathbf{x}^{0}_{1} & \mathbf{x}^{0}_{2} & \hdots & \mathbf{x}^{0}_{n_s}
    \end{bmatrix},
\end{align}
where $\mathbf{x}_{i}^{0}$ is the $i^{th}$ sample (out of a total of ${n_s}$ samples) of a multidimensional input column vector. Our weights connecting the inputs to the middle (hidden) layer and the biases associated with the hidden layer are given by
\begin{align}
    \mathbf{W}^{1} =
    \begin{bmatrix}
    \mathbf{w}^{1}_{1} \\
    \mathbf{w}^{1}_{2} \\
    \vdots      \\
    \mathbf{w}^{1}_{Q}
    \end{bmatrix},\quad
    \mathbf{B}^{1} =
    \begin{bmatrix}
    \mathbf{b}^{1}_{1} \\
    \mathbf{b}^{1}_{2} \\
    \vdots      \\
    \mathbf{b}^{1}_{Q}
    \end{bmatrix} ,
\end{align}
where $\mathbf{w}^{1}_{i}$ and $\mathbf{b}^{1}_{i}$ represent row vectors corresponding to the weights and biases related to the $i^{th}$ neuron (out of a total of $Q$ neurons) in the hidden layer. These are initialized to be small non-zero random numbers to enforce generalization. It is important to remark here that the extreme learning machine methodology prescribes biases \emph{only} for hidden layer neurons (i.e., the output layer biases $\mathbf{B}^2 = 0$). The output of the hidden layer neurons becomes
\begin{align}
    \label{eqH}
    \mathbf{H}^{\intercal} = f(\mathbf{W}^{1} \mathbf{X}^{0}  + \mathbf{B}^{1}) , 
\end{align}
where $f(\mathbf{X})$ implies a Tan-Sigmoid activation procedure on each element of a matrix $\mathbf{X}$ and the superscript $\intercal$ implies a matrix transpose. The weight matrix of the second layer may be given as
\begin{align}
    \mathbf{W}^{2} =
    \begin{bmatrix}
    \mathbf{w}^{2}_{1} & \mathbf{w}^{2}_{2} & \hdots & \mathbf{w}^{2}_{Q}
    \end{bmatrix},
\end{align}
where $\mathbf{w}^{2}_{i}$ now refers to the column vectors related to the hidden layer neurons. Our outputs of the ELM may thus be represented as
\begin{align}
    \mathbf{S}^{2} = \mathbf{W}^{2} \mathbf{H}^{\intercal},
\end{align}
as there is a linear activation in the outer layer with no bias. This output must be trained against a set of targets corresponding to each input vector given by
\begin{align}
    \mathbf{T} =
    \begin{bmatrix}
    \mathbf{t}_1 & \mathbf{t}_2 & \hdots & \mathbf{t}_{n_s}
    \end{bmatrix},
\end{align}
where $\mathbf{t}_{i}$ is a column vector of the targets corresponding to the $i^{th}$ sample. The ELM training mechanism is given as follows. In order to calculate the matrix $\mathbf{W}^{2}$, we must recognize that its optimal solution should satisfy
\begin{align}
    \mathbf{W}^{2}_{op} \mathbf{H}^{\intercal} = \mathbf{T} ,
\end{align}
or by taking a transpose of both sides
\begin{align}
    \mathbf{H} \mathbf{W}^{2^\intercal}_{op} = \mathbf{T}^{\intercal} ,
\end{align}
which leads us to the following expression for the optimal weights
\begin{align}
    \label{Trained}
    \mathbf{W}^{2^\intercal}_{op} = \mathbf{H}^{\dagger} \mathbf{T}^{\intercal} .
\end{align}
The weights $\mathbf{W}^{1}$ and biases $\mathbf{B}^{1}$ are generated initially using small random numbers. The utilization of small random numbers promotes generalization and prevents overfitting \citep{demuth2014neural}. The matrix given by $\mathbf{H}^{\dagger}$ is calculated using a generalized Moore-Penrose pseudoinverse \citep{albert1972regression,serre2002matrices}. Once the optimal weights of the second layer are obtained, our network is trained for deployment. Due to the random number values chosen for the weights in the first layer, our network is well suited to highly effective generalization of the training process due to a smaller degree of freedom of the overall ANN. The proposed direct training procedure substitutes the gradient based iterative optimization algorithms such as the Bayesian regularization method \citep{mackay1992bayesian,foresee1997gauss} in favor of the calculation of a pseudoinverse for the least-squares solution of optimal weights. This gives us the ability to obtain generalized predictive networks through training times which are several orders of magnitude lower in comparison with standard iterative techniques \citep{huang2006extreme}. Algorithms~\ref{Algo1} and \ref{Algo2} describe the training and deployment process for the single-layer ANN trained using ELM. Effectively, Step 4 in Algorithm~\ref{Algo1} represents the computational expense related to our supervised training. 
\begin{algorithm}
  \caption{Extreme Learning Machine: Training}\label{Algo1}
  \begin{algorithmic}[1]
        \State Given $\mathbf{X}^{0} \textnormal{ and } \mathbf{T}$                                               \Comment{Given inputs and targets}
        \State Initialize $\mathbf{W}^{1} \textnormal{ and } \mathbf{B}^{1}$                                      \Comment{Initialize non-zero random parameters}
        \State Calculate $\mathbf{H}^\intercal = f(\mathbf{W}^{1} \mathbf{X}^{0}  + \mathbf{B}^{1})$              \Comment{Tan-Sigmoid activation}
        \State Calculate pseudoinverse $\mathbf{H}^\dagger$                                                       \Comment{Moore-Penrose pseudoinverse}
        \State Calculate $\mathbf{W}^{2^\intercal}_{op} = \mathbf{H}^{\dagger} \mathbf{T}^{\intercal}$             \Comment{Least-squares solution for optimal weights}
  \end{algorithmic}
\end{algorithm}
\begin{algorithm}
  \caption{Extreme Learning Machine: Deployment}\label{Algo2}
  \begin{algorithmic}[1]
        \State Given $\mathbf{X}_{test}^0$                                                                  \Comment{Given testing data}
        \State Given $\mathbf{W}^1, \mathbf{B}^1 \textnormal{ and } \mathbf{W}^{2}_{op}$          \Comment{Given trained network from Algorithm 1}
        \State Obtain $\mathbf{S}^1 = \mathbf{W}^1 \mathbf{X}^0_{test}$                                               \Comment{Calculate input at Layer 1 neurons}
        \State Obtain $\mathbf{X}^1 = f(\mathbf{S}^1 + \mathbf{B}^1)$                                               \Comment{Activate to obtain output of Layer 1 neurons}
        \State Obtain $\mathbf{X}^2 = \mathbf{W}^2_{op} \mathbf{X}^1$                                   \Comment{Obtain output of the second layer: prediction}
  \end{algorithmic}
\end{algorithm}

\section{Blind Deconvolution}

The primary motivation for the development of the aforementioned training and deployment procedures for our ANN architecture is described in this section. We are motivated by the approximate deconvolution framework \citep{stolz1999approximate} which uses an iterative resubstitution procedure known as the Van Cittert iterations for reconstructing the contribution of subfilter scale content. This procedure may be represented tensorially as:
\begin{align}\label{3D_Stress_0}
  \tau_{ij} =   \bar{u}_i \bar{u}_j - \overline{u_i u_j} ,
\end{align}
where we obtain an approximation of these subfilter stresses as \citep{germano2015similarity}
\begin{align}\label{3D_Stress_1}
  \tau_{ij} =  \bar{u}_i \bar{u}_j - \overline{u_i^{*} u_j^{*}} ,
\end{align}
in which the asterisk superscript indicates an approximately deconvolved variable. An important caveat of this mechanism is a user-defined low-pass spatial filtering kernel. While the general approach is to implement various versions of Gaussian blur kernel, it is quite possible that complex flow configurations may actually exhibit natural filter shapes which are much more contorted. Fortunately, the vast wealth of data that may be collected by modern experimental techniques represents an opportunity for us to bypass this user-defined heuristic stage. In essence, blind deconvolution refers to the estimation of the underlying blur kernel without any knowledge or closed form estimate of its true nature. We must emphasize here that while there exist several algorithms that are \emph{truly} blind in nature (for instance in \cite{dabov2007color}), this investigation proposes a framework which is purely data-driven. To summarize, our data may be considered to be an aid in implicitly estimating the shape of the blur kernel through the form of a trained neural network.

Our supervised learning framework requires training data which encapsulates the relationships between the coarsened true data and its perturbed versions. Note that, as is common in data-driven strategies, we normalize all our data to a range between -1 and 1. This choice of the normalized data range is due to the fact that the Tan-Sigmoid activation function provides outputs between these limits as well. The perturbation may be introduced through a low-pass spatial filtering procedure:
\begin{align}\label{filter0}
  u'_i = G(\mathbf{x}; \sigma) u_i ,
\end{align}
which utilizes the following Gaussian kernel with standard deviation $\sigma$ (held at a default value of 1.0):
\begin{align}\label{filter}
  G(\mathbf{x}; \sigma) = \frac{1}{(\sqrt{2\pi}\sigma)^d} \exp(-\frac{\left| \mathbf{x}\right|^2}{2 \sigma^2}) ,
\end{align}
where $d$ is the number of dimensions of the field being blurred. The use of a low-pass spatially filtered field tests the ability of the proposed ANN architecture for inertial range recovery and becomes a true test of the blind deconvolution ability of a neural network.

\begin{figure}
\centering
\includegraphics[width=\textwidth]{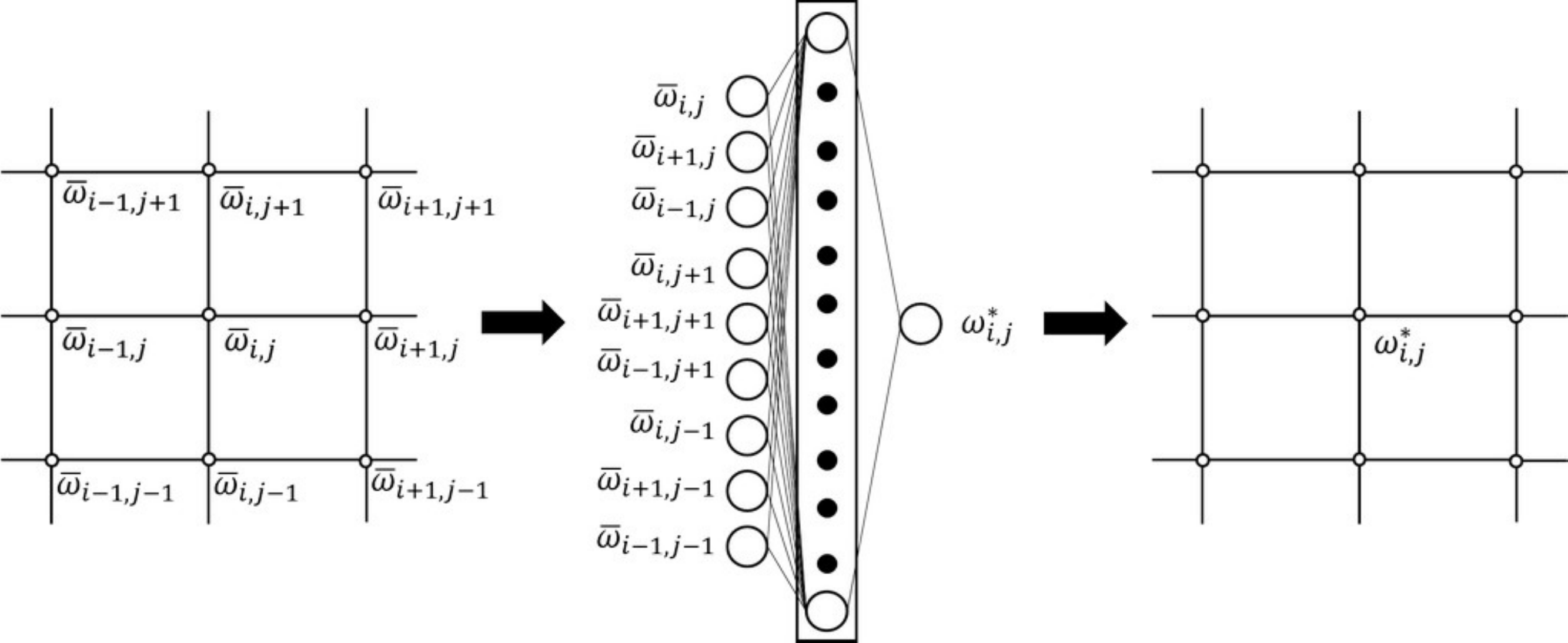}
\caption{A schematic of the proposed blind deconvolution strategy for a two-dimensional test case where we aim to recover $\omega^{*}$ from $\bar{\omega}$. Note that each hidden neuron is associated with a unique bias.}
\label{BC_Schem}
\end{figure}

Another perturbation which is examined in this investigation is the addition of a normally distributed noise with an amplitude on the order of a tenth of the maximum value of the field variable which can be expressed as:
\begin{align}\label{noise}
  u'_i = u_i + \mu \kappa ,
\end{align}
where $\kappa$ is a normally distributed random number between -1 and 1 and the coefficient $\mu$ (with a default value of 0.2) corresponds to the magnitude of the noise added to the field. The purpose of training relationships between inputs with noisy perturbations and the true solution is to test the stability of this approach for high wavenumber energy accumulation. This is particularly important for coarse-grained large eddy simulations which are susceptible to numerical overflow. Thus, our supervised learning framework utilizes $u_i'$ as our training inputs and $u_i$ are our targets. Note that we train separate networks for each type of perturbation. The ELM training mechanism leads to tractable training times even for cases with several hundred neurons in the hidden layer. The results obtained in this study have employed 100 neurons for training and testing data obtained from different realizations of field values. A schematic of the blind deconvolution methodology integrated into the one-layer ELM training based ANN is shown for a two-dimensional test case in Figure~\ref{BC_Schem}.

\section{Cross-Validation}

An important assessment of a data-driven modeling approach is a cross-validation procedure to estimate the performance of the model for situations which it has not been trained for. In other words, it is necessary to ascertain that the data-driven model is \emph{not} localized to the training data and provides a similar performance for different solution fields exhibiting similar physics but different numerical arrangements or magnitudes. Therefore, it is vital to \emph{test} the performance of the trained ANN on a set of data which it has not seen previously (during training). For the purpose of data generation, we utilize high-fidelity simulations for our previously mentioned test cases so that their flow physics may be accurately resolved. Our 3D numerical experiments are generated using $512^3$ degrees of freedom while our 2D case is generated using $2048^2$ grid points. A coarse-grained large eddy simulation is mimicked through the subsequent subsampling of these high quality data sets. Our coarse-graining procedure leaves us with a subsampled field of $64^3$ and $256^2$ degrees of freedom in the 3D and 2D test cases, respectively. 
Therefore, our coarse-graining procedure involves the selection of every eighth grid point in the high-fidelity uniform grid data.

\begin{figure}
\centering
\includegraphics[width=0.8\textwidth]{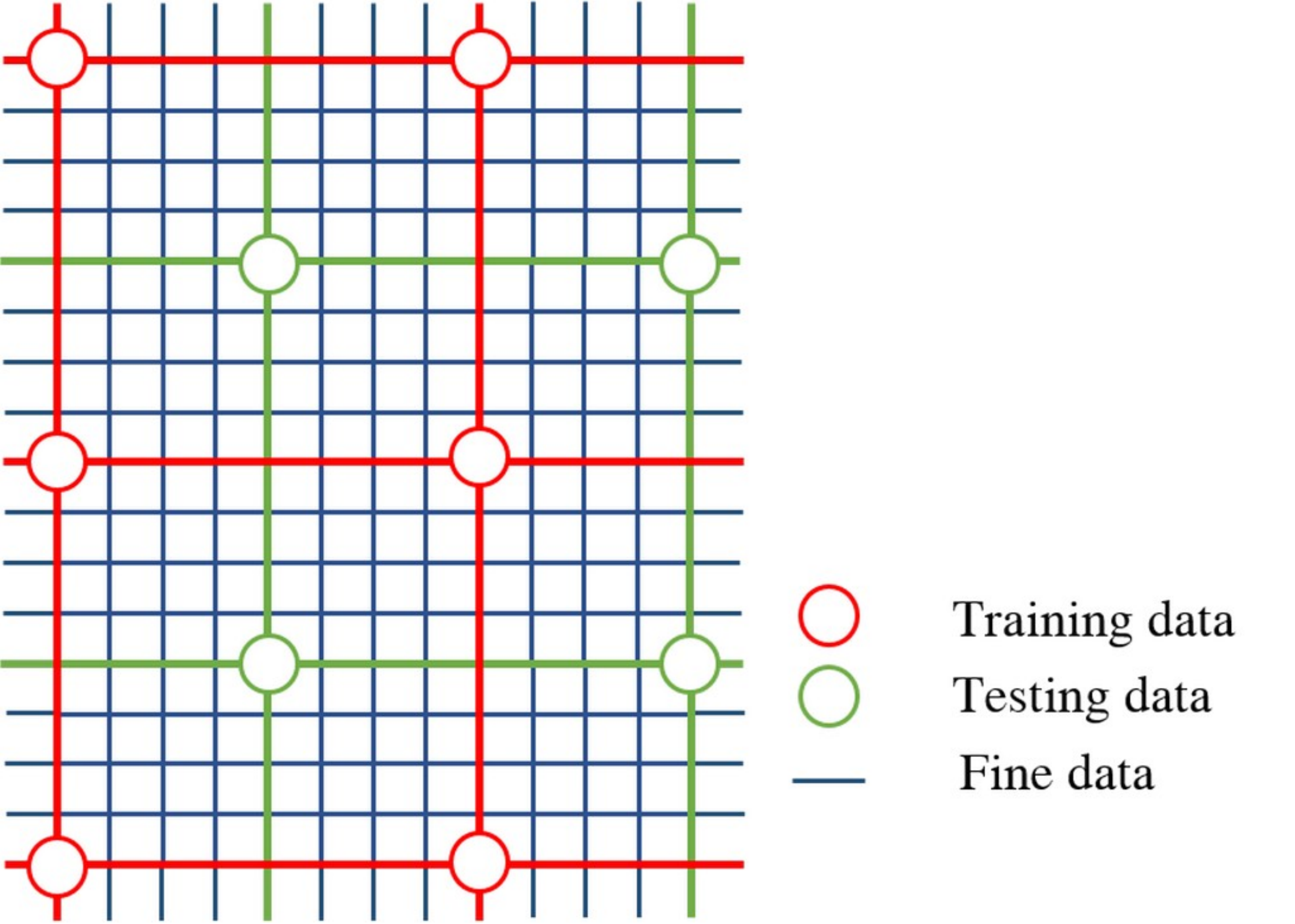}
\caption{A schematic of the spatial shifting strategy for a simplified two-dimensional grid showing two different data sets. Several different data sets may be constructed using this technique which are numerically different physically similar.}
\label{shifting_schem}
\end{figure}

We may devise several shifted data sets (each possessing $64^3$ degrees of freedom in 3D and $256^2$ degrees of freedom in 2D) but which continue to represent the physics of the fine-grained dataset. Missing data points at boundaries may be reconstructed through the use of the periodic boundary conditions for the given test cases. A simplified shifting schematic is shown in Figure~\ref{shifting_schem} where a simple technique of generating two coarse data sets from fine data is demonstrated. In a nutshell, the coarse-graining and shifting procedures allow us to devise upto 63 and 511 completely different data sets in two and three dimensions, respectively. For the purpose of cross-validation we randomly choose any four of these multiple data sets for the generation of three sets of testing data and one set of training data.

\begin{figure}
\centering
\includegraphics[width=0.8\textwidth]{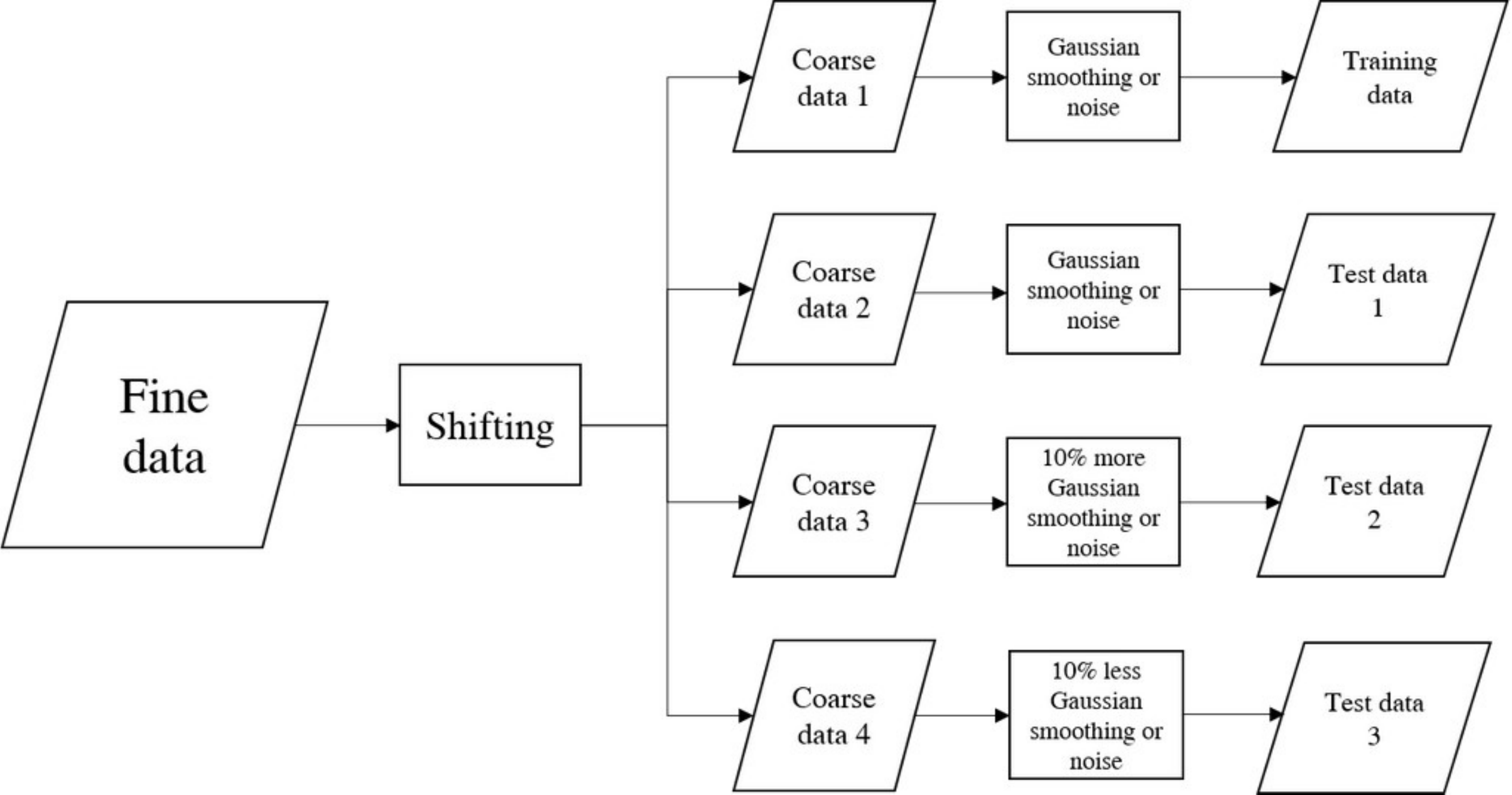}
\caption{A flow chart detailing the generation of cross-validation data sets using the shifting operation as well as different perturbations. The network is trained using Algorithm~\ref{Algo1} and is then tested on test data 1, 2 and 3.}
\label{cv_schem}
\end{figure}

Once our four different randomly generated data sets are identified, we add perturbations corresponding to the type of behavior demanded of proposed artificial neural network. To test the deconvolution ability of our network, our training data (i.e., one of the four data sets) is filtered with an appropriate Gaussian smoothing (for our inputs to the network) and unfiltered training data are utilized as outputs to the network. Three testing data sets are generated in a similar manner with one of these data being filtered with the same filter radius and the others being filtered with a 10\% larger and 10\% smaller filter radius. The trained ANN is then utilized to recover deconvolved approximations to the true field for these three test data sets. This ensures that the trends of the trained network are not due to overfitting or `data-memory' but through an implicit learning of the inverse filtering. A similar procedure is also utilized to cross-validate the regularization ability of the closure wherein the randomly chosen data sets are perturbed through Gaussian noise. A table describing the filter radii and magnitudes of noise for our training and testing data sets is shown in Table~\ref{table:CV1}. Note that these data sets are all generated from the same high fidelity solution field and correspond to a perfectly a-priori analysis. A concise summary of this cross-validation procedure is given in a flow chart in Figure~\ref{cv_schem}.


\begin{table}
\begin{center}
 \begin{tabular}{p{3cm} p{3cm}}
 \multicolumn{2}{c}{\underline{\textbf{Deconvolution}}} \\
 \underline{Data set} & \underline{Filter radius ($\sigma$)} \\
 Training data & 1.0 \\
 Test data 1 & 1.0 \\
 Test data 2 & 1.1 \\
 Test data 3 & 0.9 \\
 \end{tabular}
 \begin{tabular}{p{3cm} p{3cm}}
 \multicolumn{2}{c}{\textbf{\underline{Regularization}}} \\
 \underline{Data set} & \underline{Noise ($\mu$)} \\
 Training data & 0.2 \\
 Test data 1 & 0.2 \\
 Test data 2 & 0.22 \\
 Test data 3 & 0.18 \\
\end{tabular}
\end{center}
\caption{Cross-validation data sets for the proposed data-driven blind deconvolution closure.}
\label{table:CV1}
\end{table}


In addition to the a-priori cross-validation outlined above, we also assess our proposed architecture by utilizing training and testing data across different flow physics. This procedure is carried out to test the universal nature of the learning for the purpose of both deconvolution and regularization abilities. This may be assumed to be another challenging tier for cross-validation. We elaborate this step in further detail in Section~\ref{sec:Univ}.

\section{Results}

In this section, we outline the results of our proposed framework for three different benchmark flows showing distinct differences in physics. Our test cases are given by a two-dimensional homogeneous isotropic turbulent flow which exhibits a cascade of enstrophy from its integral to dissipative length scales, a canonical Taylor-Green decaying vortex corresponding to homogeneous isotropic turbulent flow with a cascade of kinetic energy from integral to dissipative length scales and an inviscid three-dimensional turbulent flow developed from stratified Kelvin-Helmholtz instability.

\subsection{Kraichnan Turbulence}

A periodic square domain is used to simulate a canonical flow configuration which exhibits the characteristics of decaying, homogeneous and isotropic 2D turbulence \citep{san2012high}. We utilize $2048^2$ degrees of freedom for high fidelity data which is coarsened to $256^2$ grid points to mimic a considerably coarser large eddy simulation. A Reynolds number of $Re=32,000$ is utilized to ensure a considerable scale separation and to display a prominent inertial range with its associated $k^{-3}$ Kraichnan scaling \citep{kraichnan1967inertial}. The 2D turbulence simulation is undertaken through the implementation of the Navier-Stokes equations in their vorticity-streamfunction formulation. The results shown here are obtained through the data-driven blind deconvolution of the vorticity. Our ANN closure was implemented by training a relationship between a nine point stencil consisting of the point at which the deconvolution is desired along with its immediate neighbors (i.e., see Figure~\ref{BC_Schem}). Training times were (on average) of the order of 0.01 seconds for approximately 65,000 data samples.

\begin{figure}
\centering
\mbox{
\subfigure[center][Test data 1]{
\includegraphics[width=0.44\textwidth]{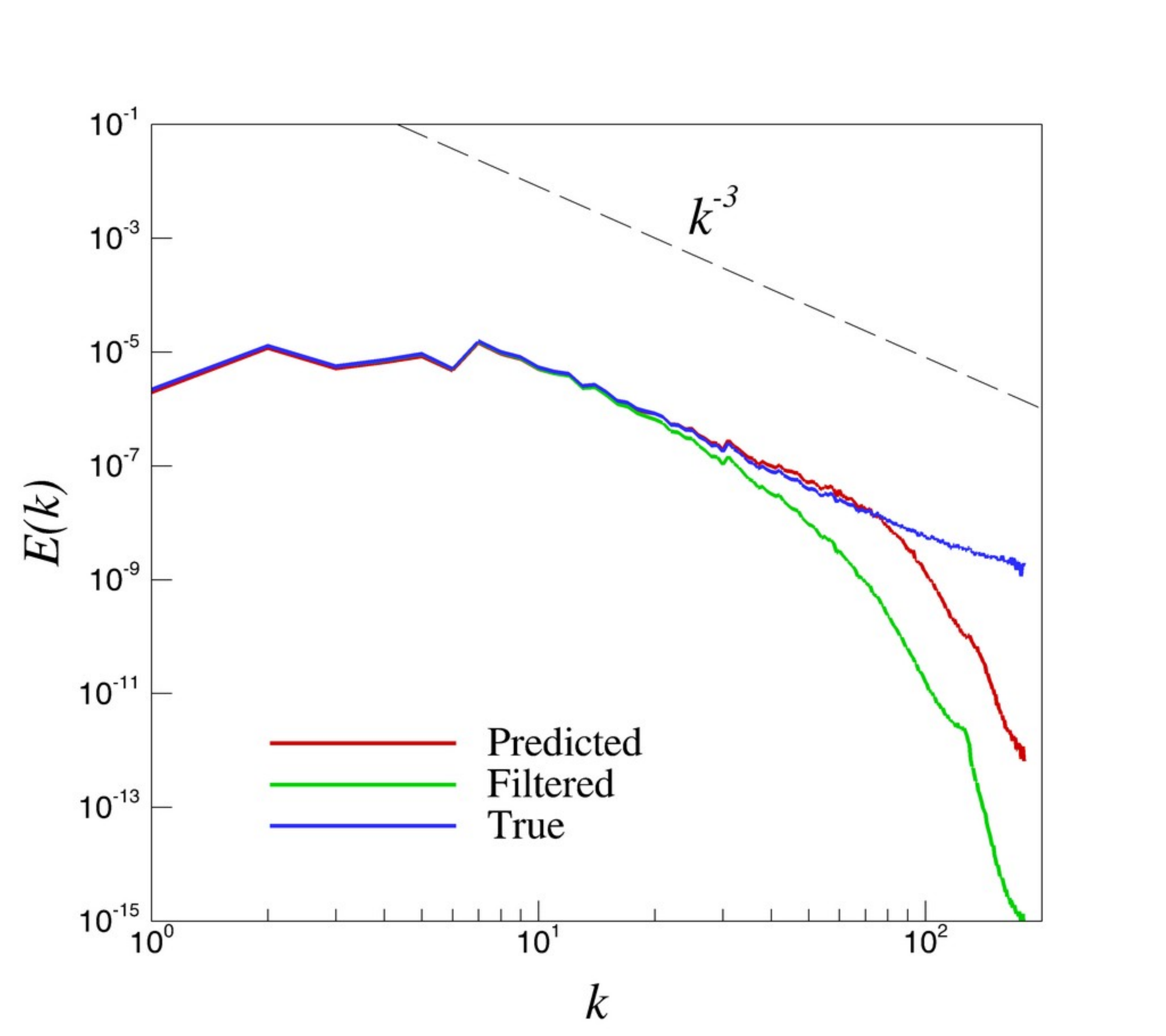}
\includegraphics[width=0.44\textwidth]{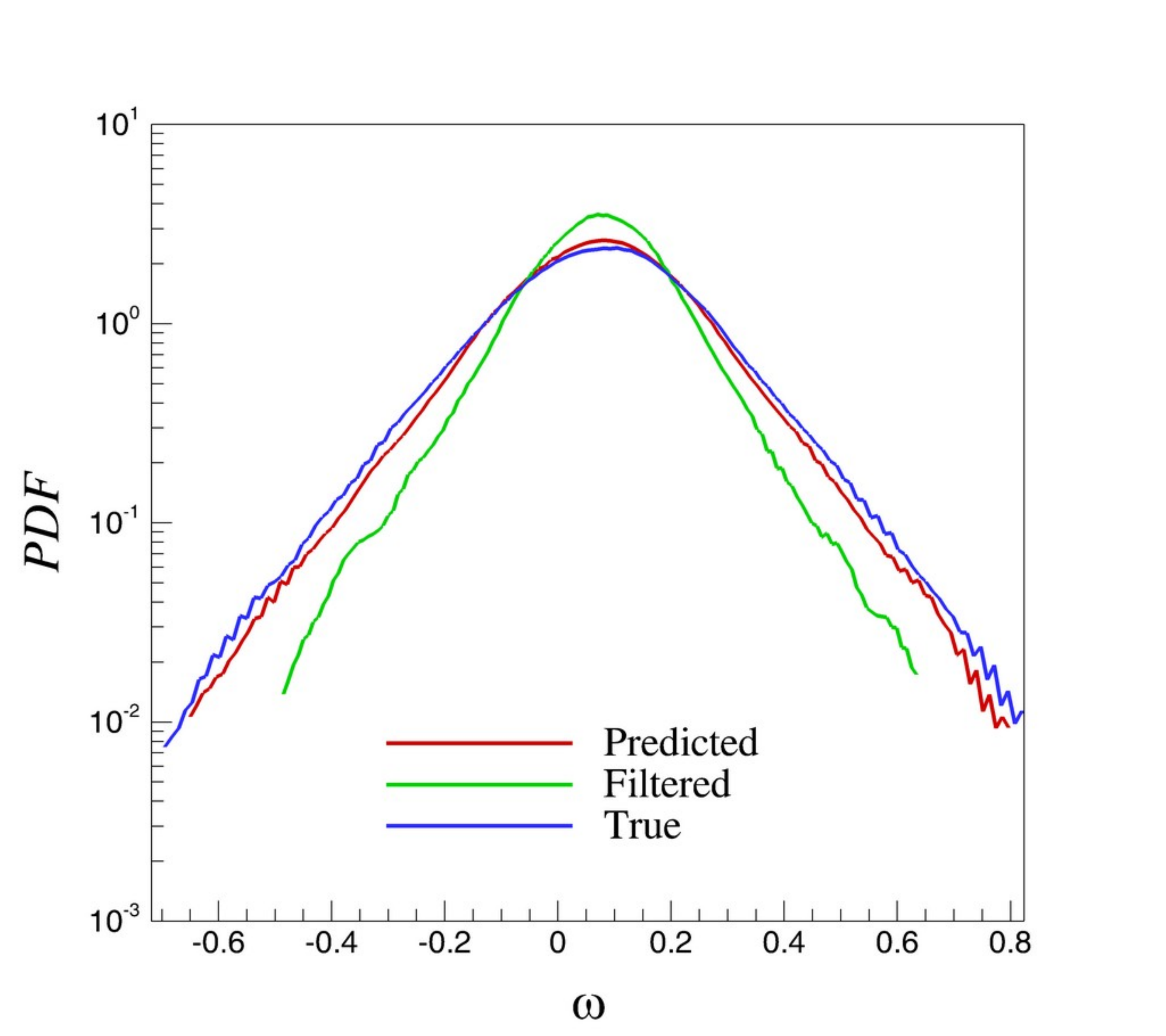}
}
}\\
\mbox{
\subfigure[Test data 2]{
\includegraphics[width=0.44\textwidth]{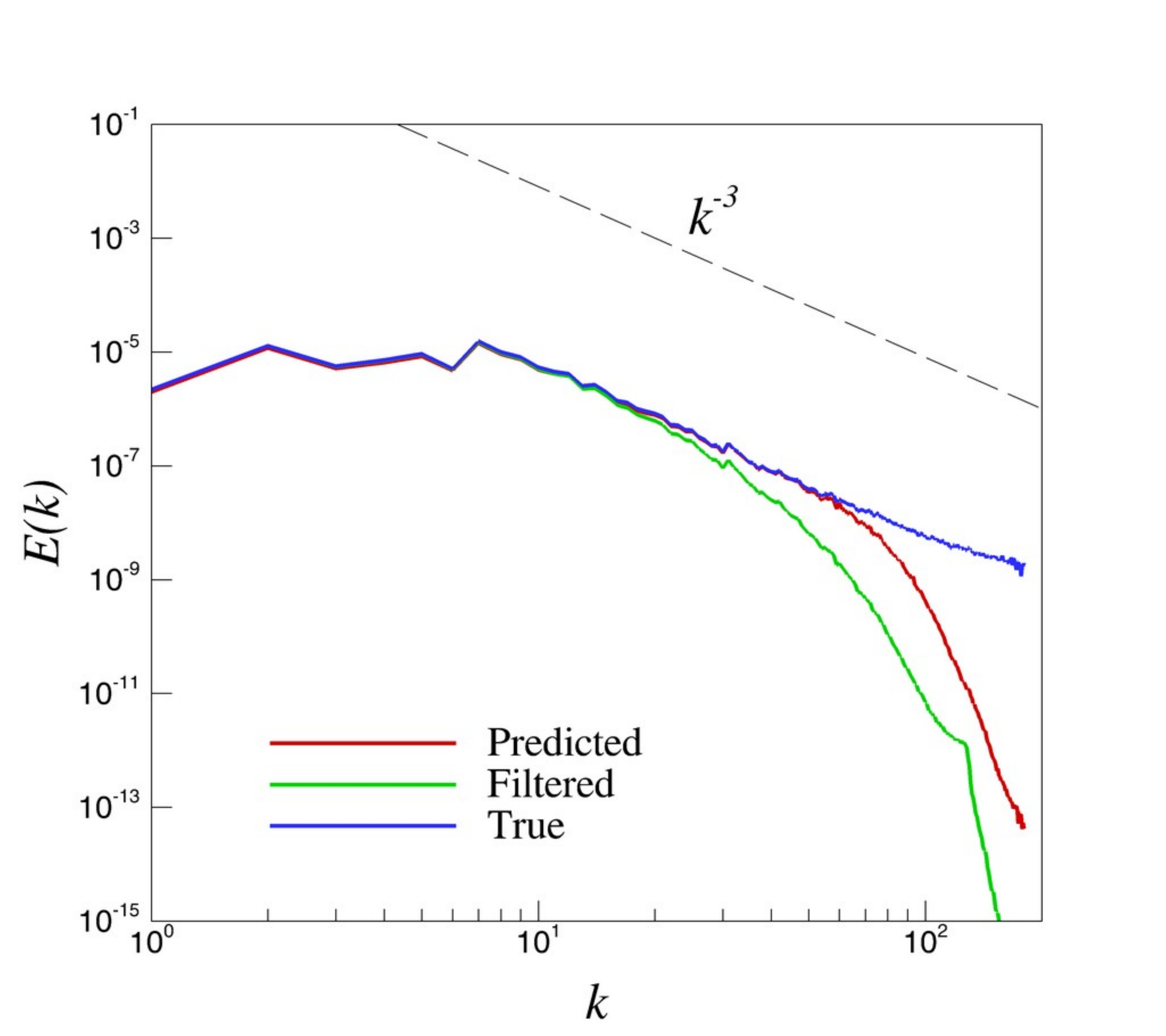}
\includegraphics[width=0.44\textwidth]{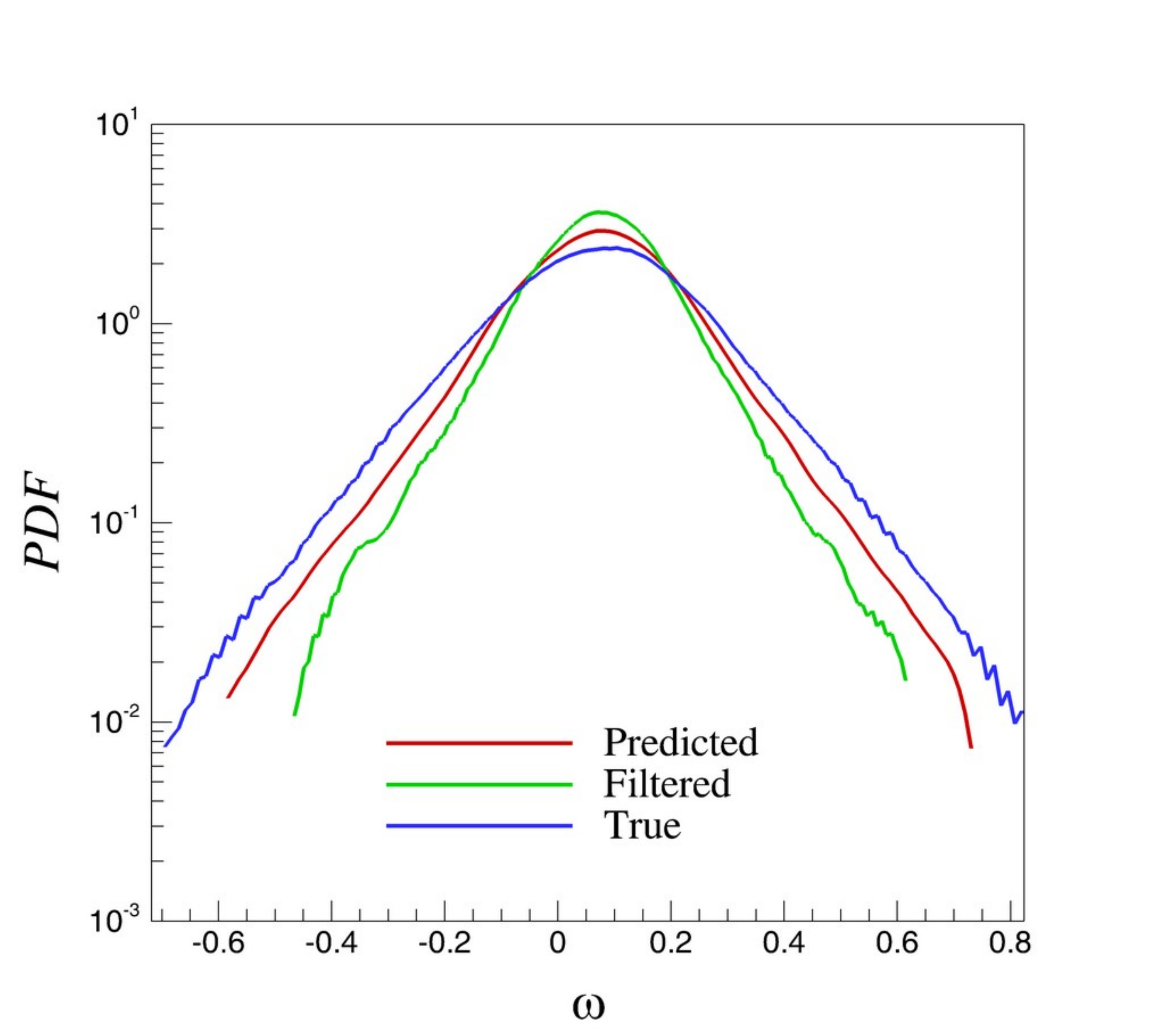}
}
}\\
\mbox{
\subfigure[Test data 3]{
\includegraphics[width=0.44\textwidth]{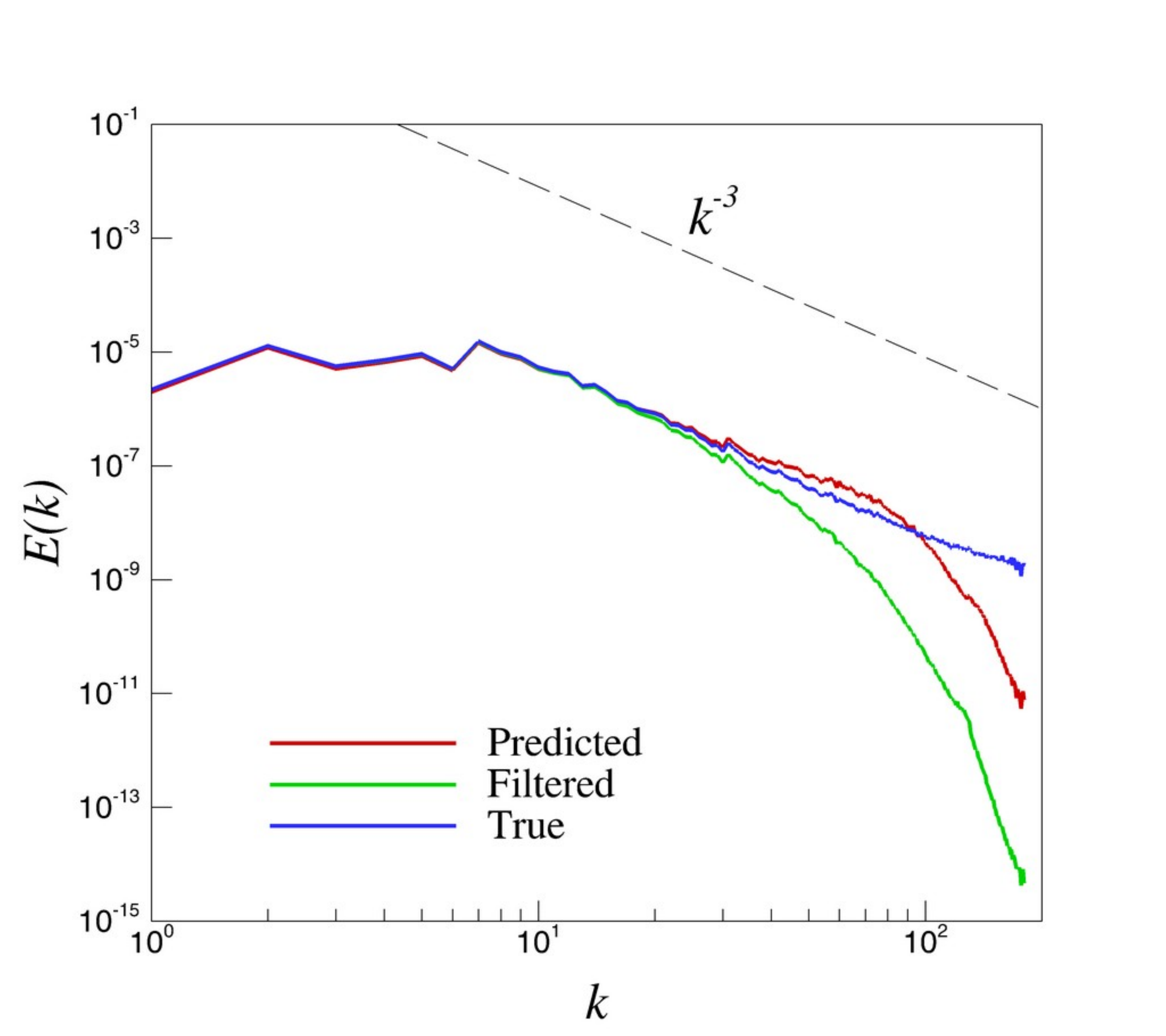}
\includegraphics[width=0.44\textwidth]{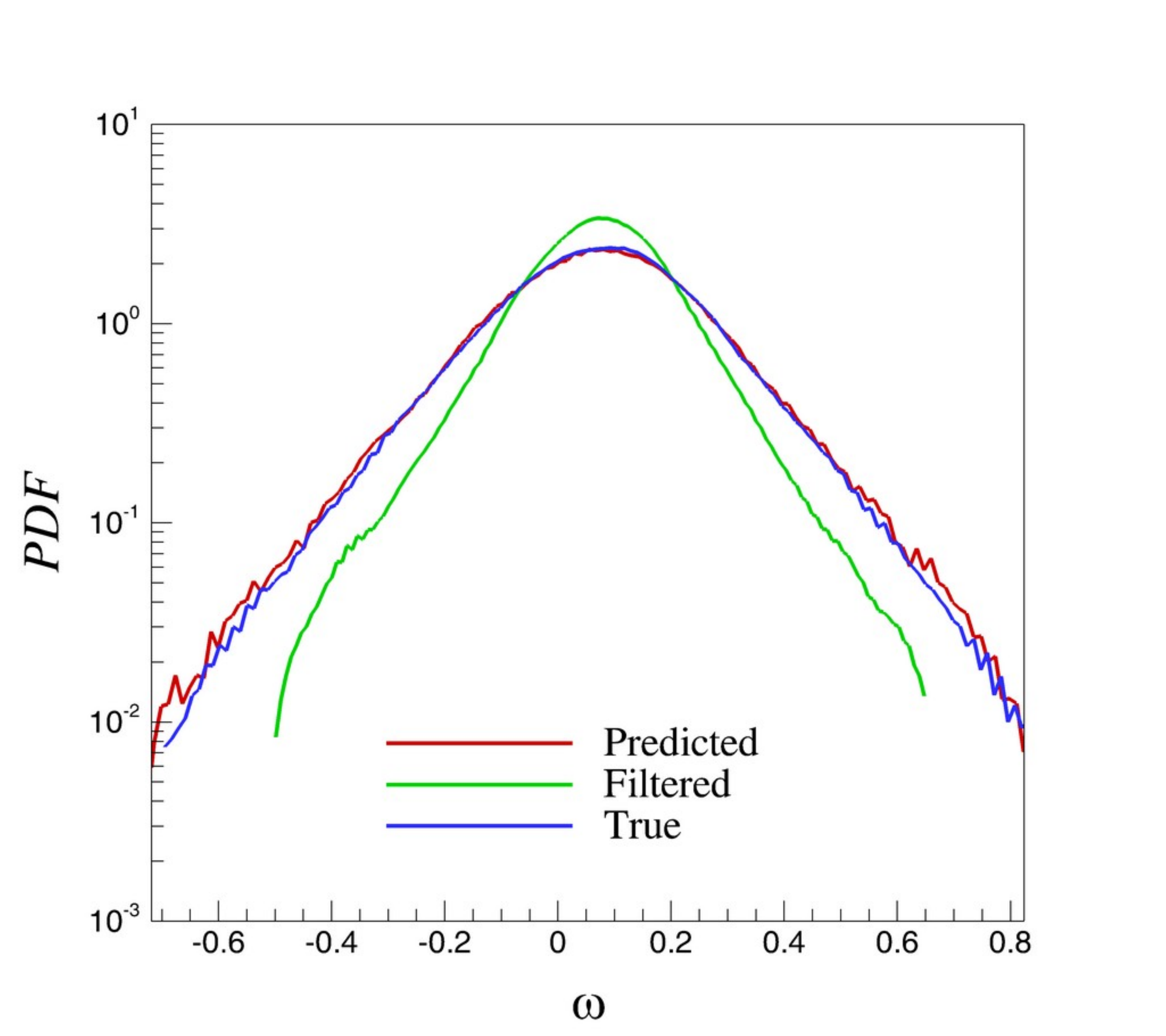}
}
}
\caption{A-priori results of the kinetic energy spectra (left) and PDF of the vorticity (right) for Kraichnan turbulence. Results for three different deconvolution test data sets shown. }
\label{fig:Spec2D_Filtered}
\end{figure}

\begin{figure}
\centering
\mbox{
\subfigure[Test data 1]{
\includegraphics[width=0.44\textwidth]{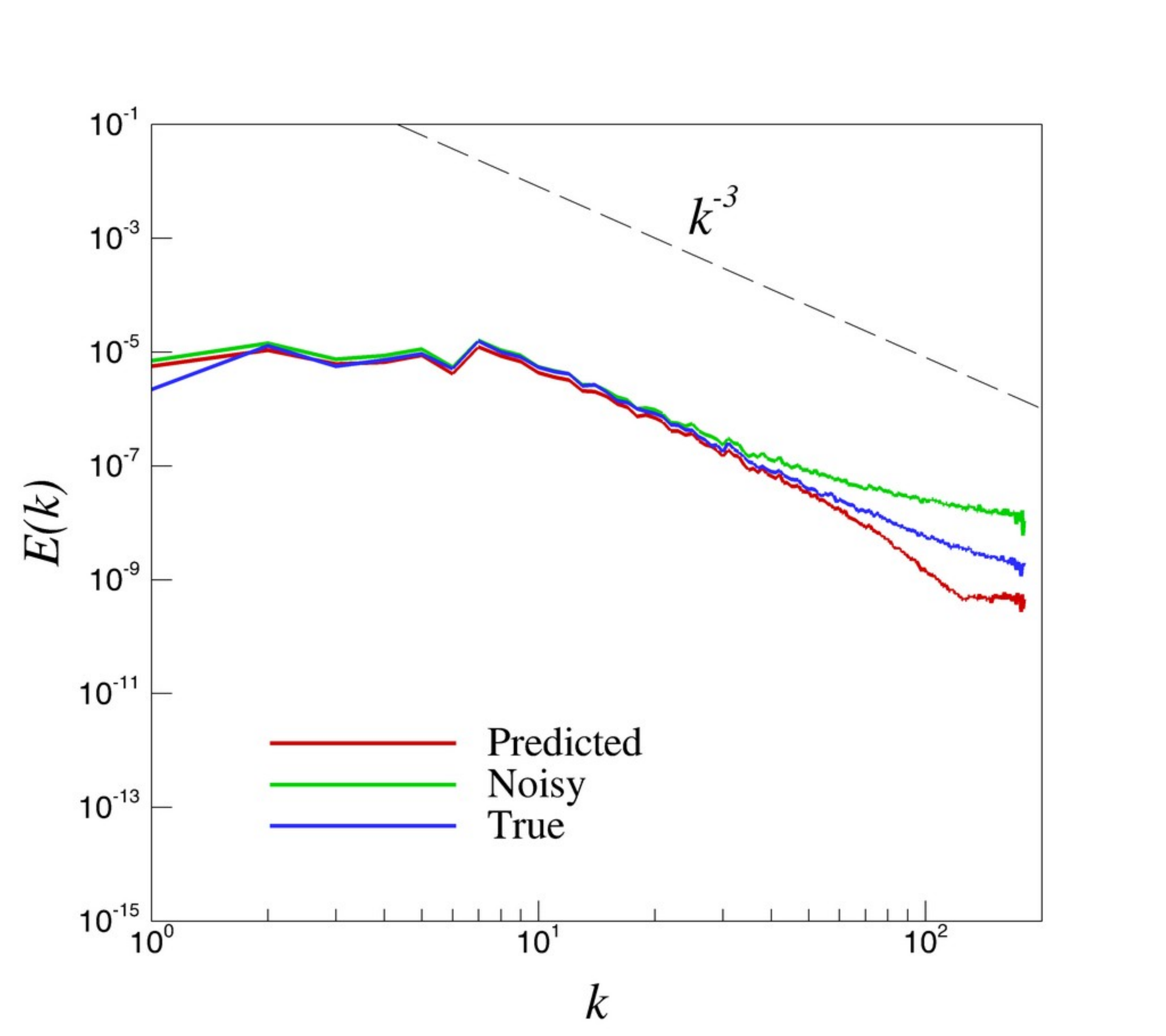}
\includegraphics[width=0.44\textwidth]{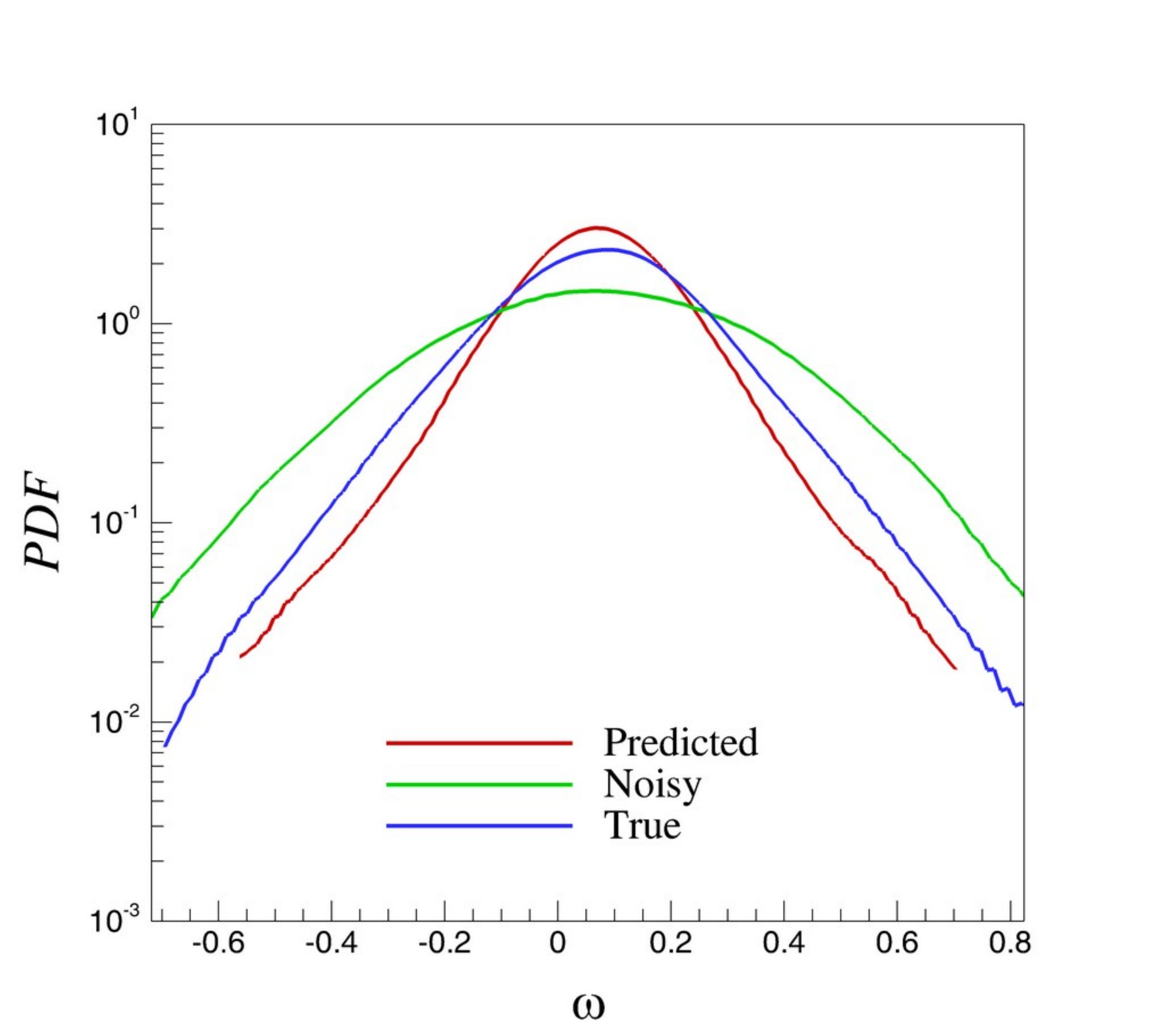}
}
}\\
\mbox{
\subfigure[Test data 2]{
\includegraphics[width=0.44\textwidth]{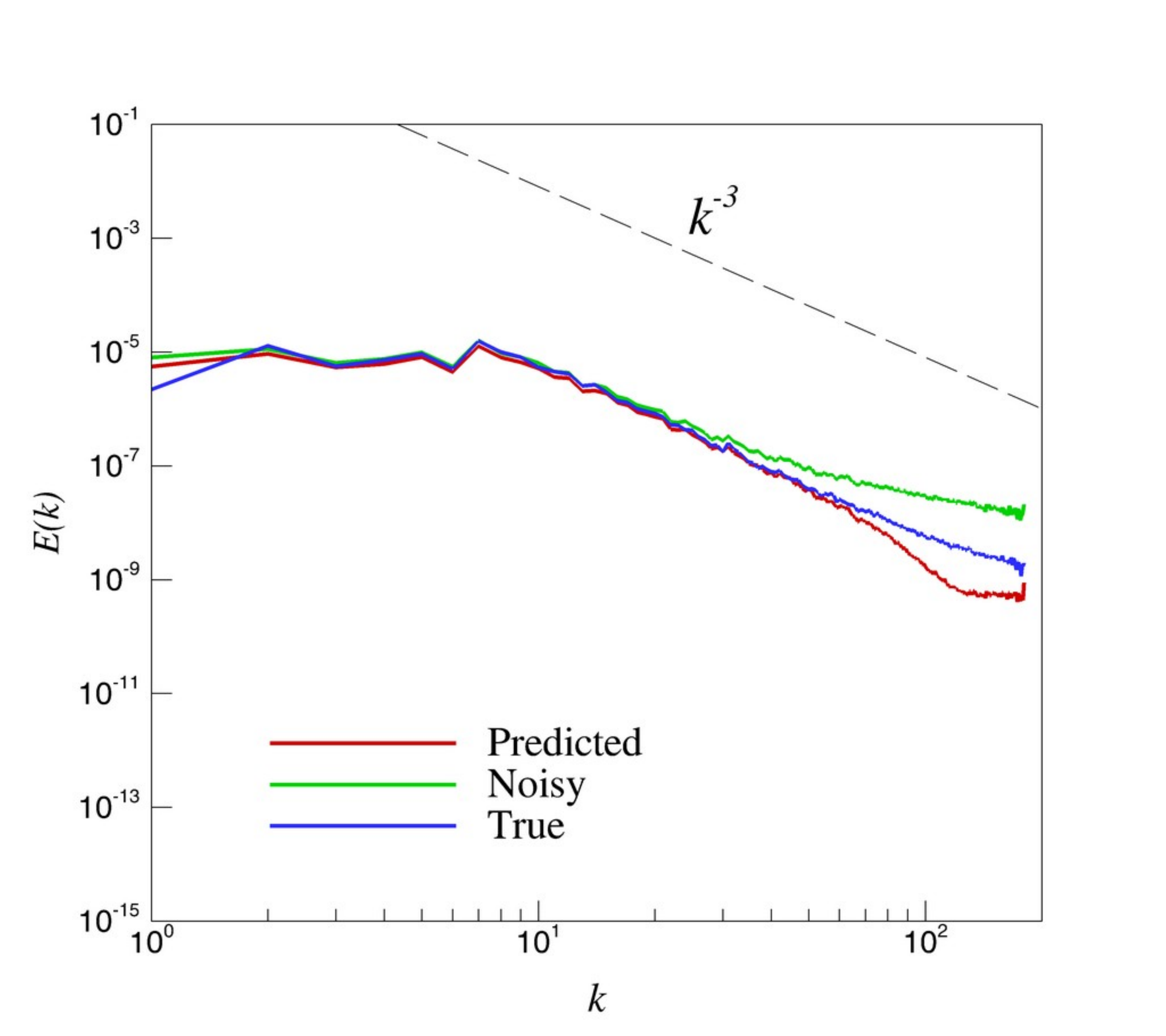}
\includegraphics[width=0.44\textwidth]{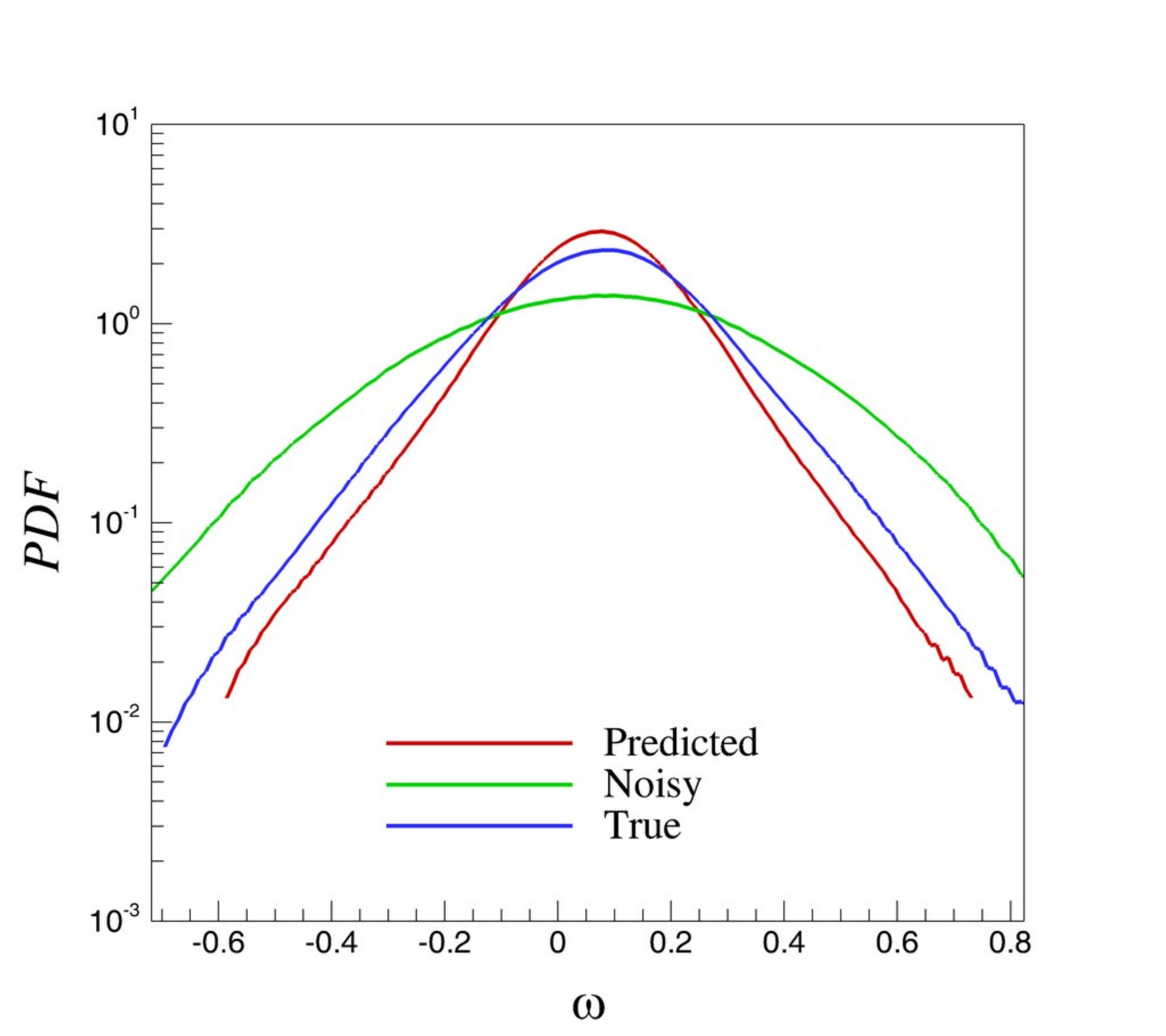}
}
}\\
\mbox{
\subfigure[Test data 3]{
\includegraphics[width=0.44\textwidth]{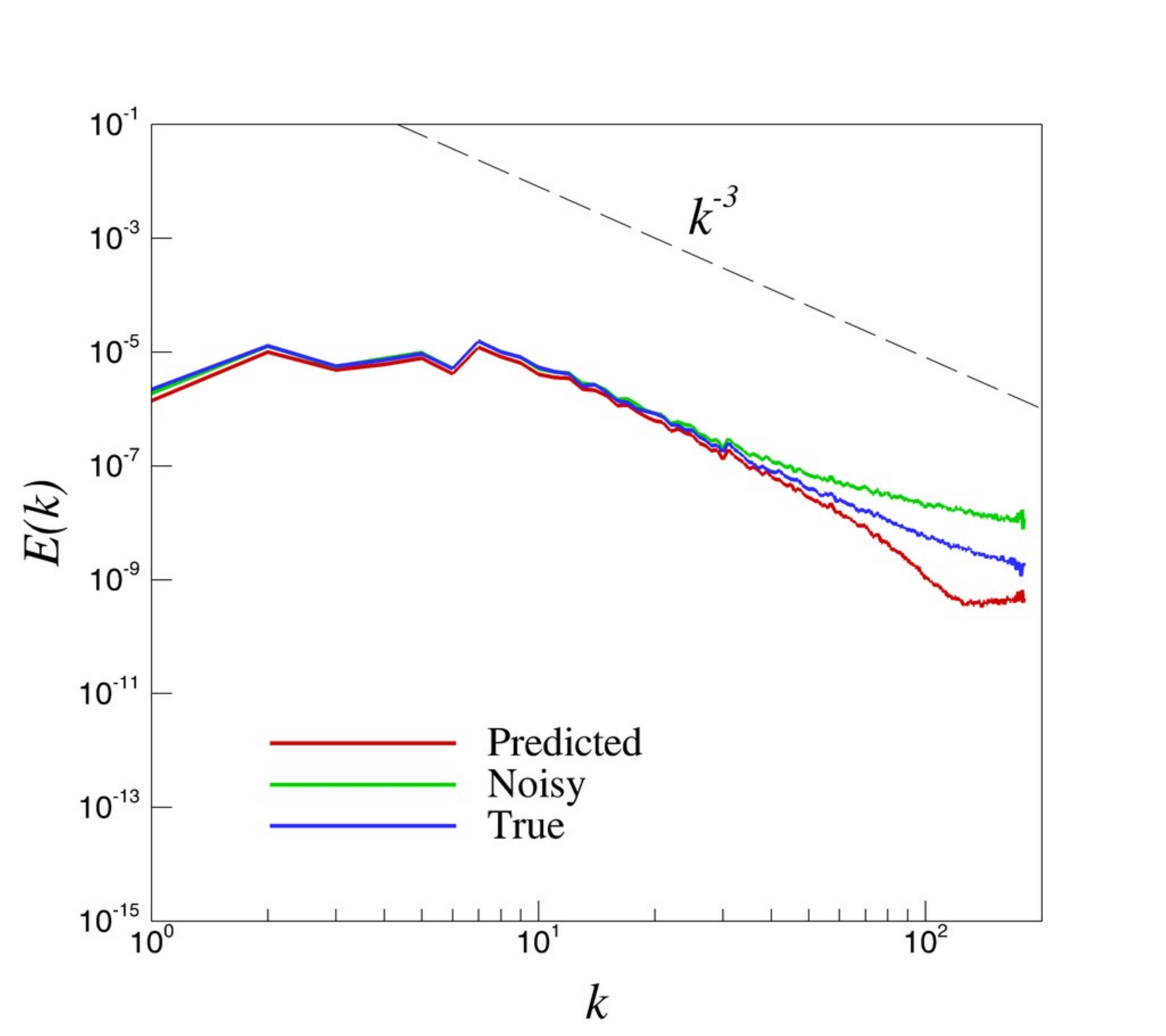}
\includegraphics[width=0.44\textwidth]{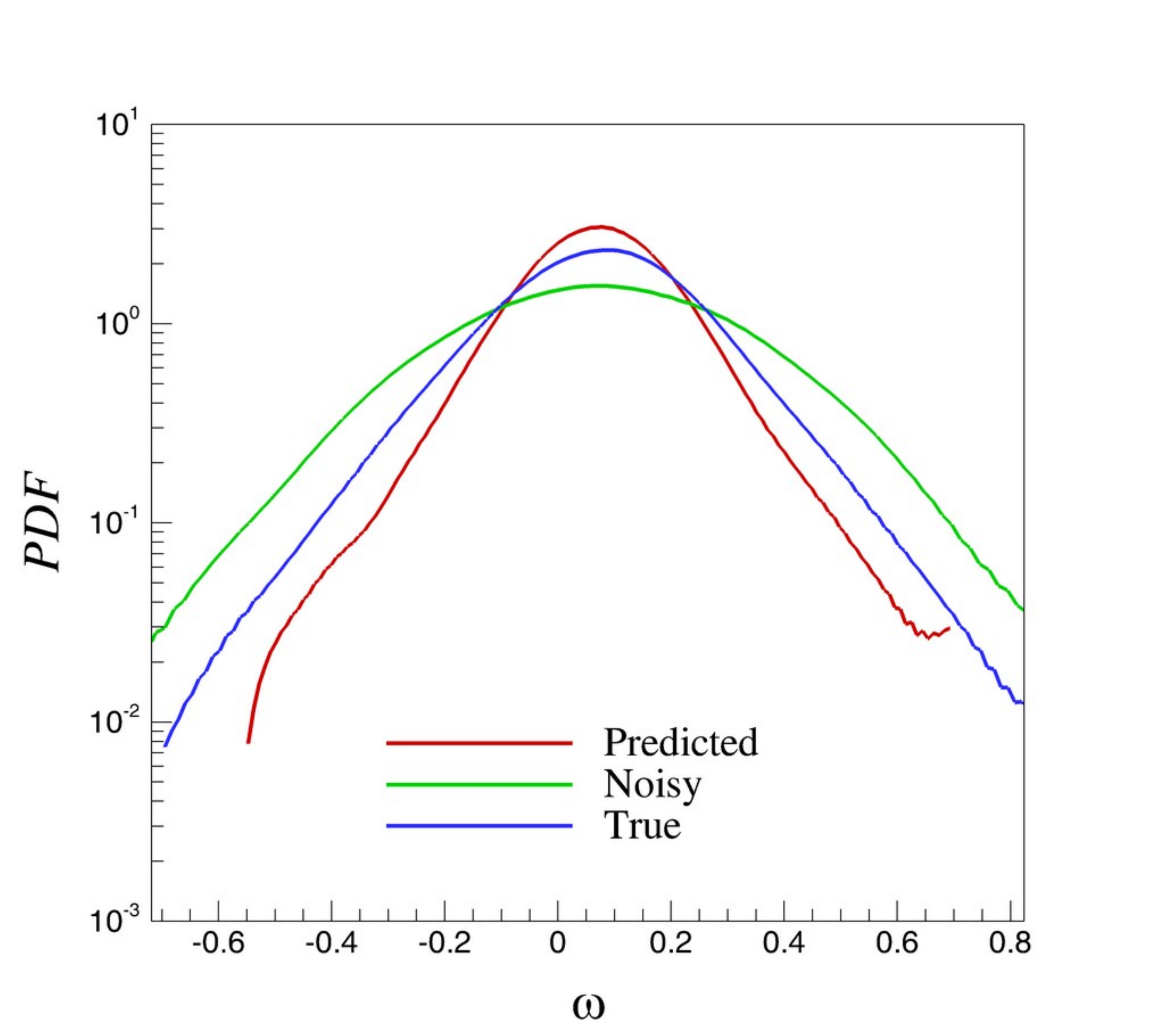}
}
}
\caption{A-priori results of the kinetic energy spectra (left) and PDF of the vorticity (right) for Kraichnan turbulence. Results for three different regularization test data sets shown.}
\label{fig:Spec2D_Noised}
\end{figure}

Figure~\ref{fig:Spec2D_Filtered} shows the performance of the proposed methodology in terms of statistical assessments given by angle averaged kinetic energy spectra and probability density functions (PDF) of the vorticity field when the deconvolution aspect of the proposed framework is tested. For subfilter scale reconstruction assessment, it can readily be observed that the proposed architecture manages to capture a far greater region of the inertial range in accordance with the $k^{-3}$ scaling. This may also be observed from the PDF comparisons where the narrow distribution caused by the low-pass spatial filtering is successfully flattened to its original spread. A similar performance is exhibited when the testing data exhibits slightly different physics. As mentioned previously, this is simulated through the utilization of slightly larger and smaller filter radii to mimic physical behavior in a local neighboring range around the training data. The proposed framework successfully displays a similar performance for all the testing data versions thereby demonstrating a good generalization. Field reconstruction attempts from noisy data display marginal benefits in terms of inertial range capture (as shown in Figure~\ref{fig:Spec2D_Noised}) but the regularizing nature of the ELM training ensures that the cut-off length scale pile up of energy (or aliasing error) is successfully stabilized. This behavior is also replicated for different testing data. As mentioned previously, this is very promising for coarse-grained large eddy simulations particularly for non-conservative underlying numerical schemes.

\begin{figure}
\centering
\mbox{
\subfigure[True]{\includegraphics[width=0.3\textwidth]{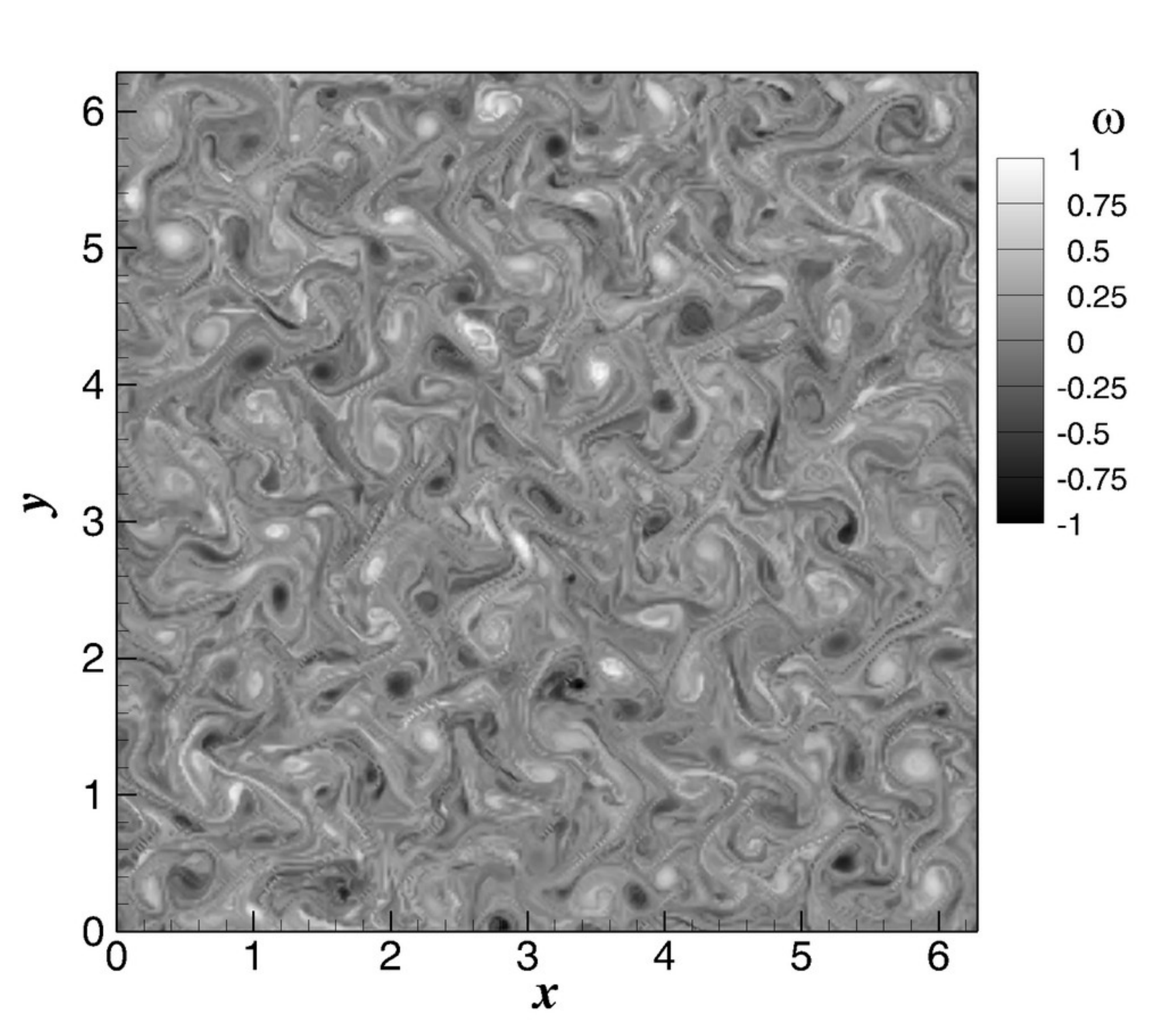}}
\subfigure[Filtered]{\includegraphics[width=0.3\textwidth]{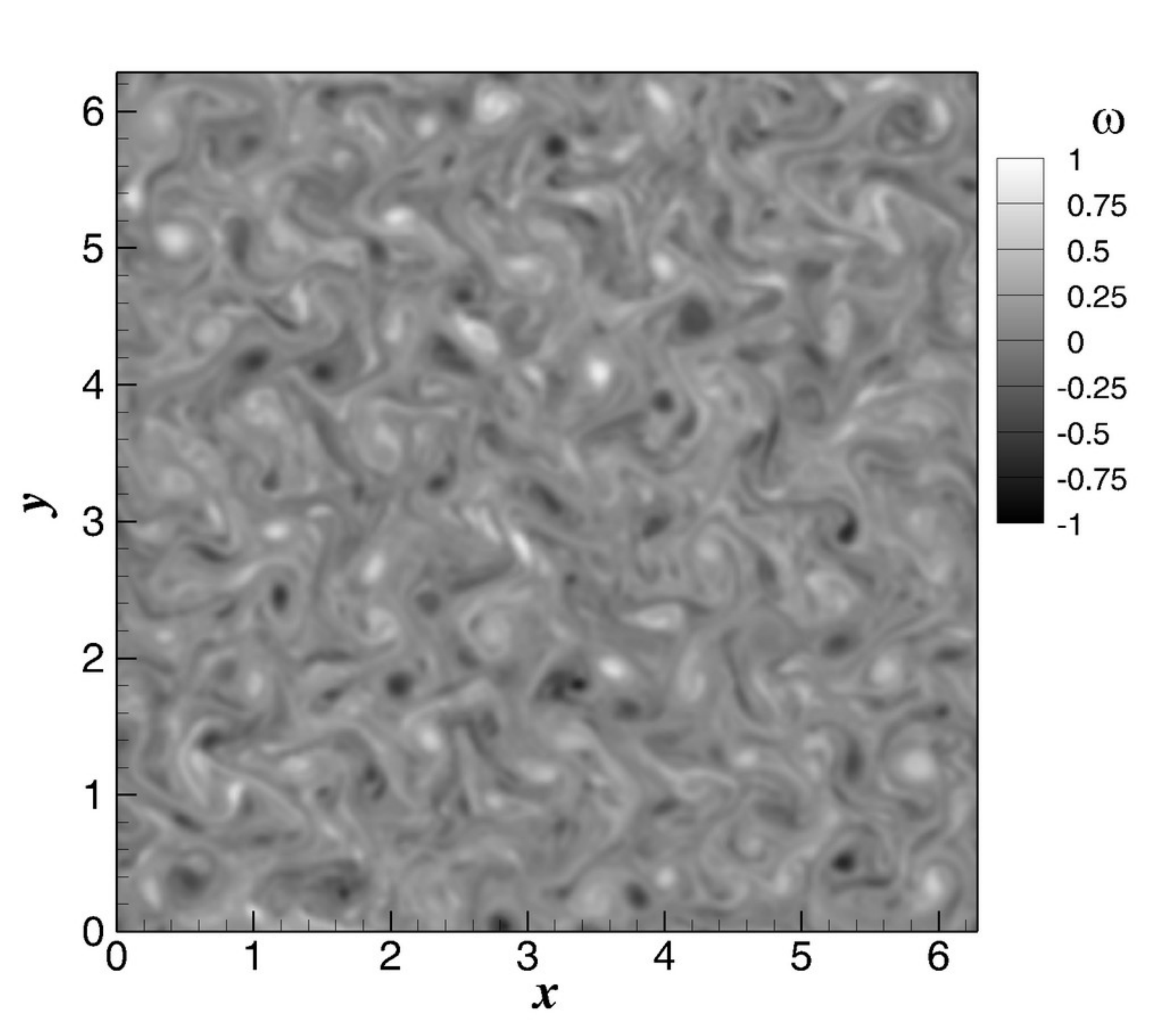}}
\subfigure[Recovered]{\includegraphics[width=0.3\textwidth]{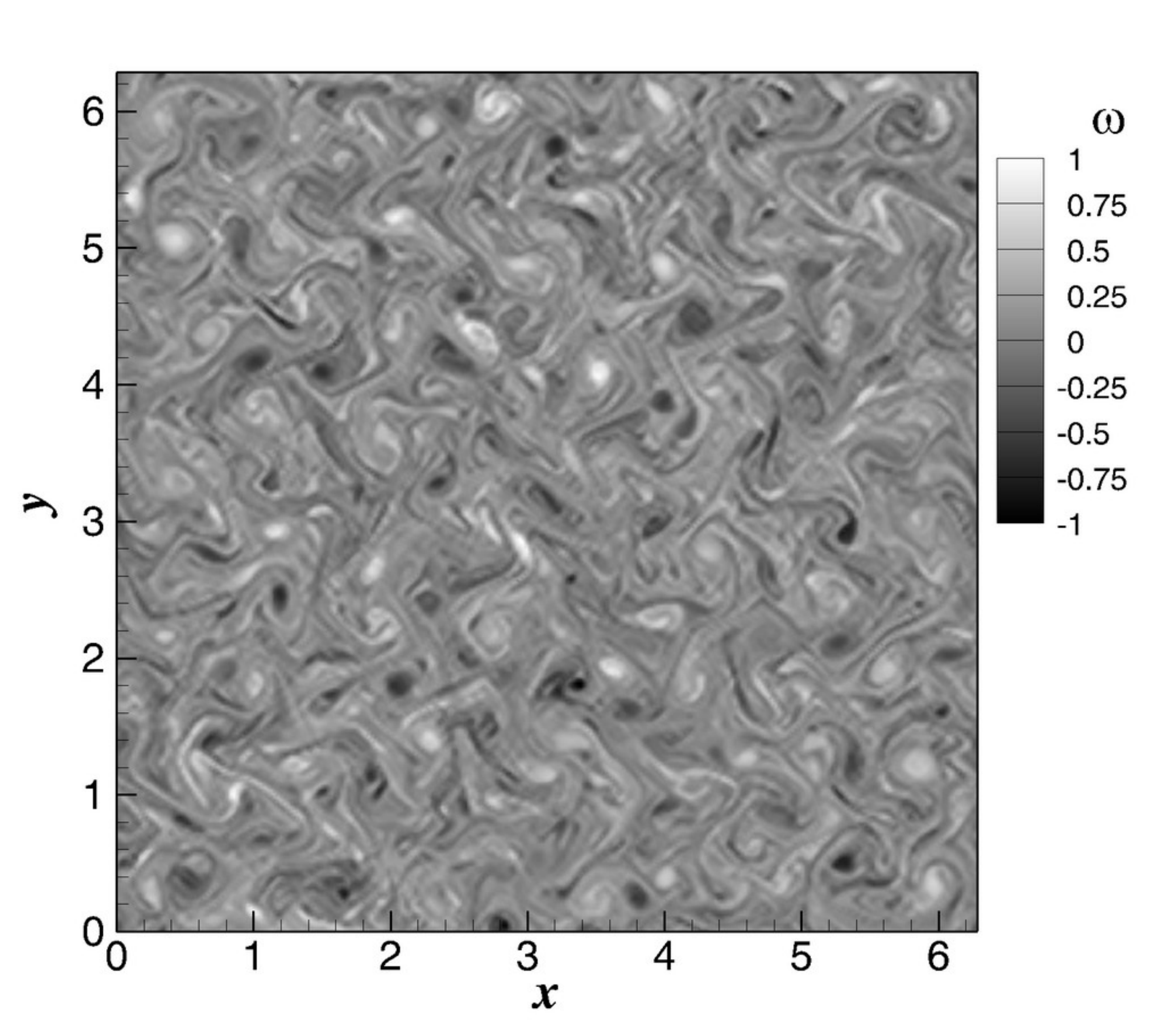}}
}
\caption{A-priori results for vorticity recovery from low-pass spatially filtered inputs for Kraichnan turbulence. Data shown for deconvolution test data 1: (a) true coarse-grained fields, (b) coarse-grained fields with Gaussian smoothing, and (c) coarse-grained fields reconstructed using proposed framework.}
\label{fig:field2D_1}
\end{figure}

\begin{figure}
\centering
\mbox{
\subfigure[True]{\includegraphics[width=0.3\textwidth]{2D_True.pdf}}
\subfigure[Noisy]{\includegraphics[width=0.3\textwidth]{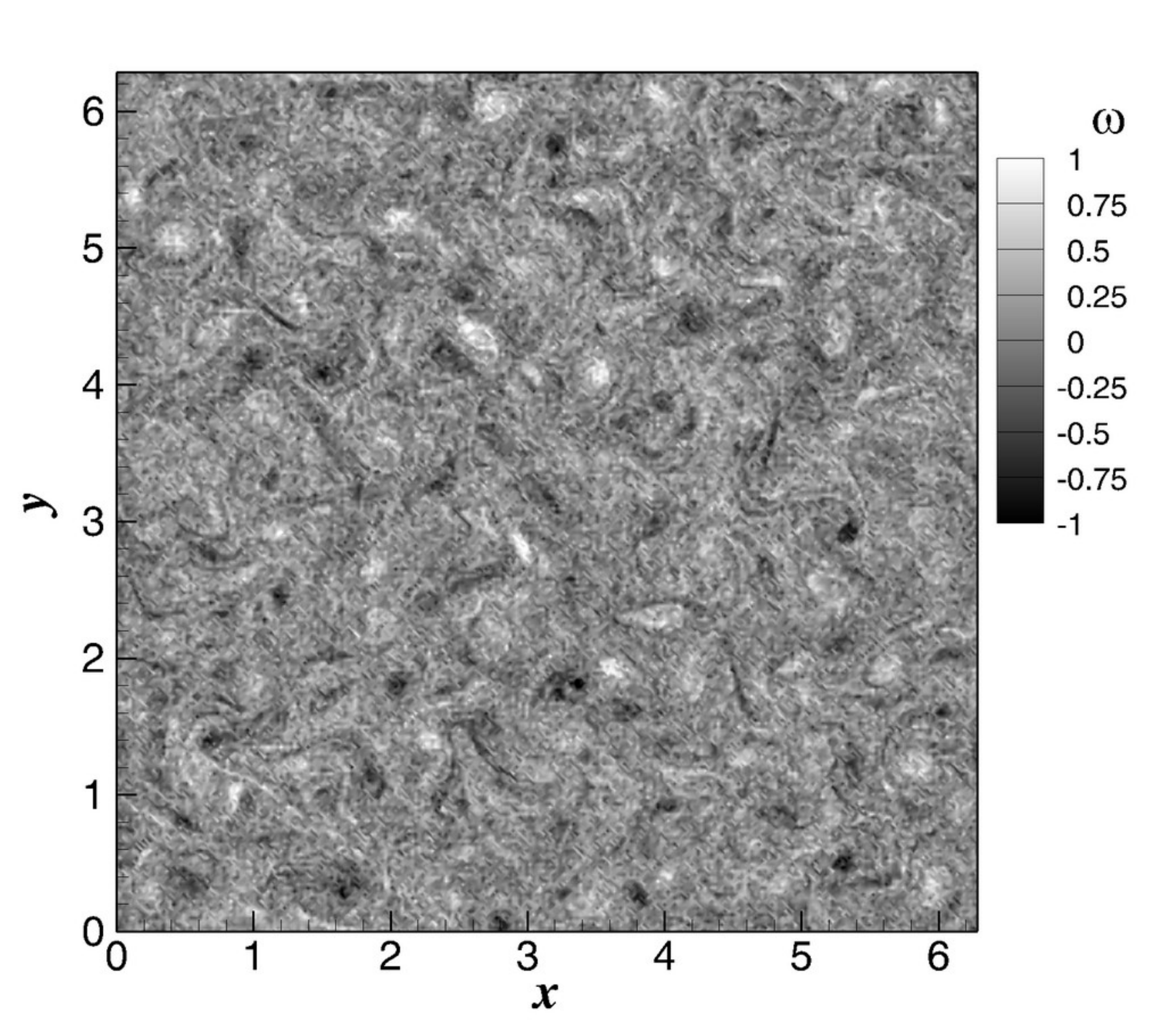}}
\subfigure[Recovered]{\includegraphics[width=0.3\textwidth]{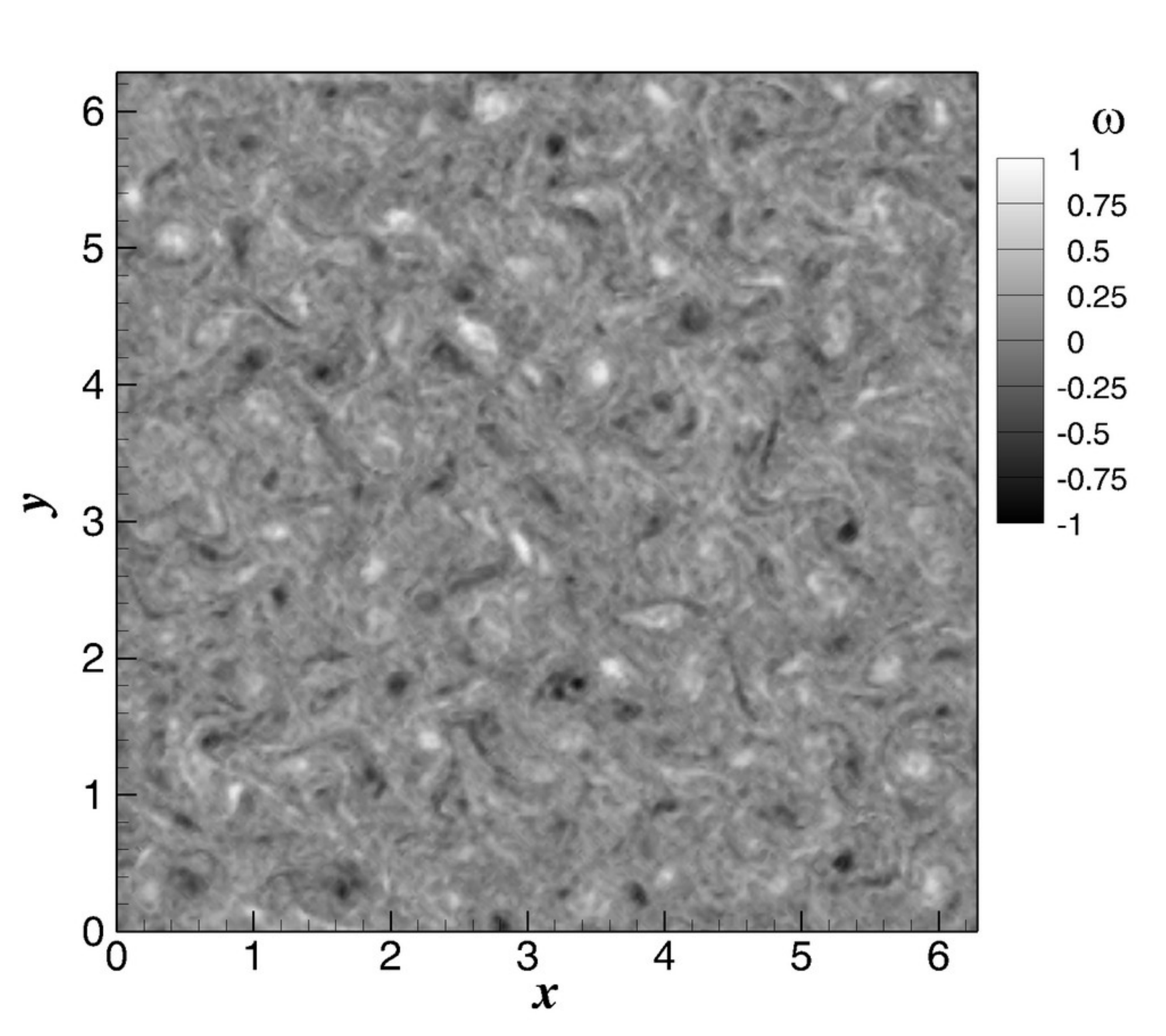}}
}
\caption{A-priori results for vorticity recovery from noisy perturbation inputs for Kraichnan turbulence. Data shown for regularization test data 1: (a) true coarse-grained fields, (b) coarse-grained fields with Gaussian noise, and (c) coarse-grained fields reconstructed using proposed framework.}
\label{fig:field2D_2}
\end{figure}

Figure~\ref{fig:field2D_1} and Figure~\ref{fig:field2D_2} visually represent the results of our proposed architecture. The enhancement of smaller features (due to a higher retention of the inertial range) in the case of the filtered inputs is clearly visible. On the other hand, a blurring effect can be discerned for the noisy perturbations which correspond to the stabilization effect of the framework. The distribution of normalized vorticity values seen in the PDF diagrams show a significant reconstruction of the true trends using an optimally trained ANN. A quantitative description of the performance of our data-driven deconvolution and regularization architecture for the Kraichnan turbulence test case is shown in Table \ref{table:CV2} where mean-squared-error values of the perturbed and reconstructed fields (with respect to the true field) are tabulated. It can be seen that the proposed architecture performs remarkably well for the reduction of mean-squared-error for all three testing datasets.

\begin{table}
  \centering
  \begin{tabular}{p{3cm} p{3cm} p{3cm}}
    \multicolumn{3}{c}{\textbf{Kraichnan turbulence}} \\
    \hline
    \multicolumn{3}{c}{\underline{\textbf{Deconvolution}}} \\
    \addlinespace[0.1cm]
    Dataset & Filtered & Deconvolved \\
    Test data 1 & $1.02 \times 10^{-2}$ & $4.38 \times 10^{-3}$\\
    Test data 2 & $1.14 \times 10^{-2}$ & $5.73 \times 10^{-3}$\\
    Test data 3 & $9.11 \times 10^{-3}$ & $3.71 \times 10^{-3}$\\
    \hline
  \end{tabular}
  \begin{tabular}{p{3cm} p{3cm} p{3cm}}
    \multicolumn{3}{c}{\underline{\textbf{Regularization}}} \\
    \addlinespace[0.1cm]
    Dataset & Noised & Regularized \\
    Test data 1 & $4.03 \times 10^{-2}$ & $1.26 \times 10^{-2}$\\
    Test data 2 & $4.88 \times 10^{-2}$ & $1.38 \times 10^{-2}$\\
    Test data 3 & $3.25 \times 10^{-2}$ & $1.14 \times 10^{-2}$\\
    \hline
  \end{tabular}
  \caption{Mean-squared-error values for deconvolved and regularized fields obtained from the proposed architecture. Data shown from the two-dimensional Kraichnan turbulence test case. Note that the mean-squared-error values are obtained from the vorticity magnitudes of the field.}\label{table:CV2}
\end{table}

\subsection{Kolmogorov Turbulence}

To further validate our claims about the potential of the proposed approach, the Taylor-Green vortex (TGV) problem was simulated in a periodic box l, following \cite{bull2015simulation}, to exhibit the properties of decaying isotropic homogeneous 3D turbulence with the Kolmogorov scaling given by $k^{-5/3}$ \citep{frisch1996turbulence}. In a manner similar to the 2D turbulence test case, high fidelity DNS data for $Re=1600$ was generated on a uniform grid with $512^3$ degrees of freedom from which coarser data at a resolution of $64^3$ grid points was extracted for our studies. Three networks are trained for each velocity component and each network is trained using approximately 260,000 samples with inputs consisting of a stencil of 27 points (i.e., a three dimensional version of the stencil described for Kraichnan turbulence) which represents a much larger computational task than the previous test case. However, the ELM approach is able to successfully train these three networks in under 10 seconds.

\begin{figure}
\centering
\mbox{
\subfigure[Test data 1]{
\includegraphics[width=0.44\textwidth]{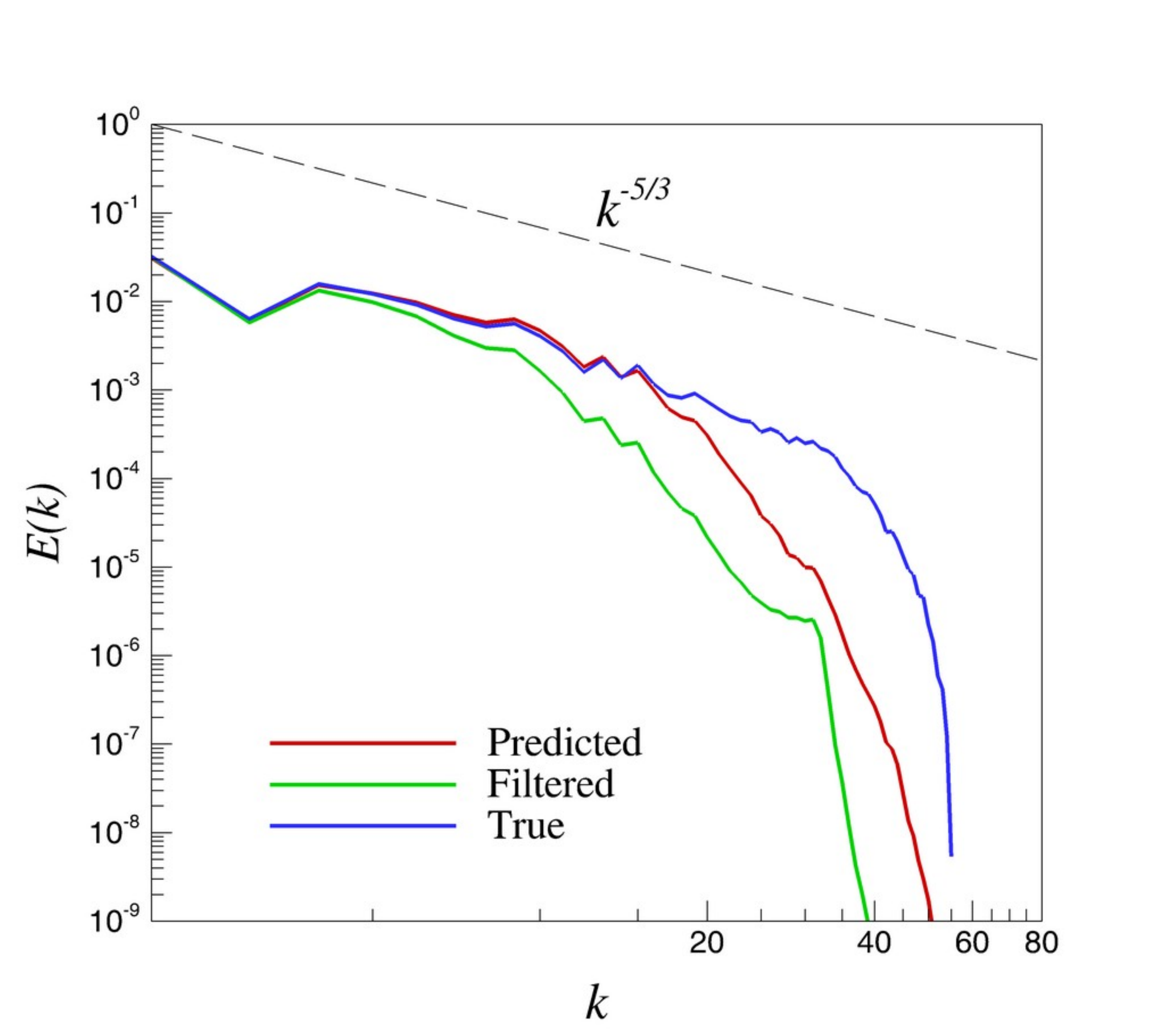}
\includegraphics[width=0.44\textwidth]{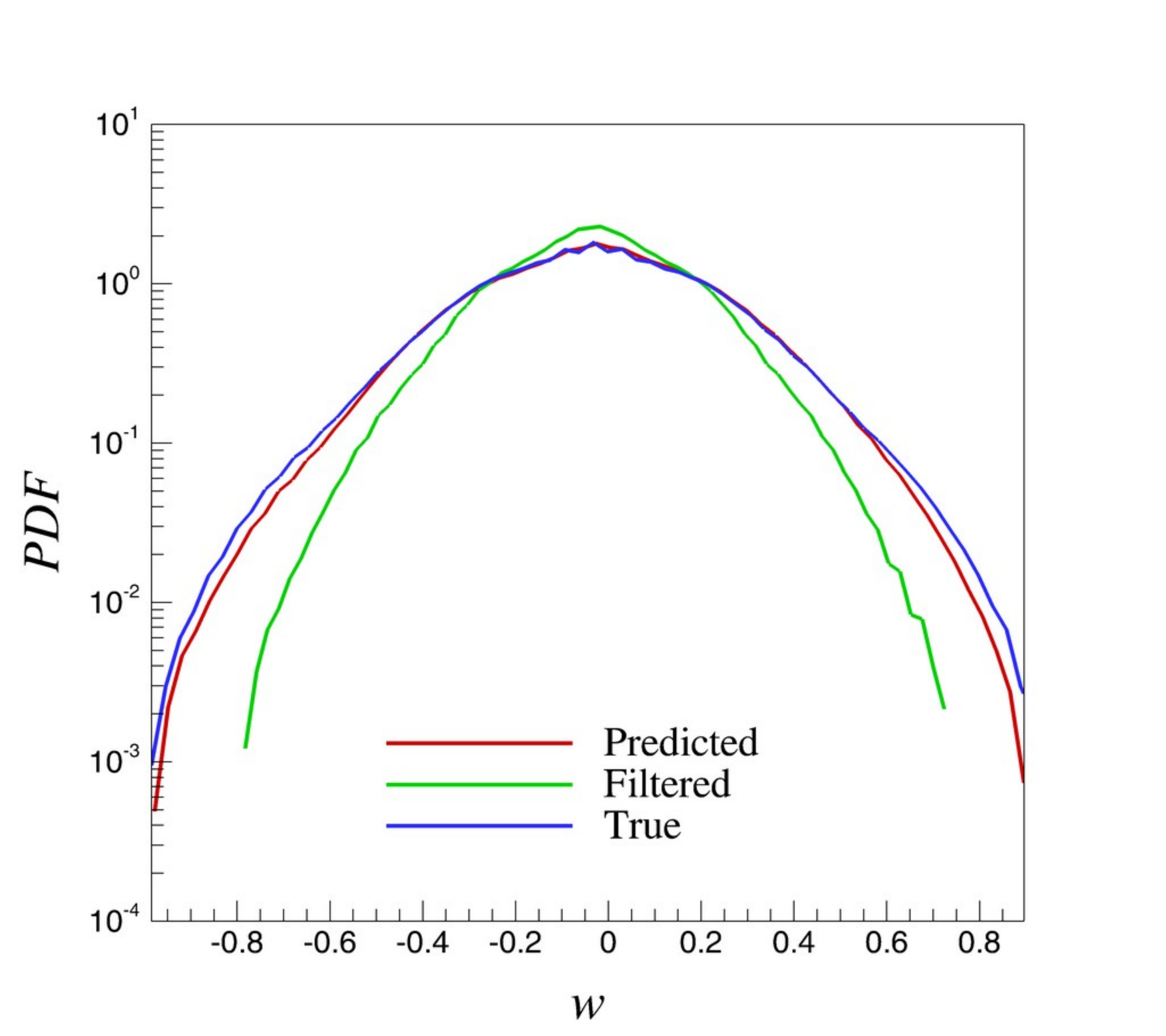}
}
}\\
\mbox{
\subfigure[Test data 2]{
\includegraphics[width=0.44\textwidth]{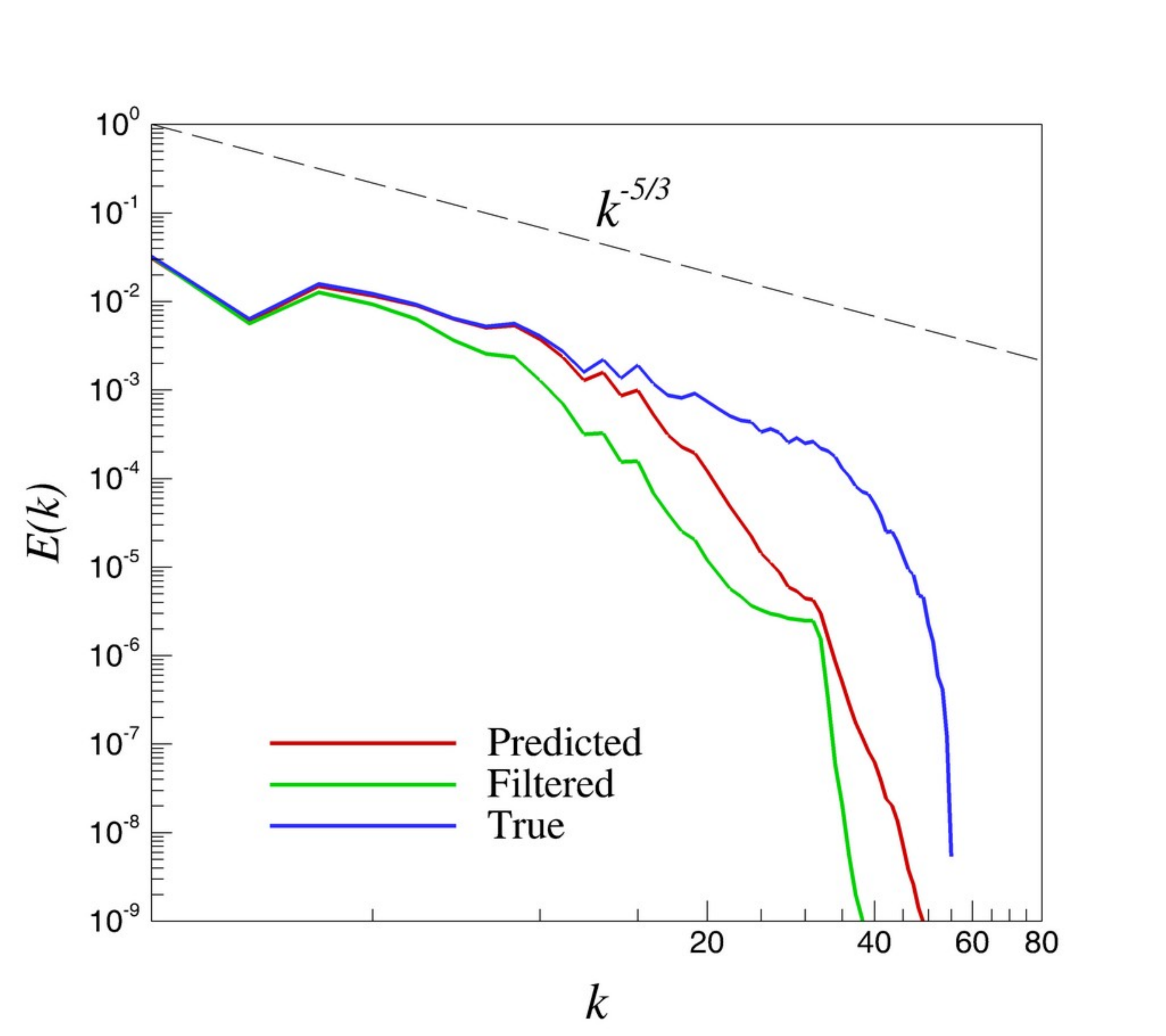}
\includegraphics[width=0.44\textwidth]{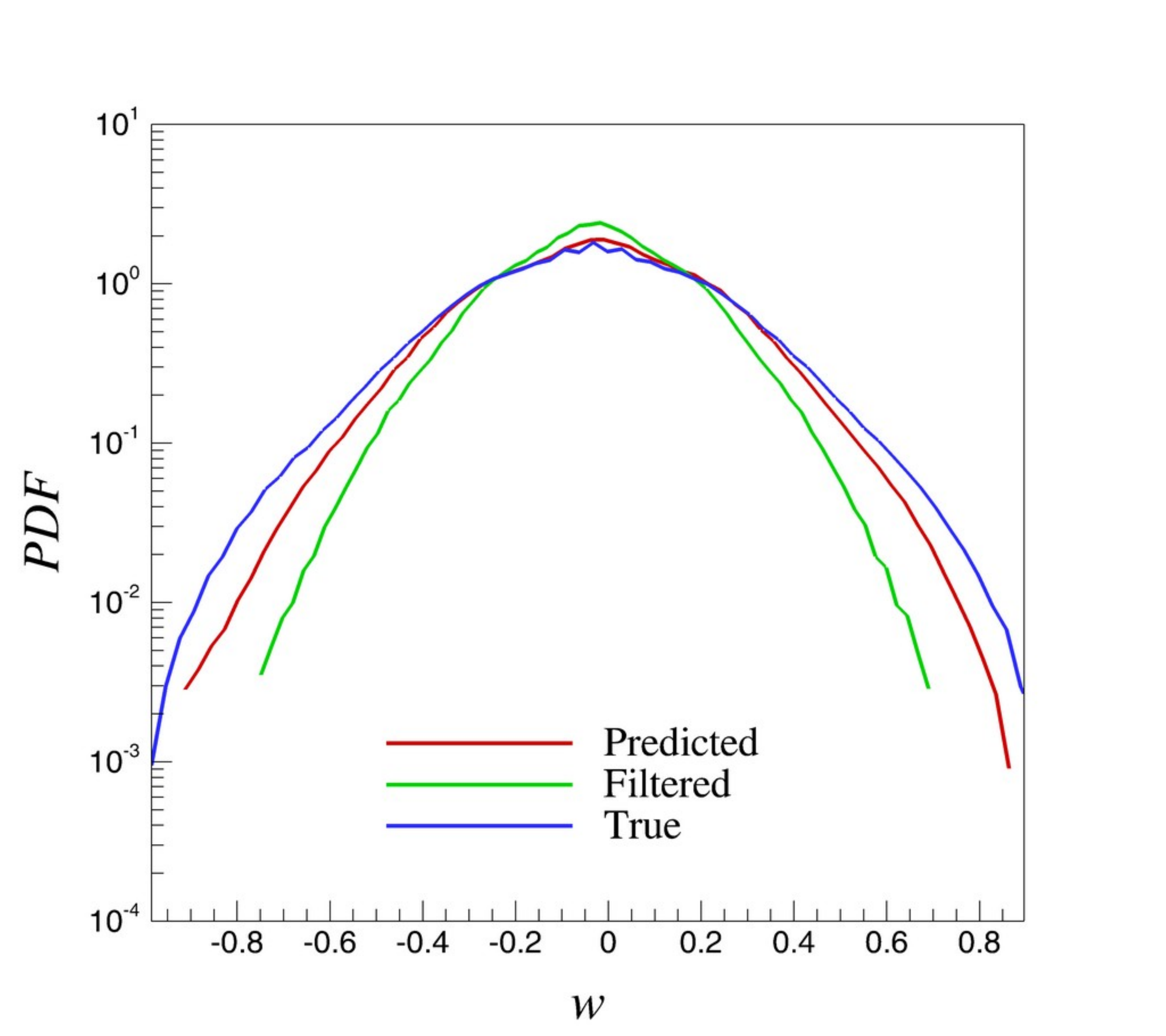}
}
}\\
\mbox{
\subfigure[Test data 3]{
\includegraphics[width=0.44\textwidth]{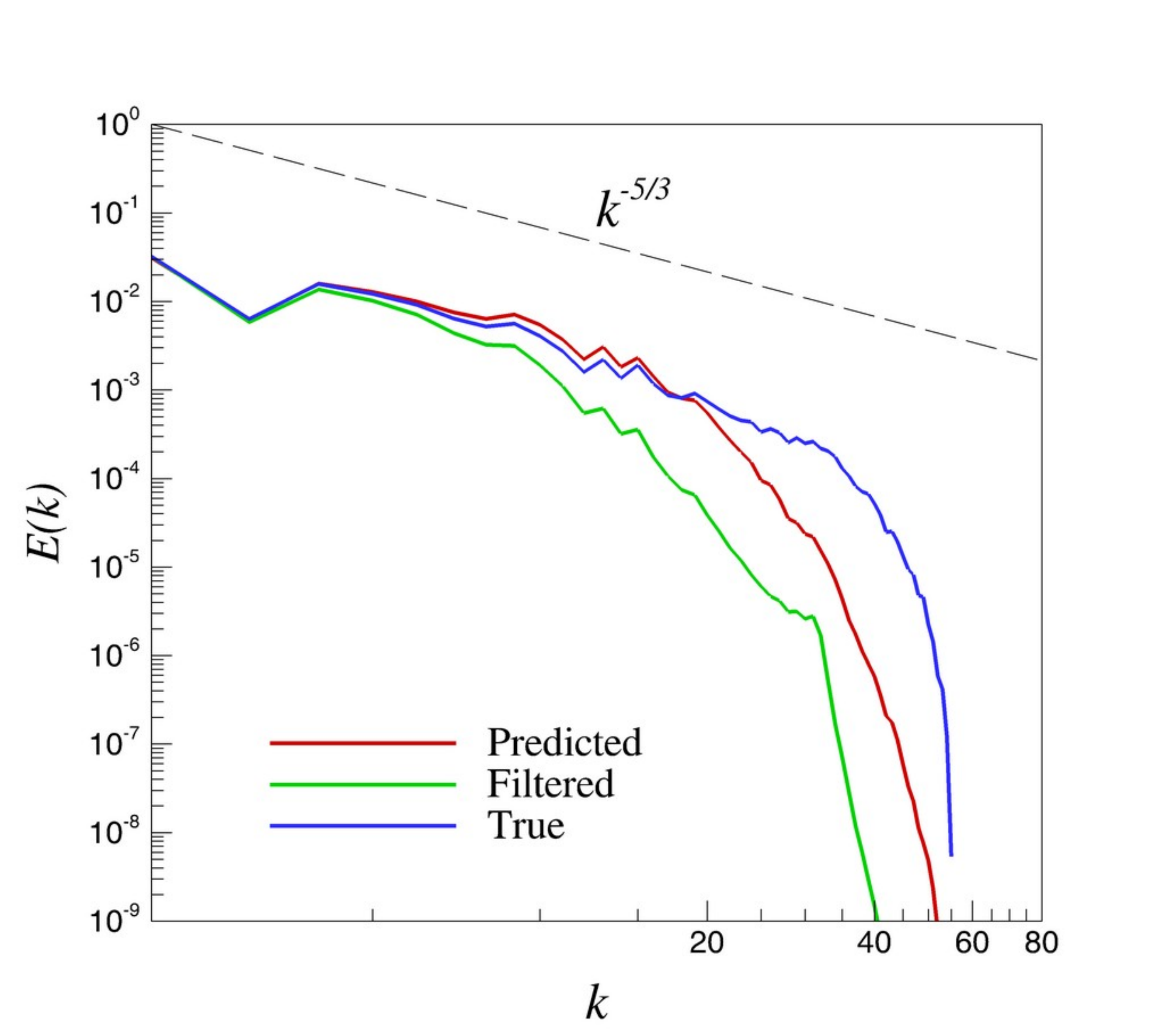}
\includegraphics[width=0.44\textwidth]{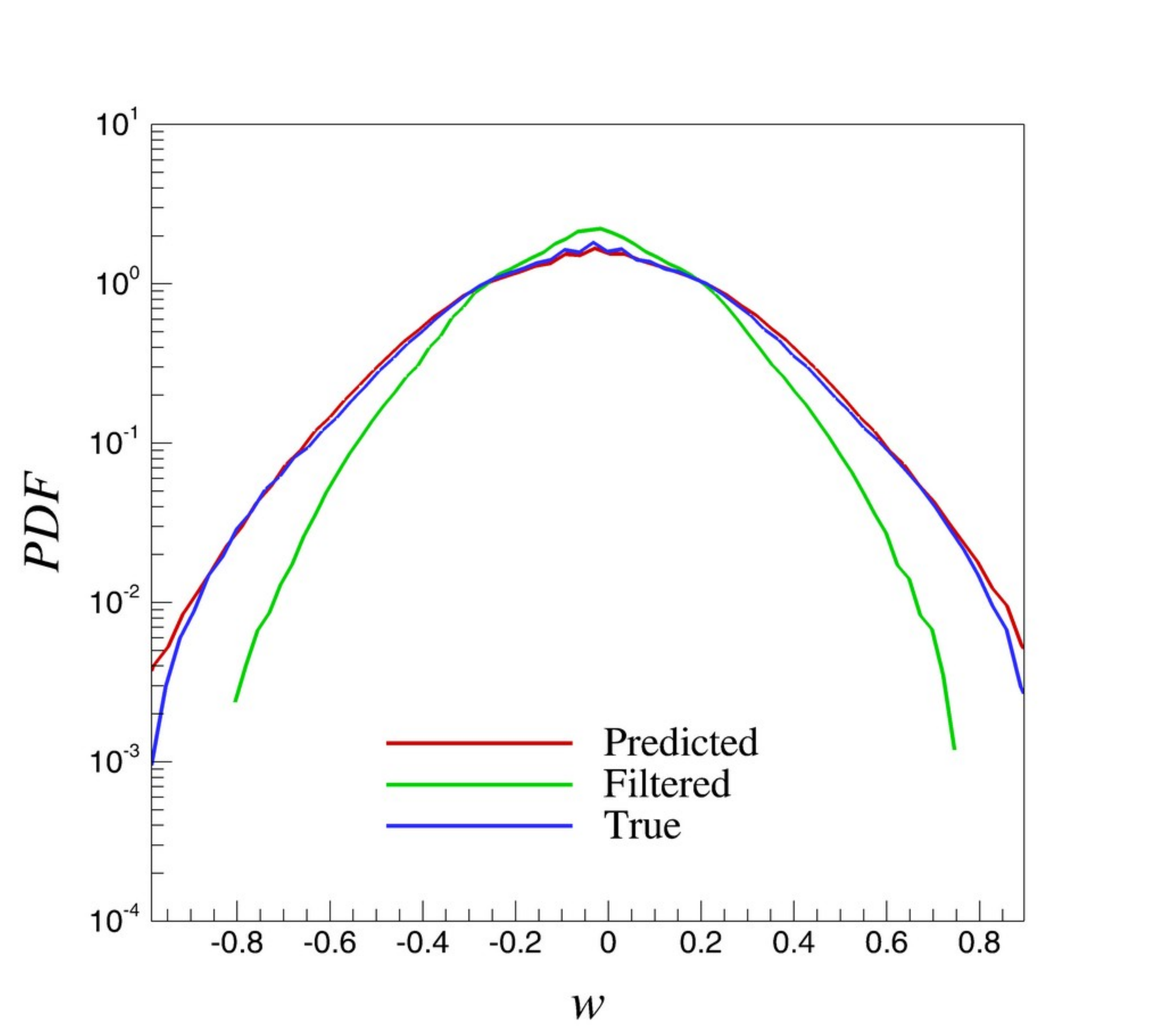}
}
}
\caption{A-priori results of the kinetic energy spectra (left) and PDF of the $z$ component of velocity (right) for Kolmogorov turbulence. Results for three different deconvolution test data sets shown. }
\label{fig:Spec3D_Filtered}
\end{figure}

Figure~\ref{fig:Spec3D_Filtered} demonstrates our statistical assessments for the low-pass spatially filtered test cases. As per our validation mechanism, we display results from three different versions of testing data where similar results are obtained qualitatively. Our proposed framework recovers a significant enhancement in the approximation of the true inertial range in the spherically averaged kinetic energy spectrum. It can also be seen that the peak and the tail of the true probability distribution function of the $z$ component of velocity is captured quite accurately from the modified distribution obtained through low-pass spatial filtering. The PDF comparison between true and recovered fields is another indication of the ability of the proposed ANN architecture for reconstructing the true underlying trends from the filtered input data. It may be observed that the proposed architecture performs better for test data 3 as against test data 2 (although both trends are positive) since much more subfilter scale information must be reconstructed than was trained for. This is possibly due to the data being rather nonlinear in physical space as against the linear parameter space we attempt to explore. Essentially, increasing the filter radius by 10\% leads to some physical behavior that the framework has not fully been exposed to in training. Addressing issues of predicting multiple perturbations using a singly trained network will require in-depth investigation of sampling strategies. Vitally, the behavior exhibited by our proposed framework is consistent with our observations for the Kraichnan turbulence test case.

\begin{figure}
\centering
\mbox{
\subfigure[Test data 1]{
\includegraphics[width=0.44\textwidth]{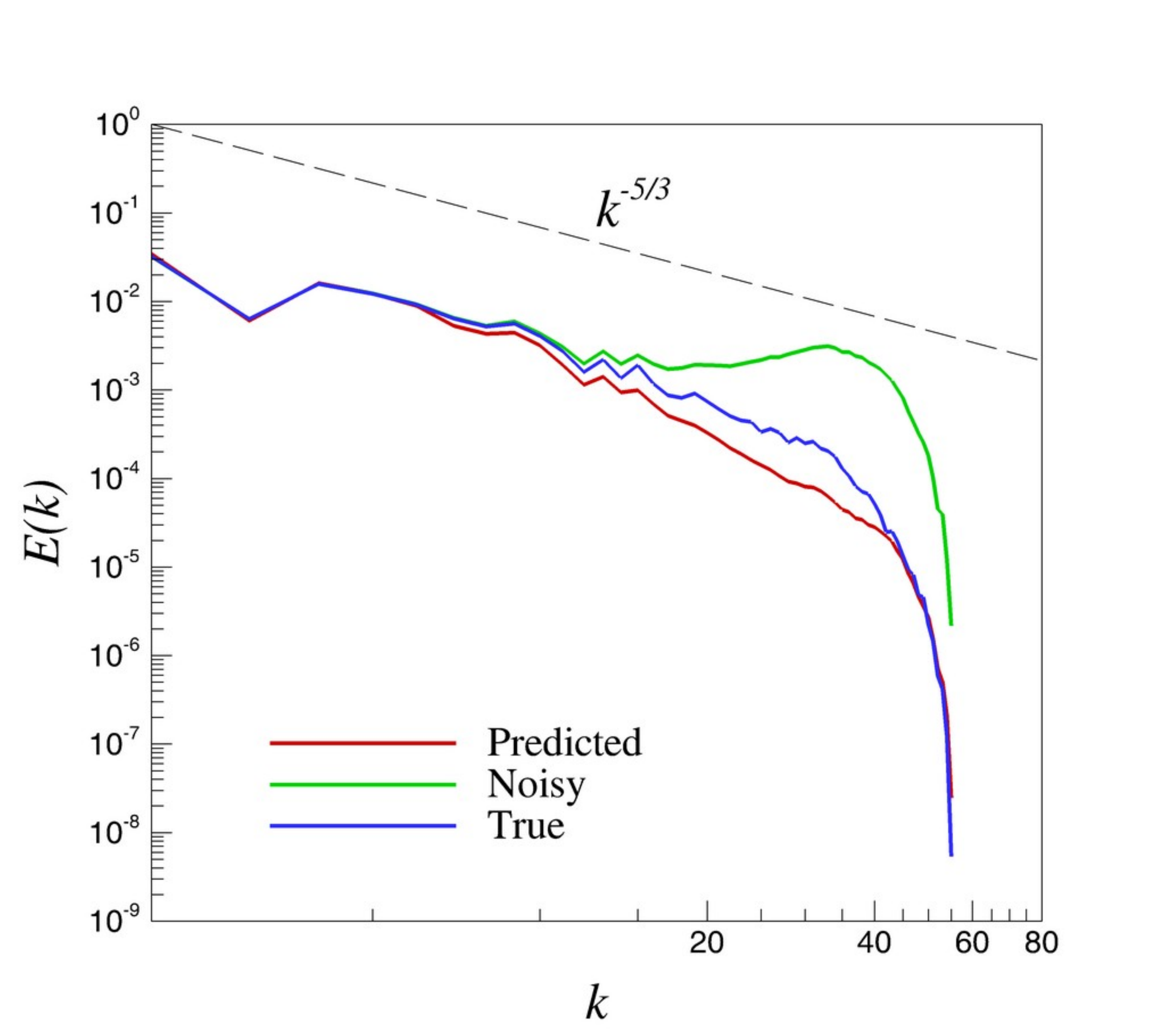}
\includegraphics[width=0.44\textwidth]{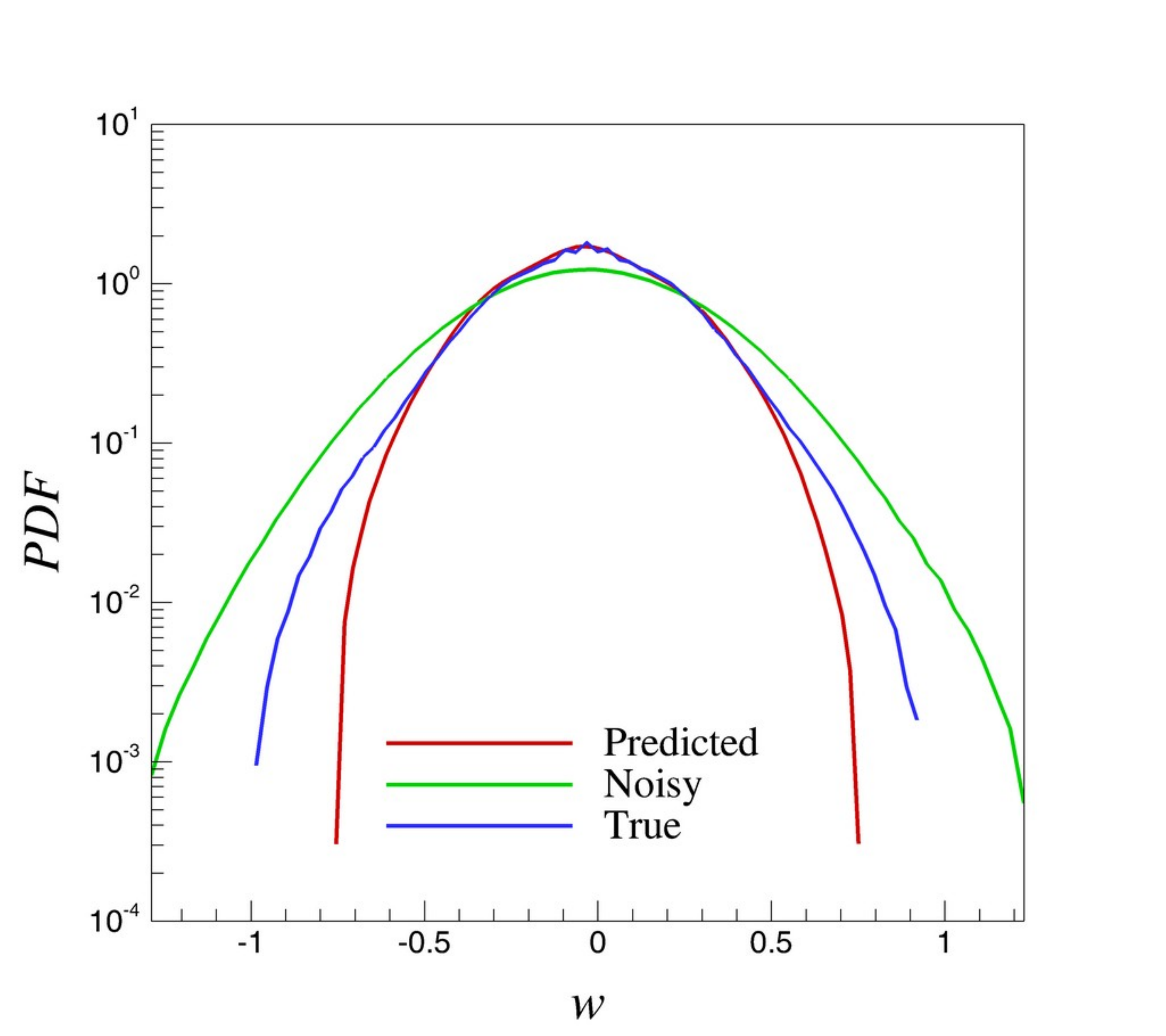}
}
}\\
\mbox{
\subfigure[Test data 2]{
\includegraphics[width=0.44\textwidth]{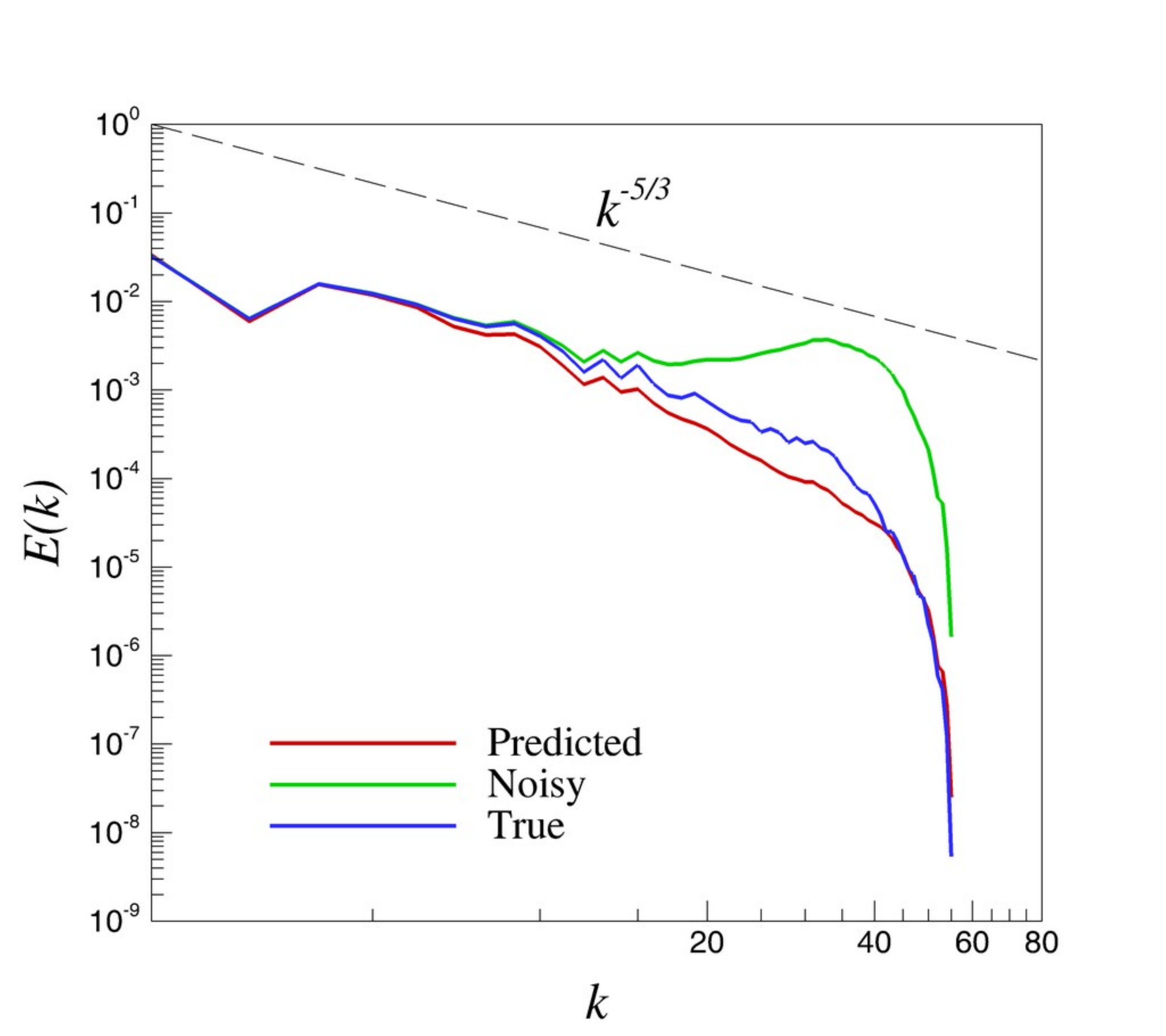}
\includegraphics[width=0.44\textwidth]{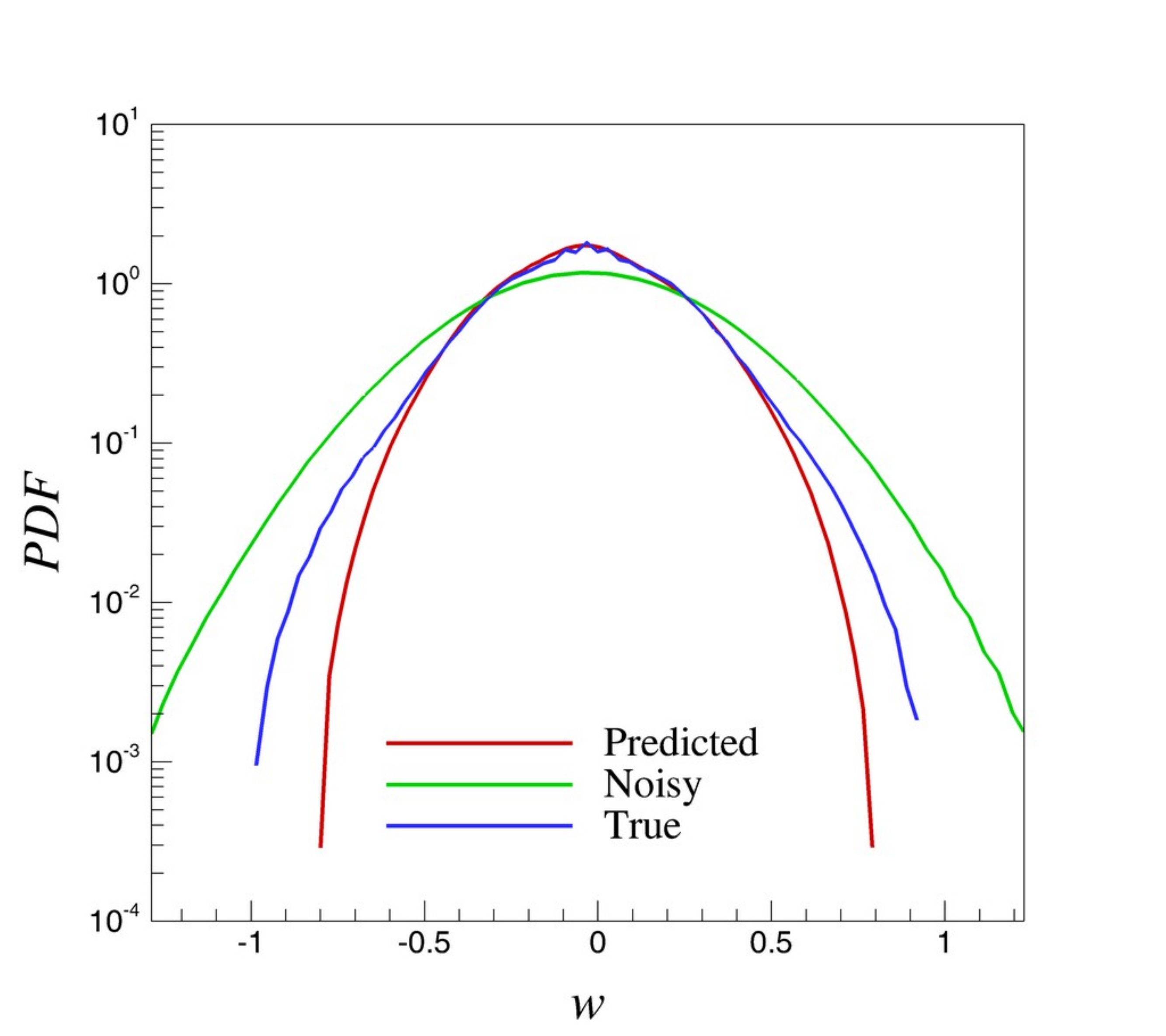}
}
}\\
\mbox{
\subfigure[Test data 3]{
\includegraphics[width=0.44\textwidth]{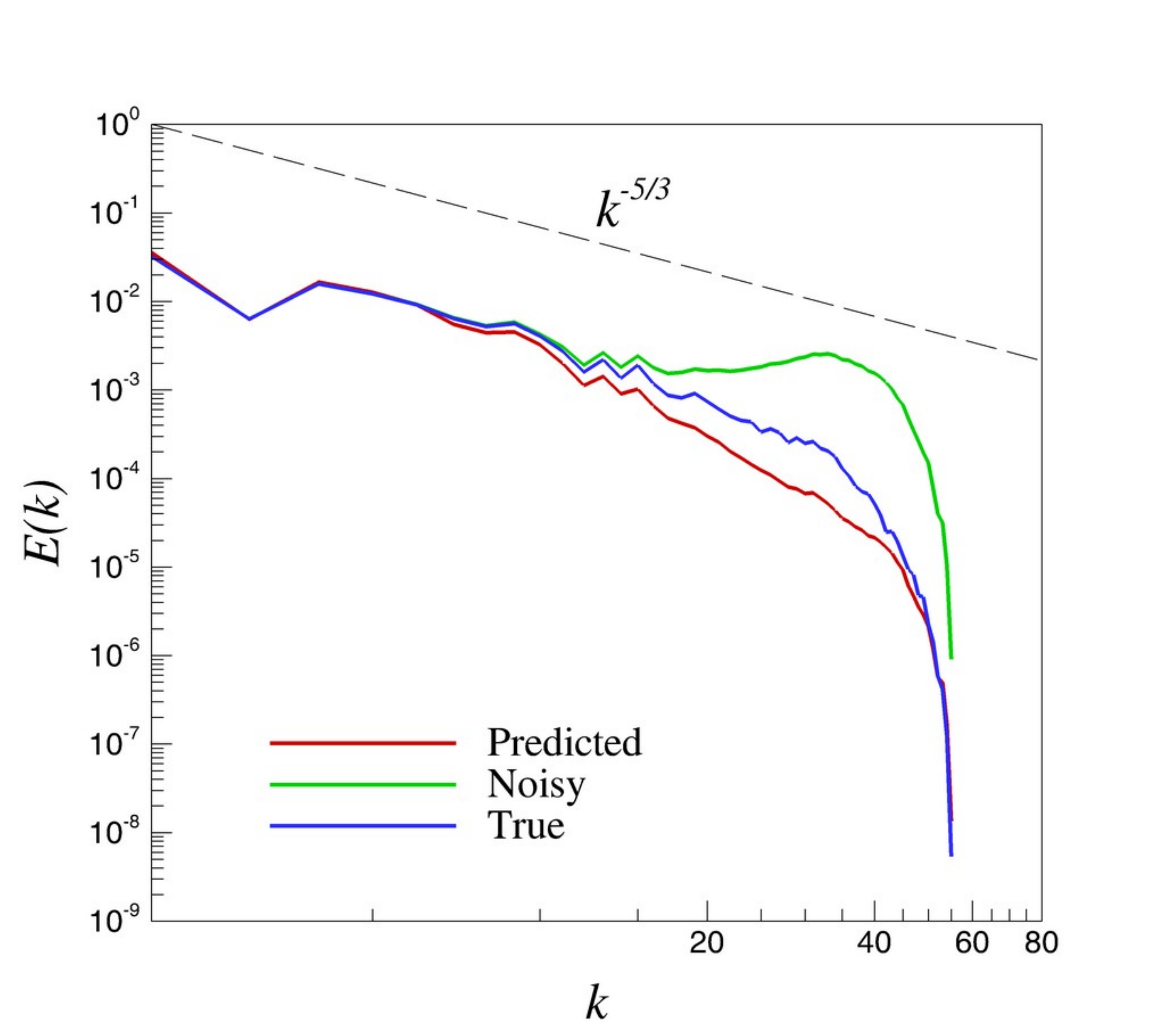}
\includegraphics[width=0.44\textwidth]{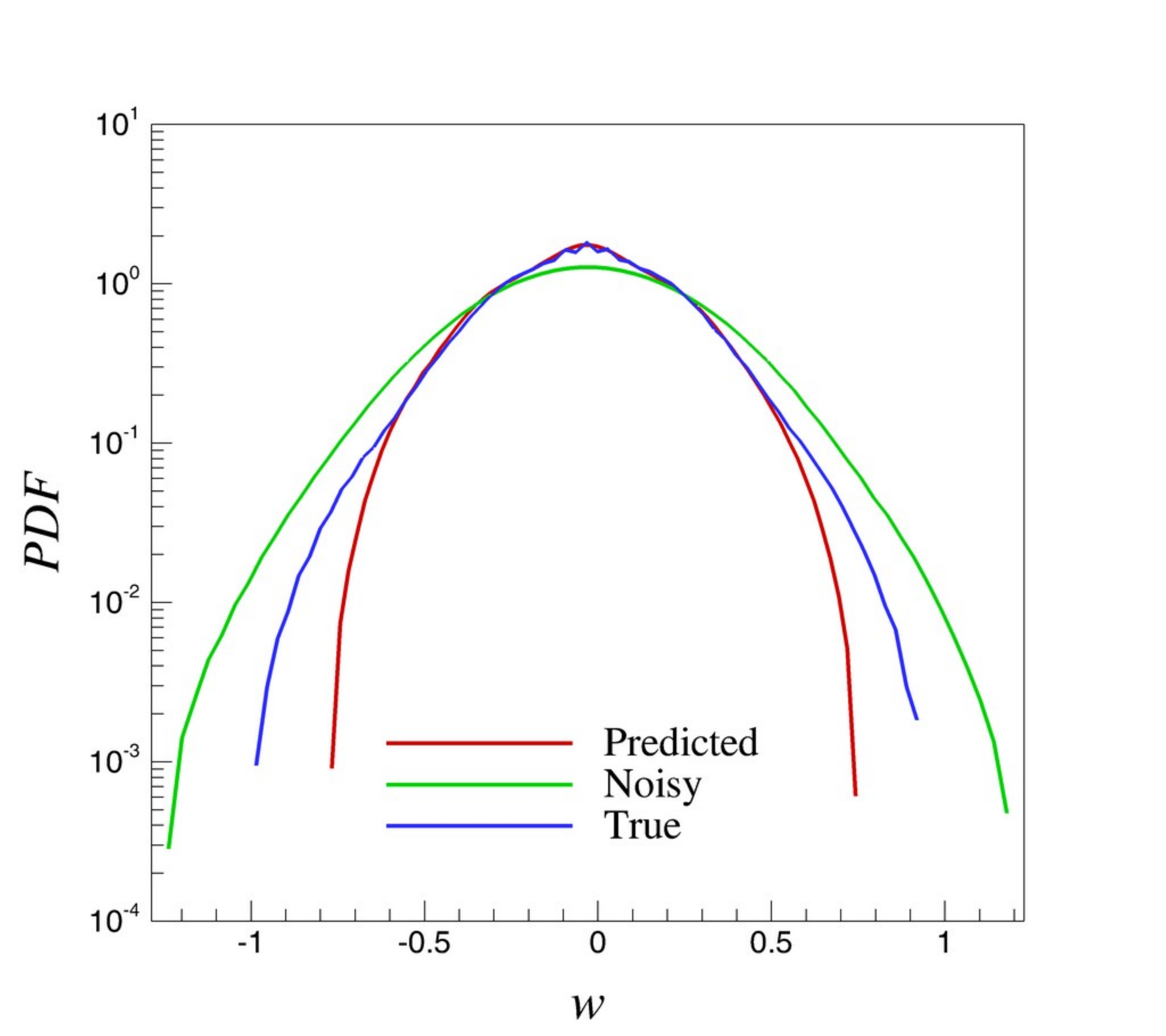}
}
}
\caption{A-priori results of the kinetic energy spectra (left) and PDF of the $z$ component of velocity (right) for Kolmogorov turbulence. Results for three regularization different test data sets shown. }
\label{fig:Spec3D_Noised}
\end{figure}

When trained to estimate the relationship between noisy inputs and true outputs, the proposed architecture demonstrates an exceptional generalization ability through the reduction of high wavenumber content as shown in Figure~\ref{fig:Spec3D_Noised}. While inertial range enhancement is once again marginal, a general improvement in the statistical trends of the recovered variable is observed through the PDF where the peaks of the true PDF are recovered accurately. However, it is seen that an accurate capture of the tails of the PDF of the true field remains elusive for this particular architecture. Studies are underway to determine if the architecture of the ANN may aid in improving this inaccuracy since a larger number of neurons may offset the rather stringent generalization behavior of its current implementation. This statistical generalization ability is also observed in the 2D test case where tails of the PDF tend to be captured with lower accuracy.

\begin{figure}
\centering
\mbox{
\subfigure[True]{\includegraphics[width=0.3\textwidth]{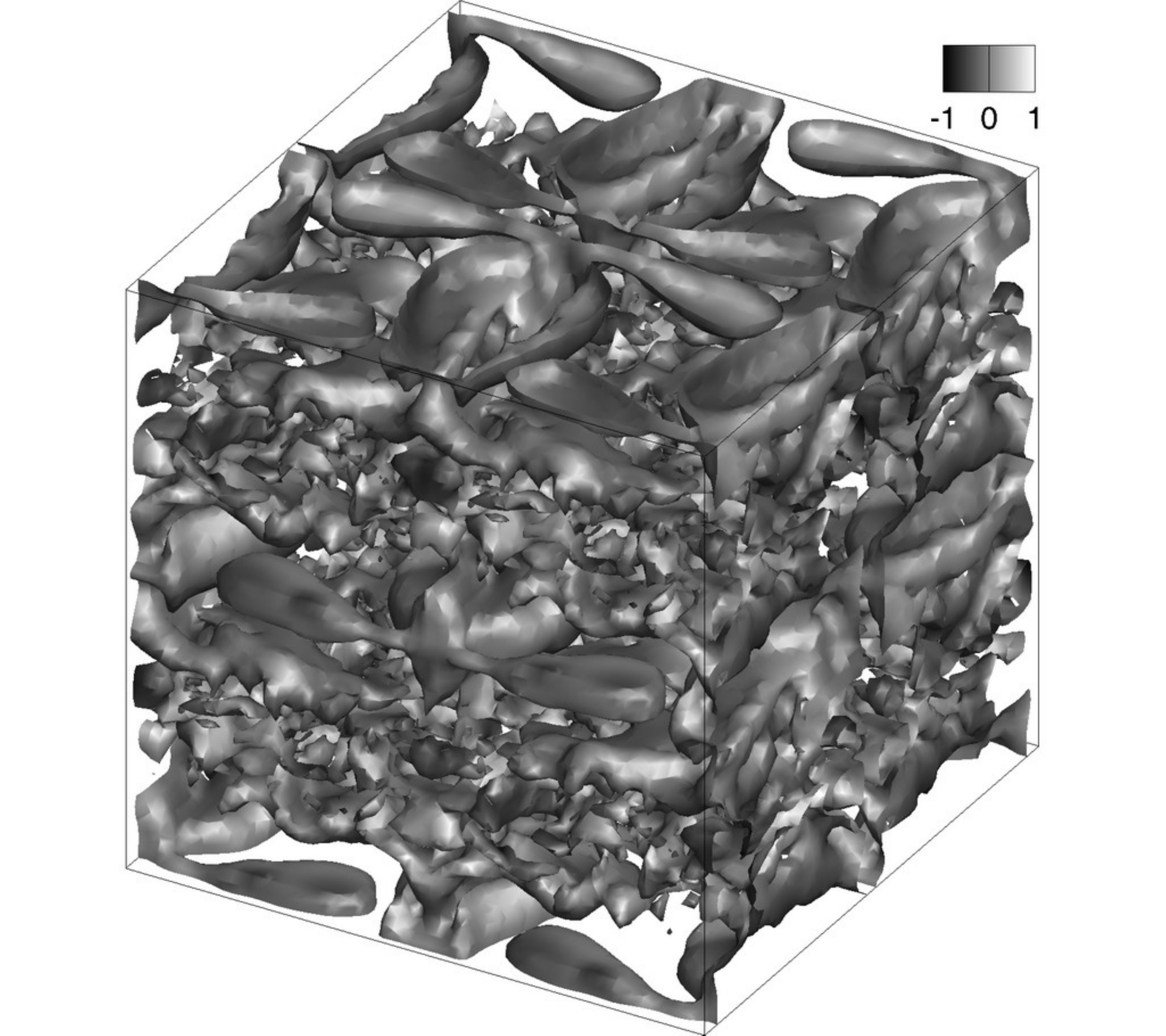}}
\subfigure[Filtered]{\includegraphics[width=0.3\textwidth]{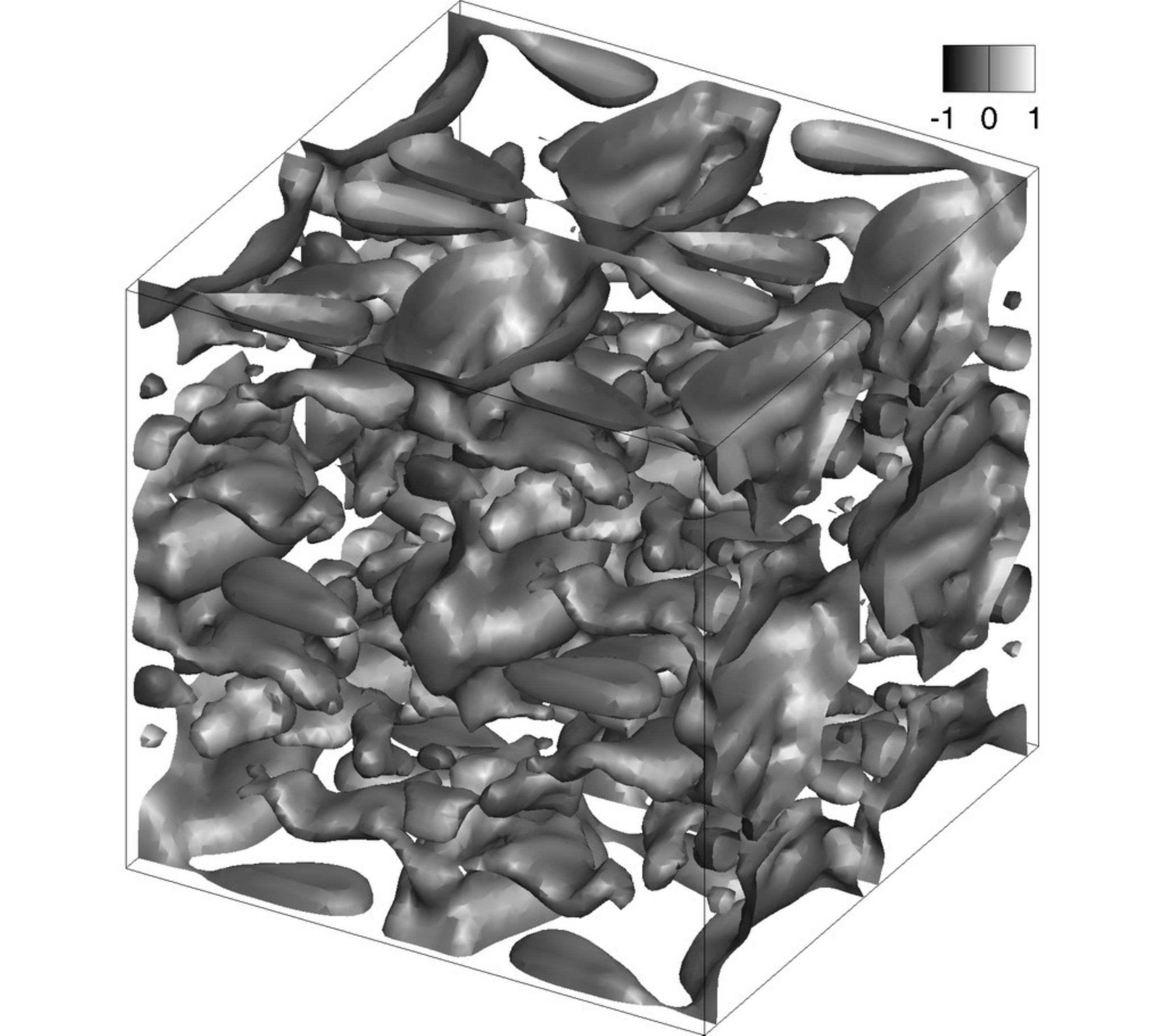}}
\subfigure[Recovered]{\includegraphics[width=0.3\textwidth]{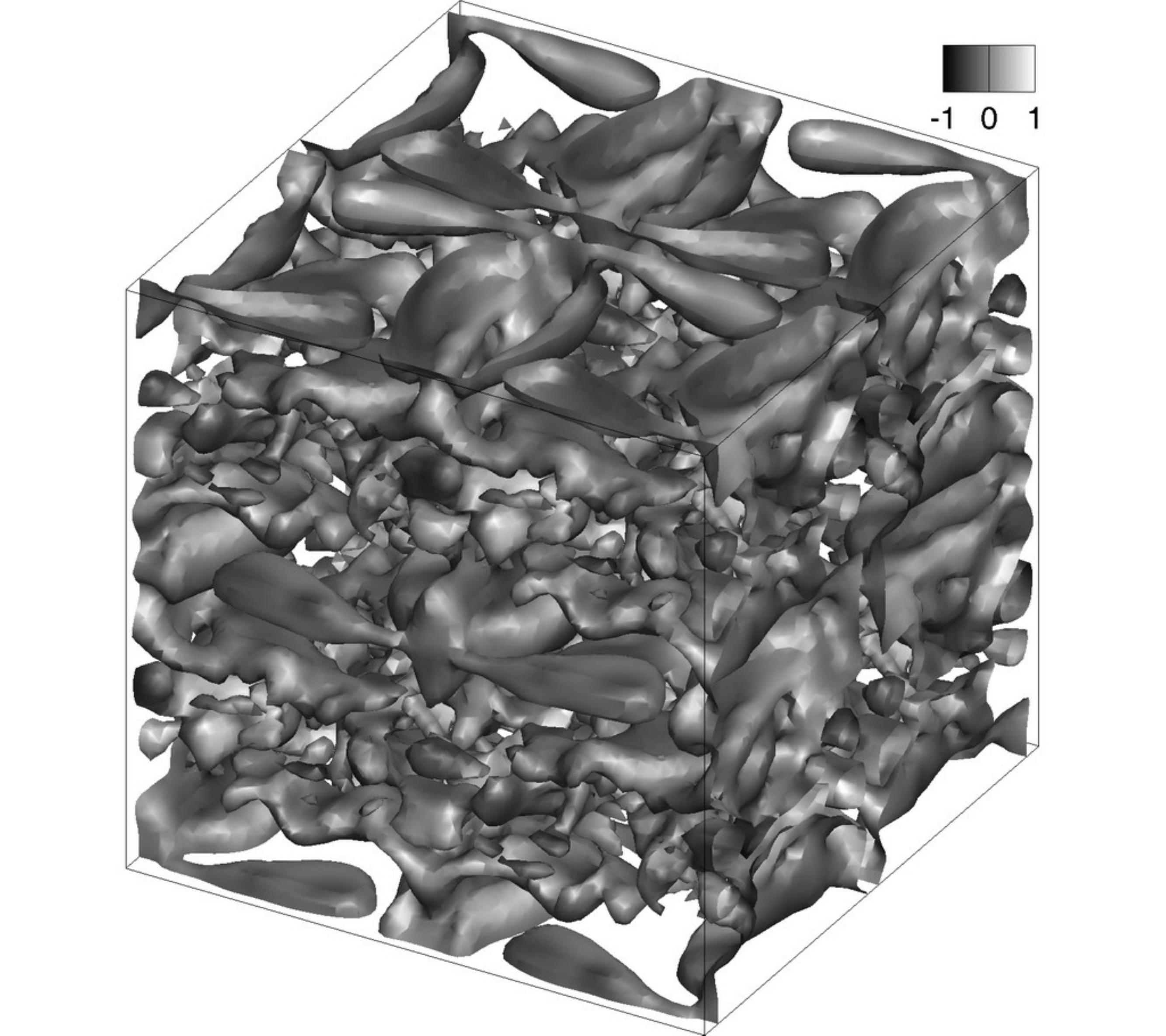}}
}\\
\caption{A-priori results for velocity field recovery from low-pass spatially filtered perturbations for Kolmogorov turbulence. Isosurfaces for $x$-component of the velocity colored by $z$-component are shown. Data shown for deconvolution test data 1: (a) true coarse-grained fields (b) coarse-grained fields with Gaussian smoothing, and (c) coarse-grained fields reconstructed using proposed framework.}
\label{fig:field3D_1}
\end{figure}

\begin{figure}
\centering
\mbox{
\subfigure[True]{\includegraphics[width=0.3\textwidth]{3D_True.pdf}}
\subfigure[Noisy]{\includegraphics[width=0.3\textwidth]{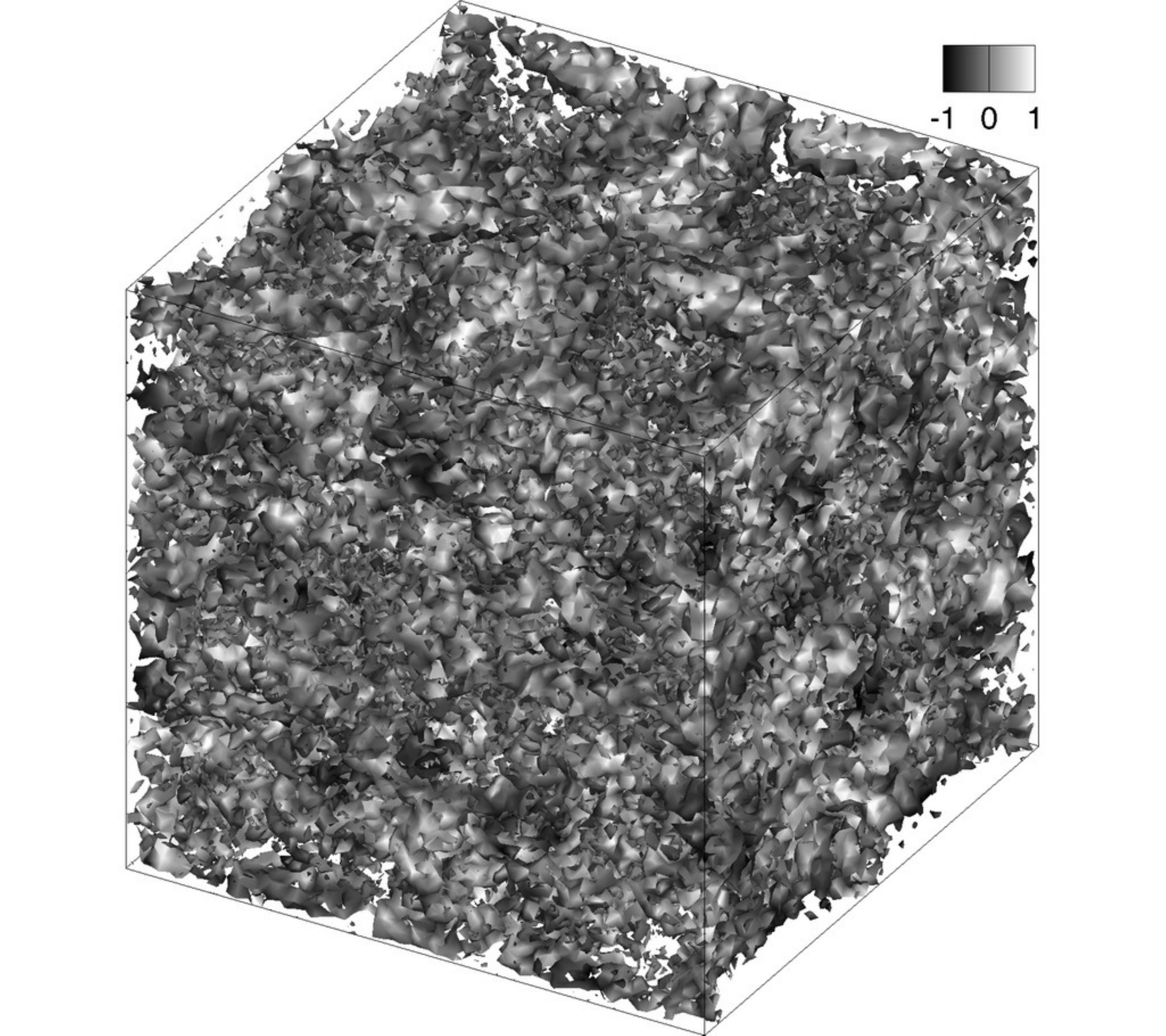}}
\subfigure[Recovered]{\includegraphics[width=0.3\textwidth]{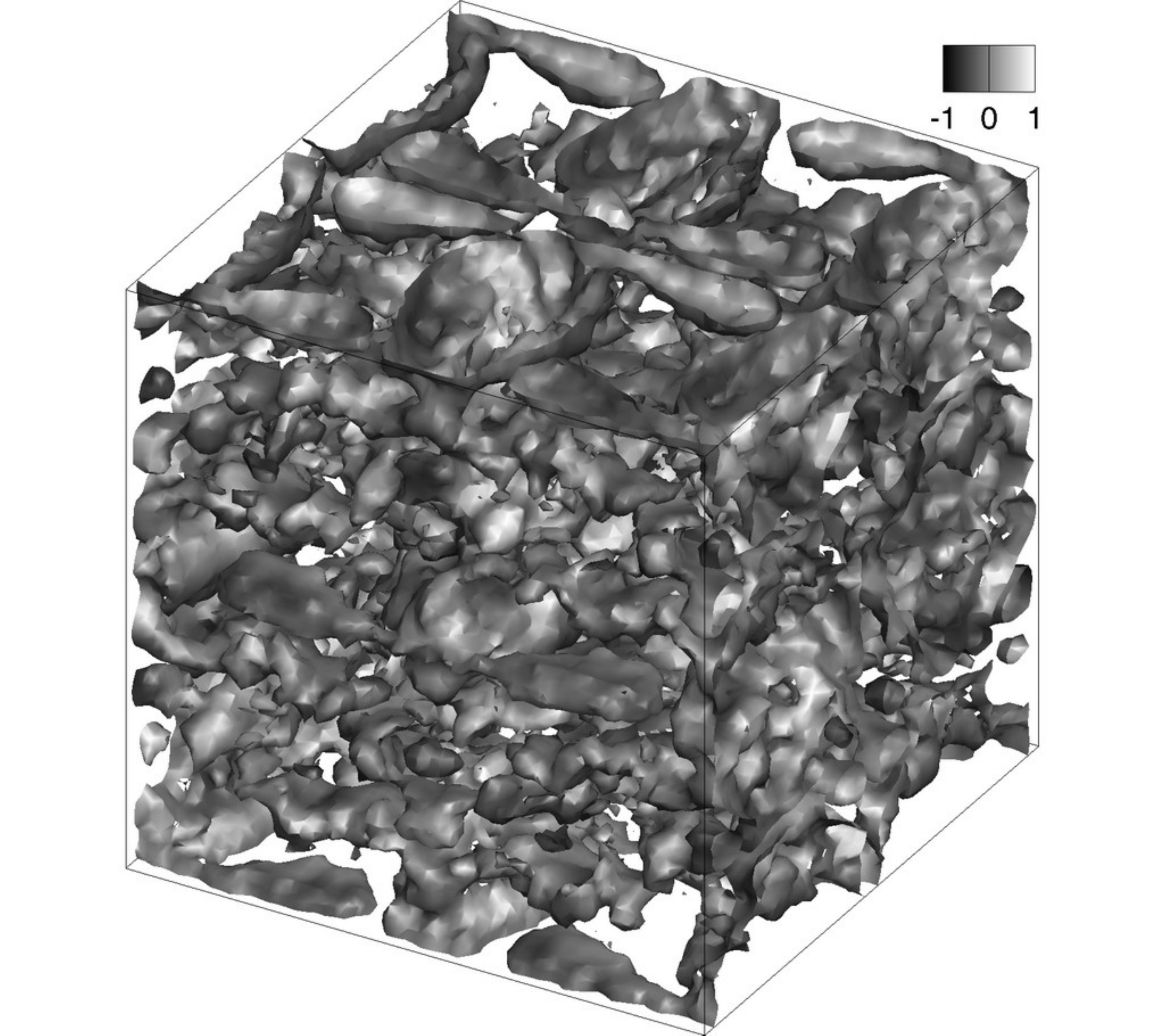}}
}
\caption{A-priori results for velocity field recovery from noisy perturbation inputs for Kolmogorov turbulence. Isosurfaces for $x$-component of the velocity colored by $z$-component are shown. Data shown for regularization test data 1: (a) true coarse-grained fields, (b) coarse-grained fields with Gaussian noise, and (c) coarse-grained fields reconstructed using proposed framework.}
\label{fig:field3D_2}
\end{figure}

We clarify that the trends in the PDF of other primitive variables are similar. The blind deconvolution ability of the proposed methodology is shown in Figure~\ref{fig:field3D_1} where one can clearly observe the presence of finer structures in the reconstructed field. Also, an examination of the noisy inputs in Figure~\ref{fig:field3D_2} show a remarkable reconstruction of data. It is apparent that the proposed approach proves adept at a partial deconvolution of the field when trained for the relationship using filtered inputs. This represents a viable opportunity for integrating data-driven closures into mainstream LES methodologies in combination with knowledge from first principles. In fact, it may even be possible to devise turbulence closures where phenomenological or heuristic arguments can be eschewed completely in favor of data-driven subfilter scale estimations. This holds promise for inhomogeneous flows where distinct deviations are observed from Gaussian statistics in terms of the location of the integral length scales. We may effectively let experimental data determine the shape of a particular filter in a particular region of the flow. Machine learning inspired blind deconvolution may thus hold the key to a greater enhancement of the \emph{uniformity} of an applicable turbulence closure. Table \ref{table:CV3} offers a quantitative insight into the performance of the proposed closure where it is once again observed that the framework is sufficiently capable in reducing the mean-squared-error of the perturbed fields for all testing datasets.

\begin{table}
  \centering
  \begin{tabular}{p{3cm} p{3cm} p{3cm}}
    \multicolumn{3}{c}{\textbf{Kolmogorov turbulence}} \\
    \hline
    \multicolumn{3}{c}{\underline{\textbf{Deconvolution}}} \\
    \addlinespace[0.1cm]
    Dataset & Filtered & Deconvolved \\
    Test data 1 & $9.56 \times 10^{-3}$ & $3.57 \times 10^{-3}$\\
    Test data 2 & $1.16 \times 10^{-2}$ & $4.64 \times 10^{-3}$\\
    Test data 3 & $8.10 \times 10^{-3}$ & $3.04 \times 10^{-3}$\\
    \hline
  \end{tabular}
  \begin{tabular}{p{3cm} p{3cm} p{3cm}}
    \multicolumn{3}{c}{\underline{\textbf{Regularization}}} \\
    \addlinespace[0.1cm]
    Dataset & Noised & Regularized \\
    Test data 1 & $4.20 \times 10^{-2}$ & $8.28 \times 10^{-3}$\\
    Test data 2 & $5.06 \times 10^{-2}$ & $9.18 \times 10^{-3}$\\
    Test data 3 & $3.39 \times 10^{-2}$ & $7.58 \times 10^{-2}$\\
    \hline
  \end{tabular}
  \caption{Mean-squared-error values for deconvolved and regularized fields obtained from the proposed architecture. Data shown from the three-dimensional Kolmogorov turbulence test case. Note that the mean-squared-error values are obtained from the $z$ component of the velocity field.}\label{table:CV3}
\end{table}

\begin{figure}
\centering
\mbox{
\subfigure[$\tau_{11}$]{\includegraphics[width=0.44\textwidth]{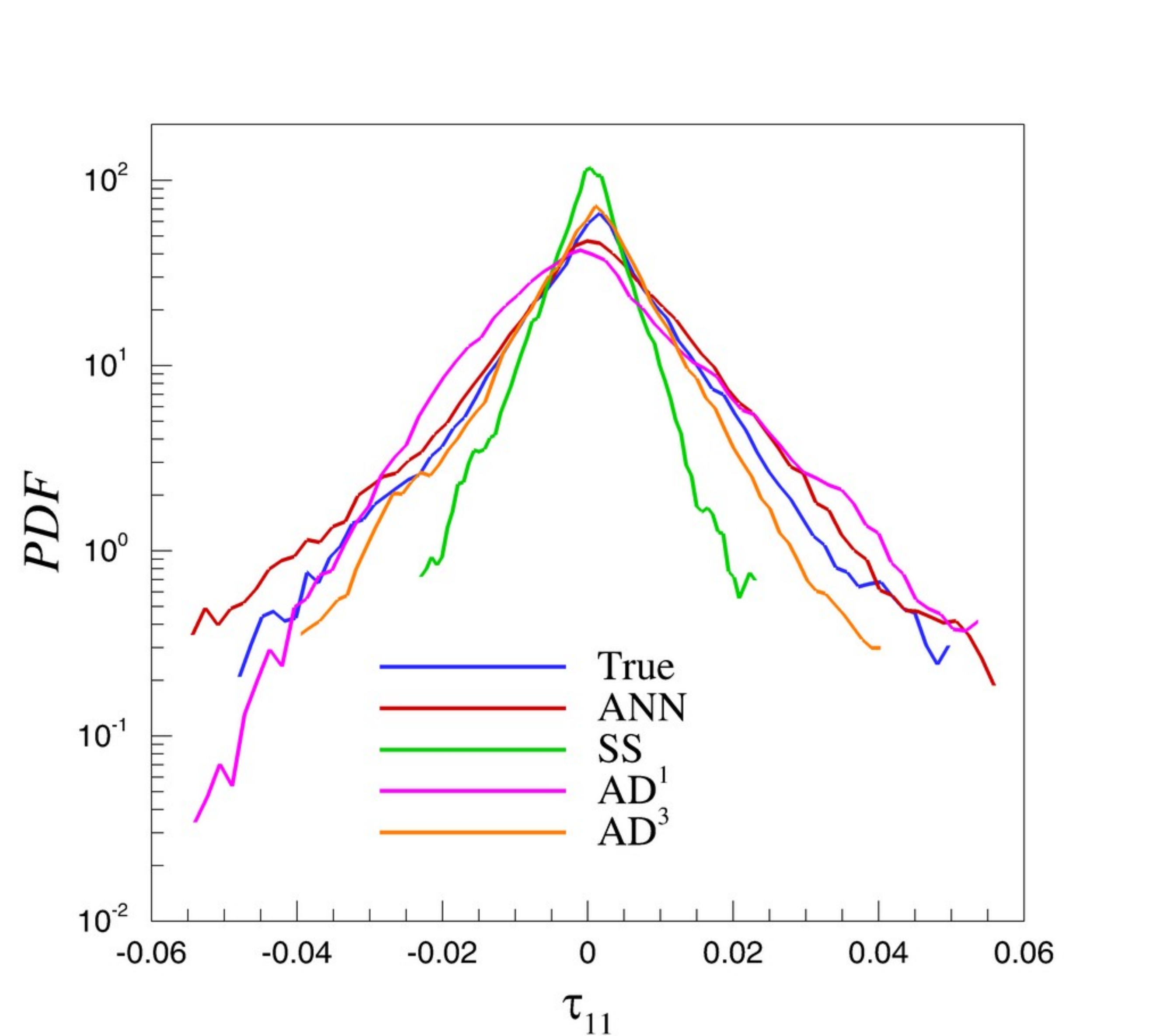}}
\subfigure[$\tau_{12}$]{\includegraphics[width=0.44\textwidth]{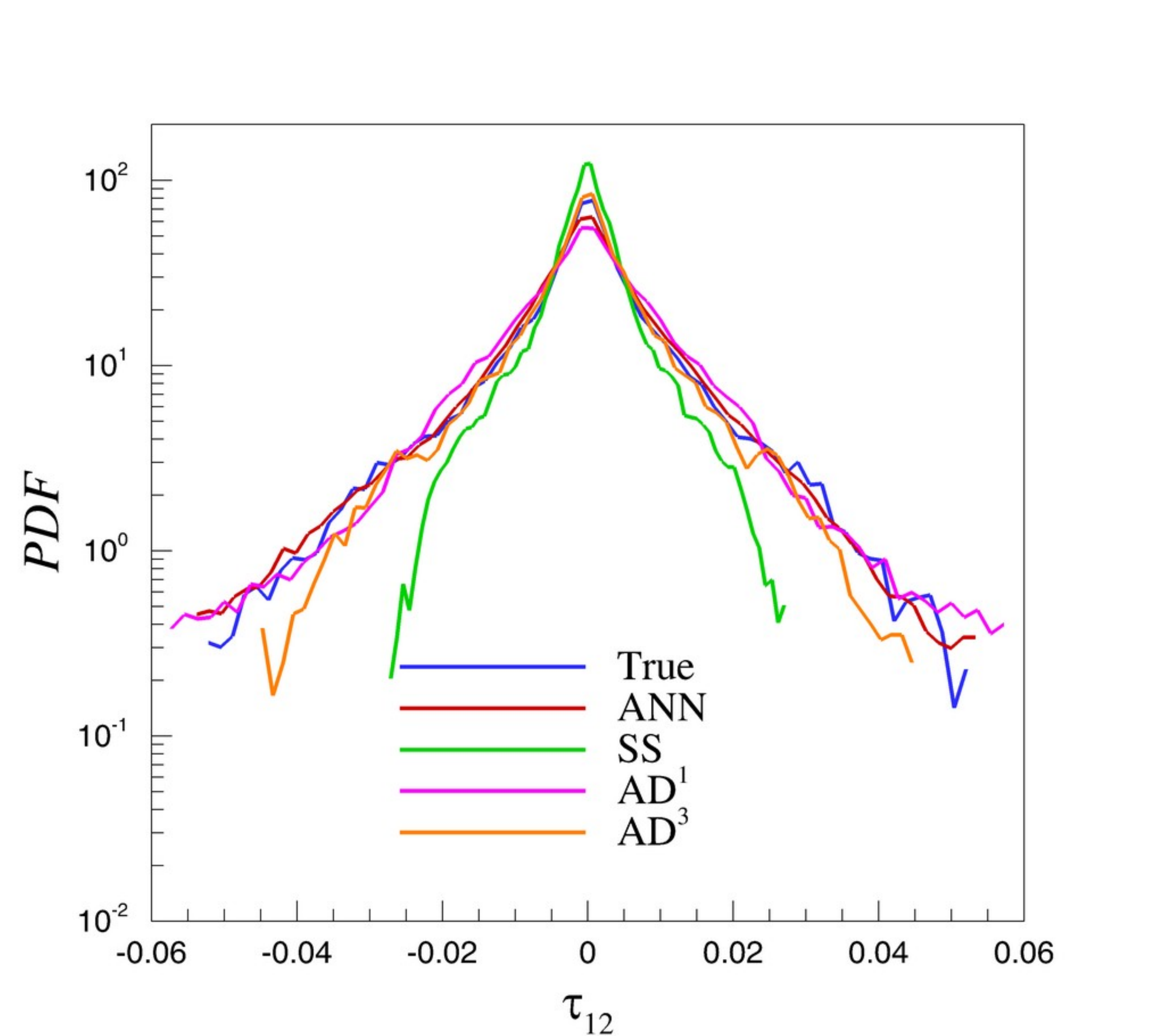}}
}\\
\mbox{
\subfigure[$\tau_{13}$]{\includegraphics[width=0.44\textwidth]{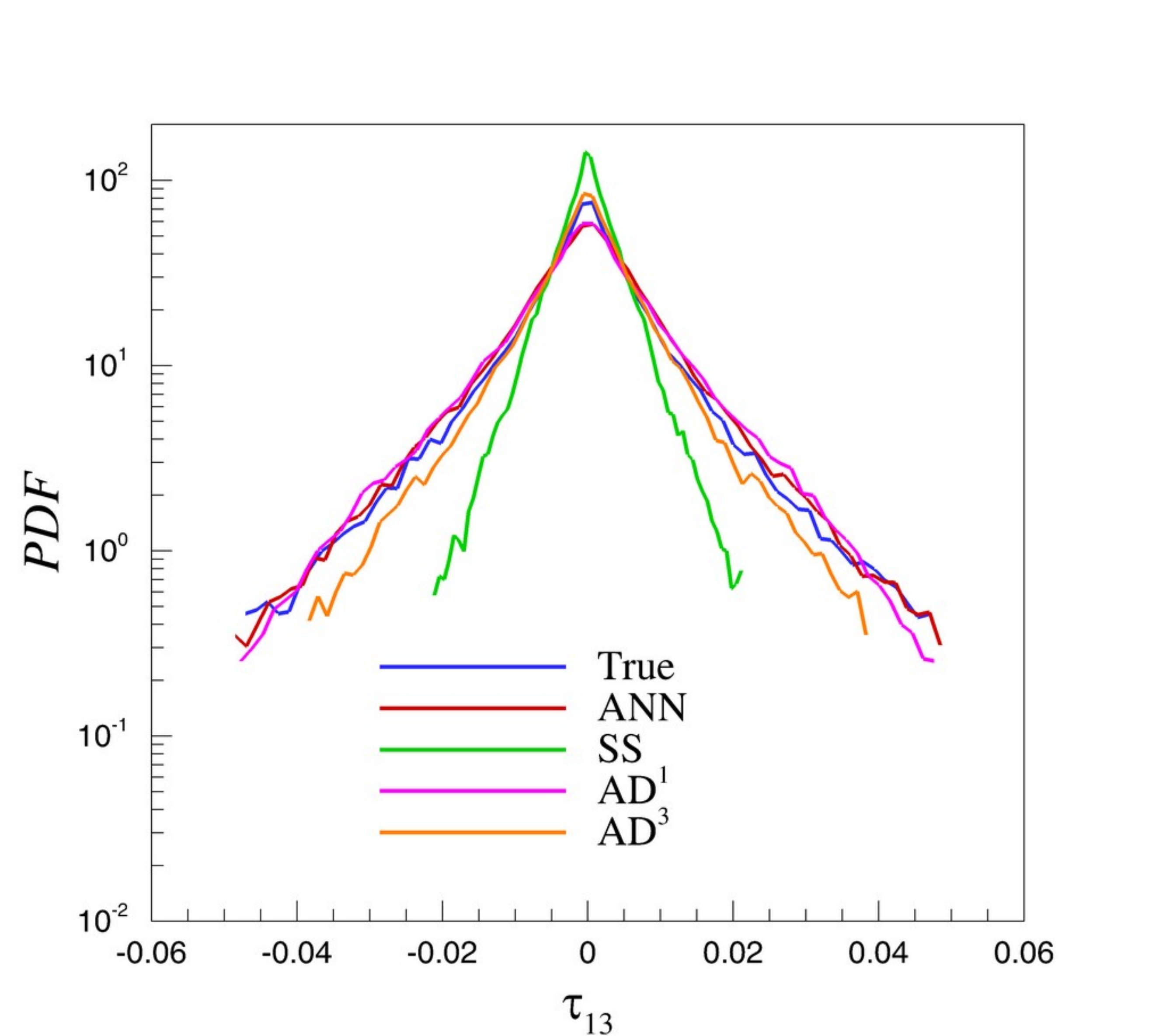}}
\subfigure[$\tau_{22}$]{\includegraphics[width=0.44\textwidth]{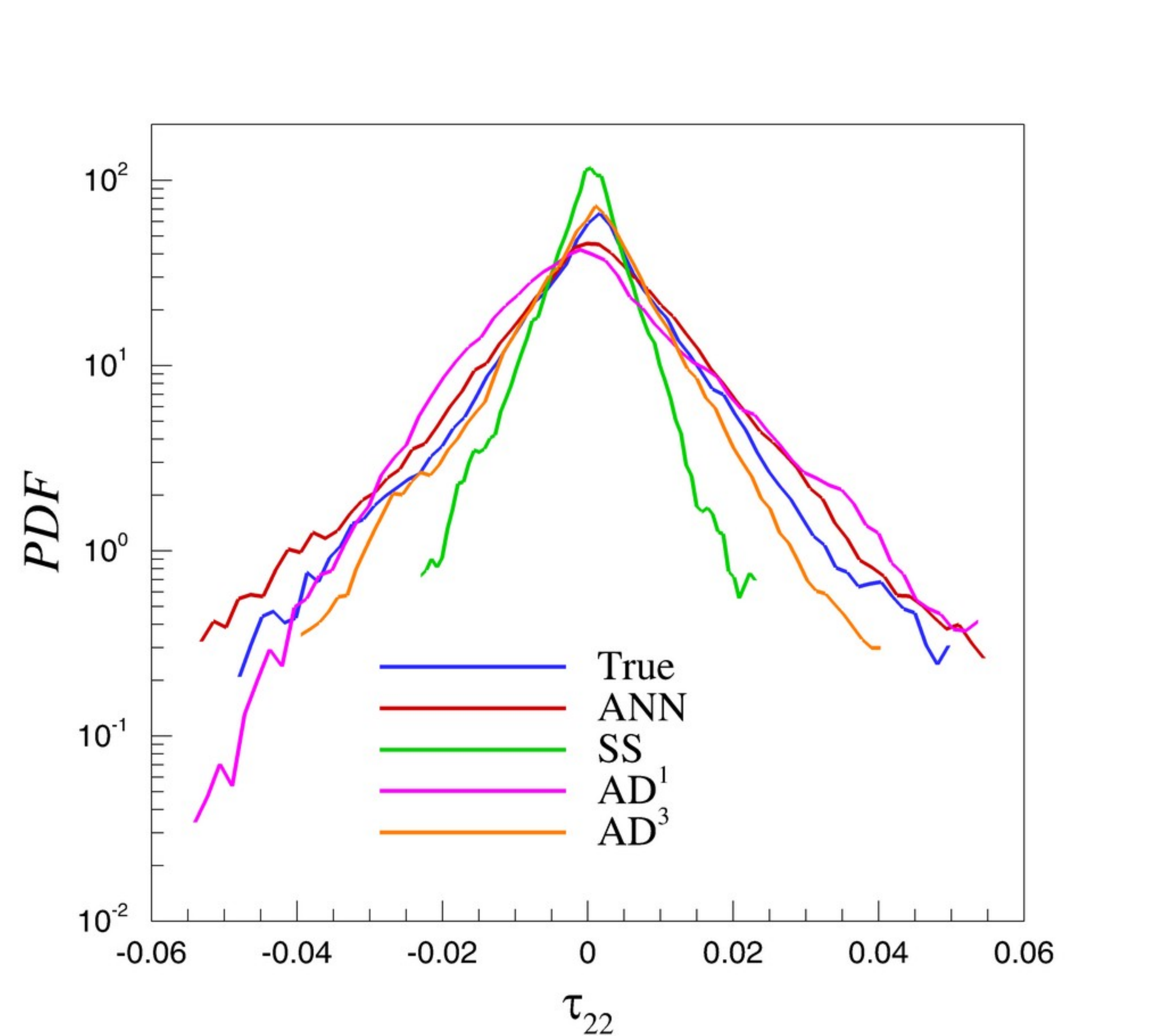}}
}
\\
\mbox{
\subfigure[$\tau_{32}$]{\includegraphics[width=0.44\textwidth]{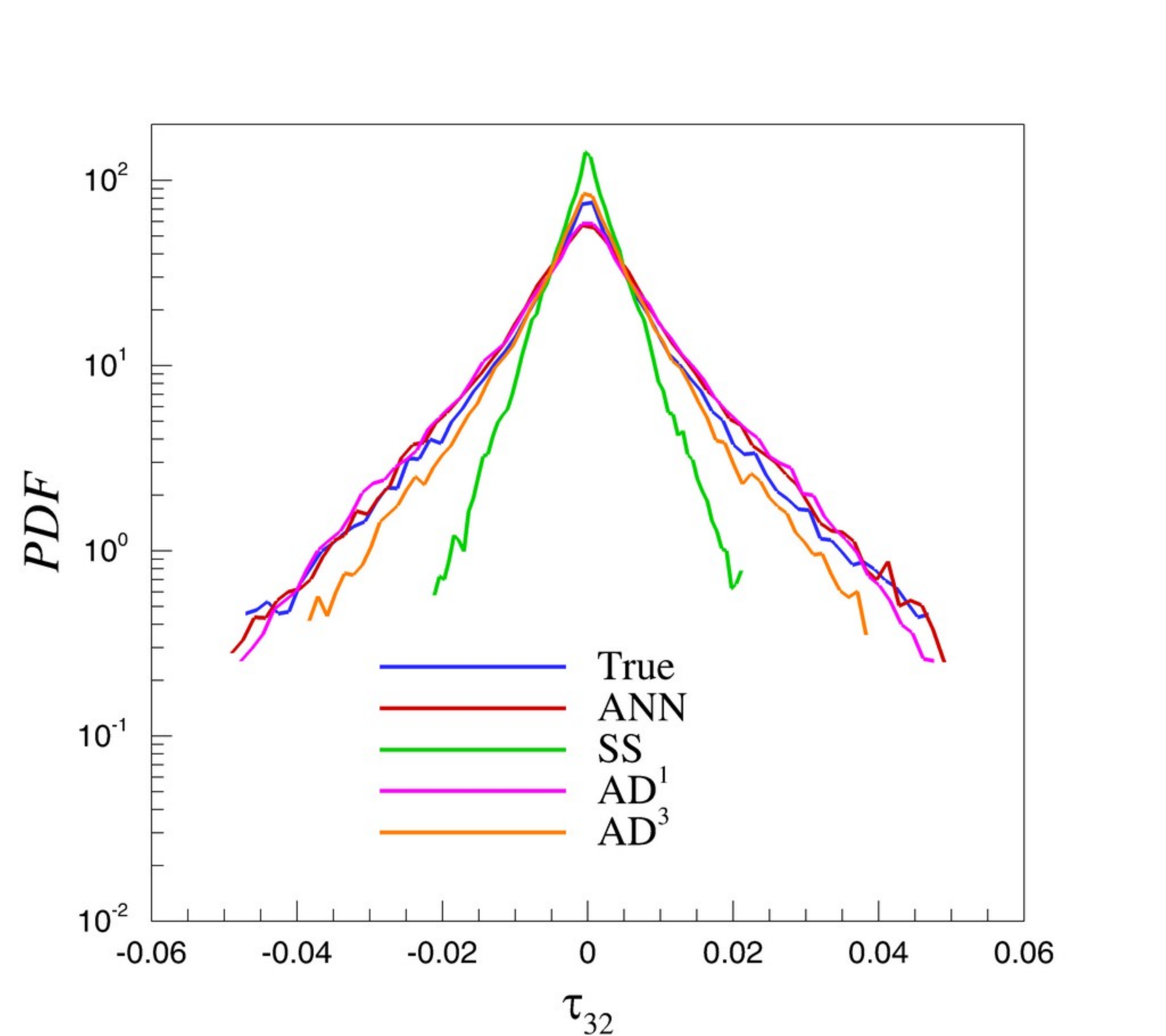}}
\subfigure[$\tau_{33}$]{\includegraphics[width=0.44\textwidth]{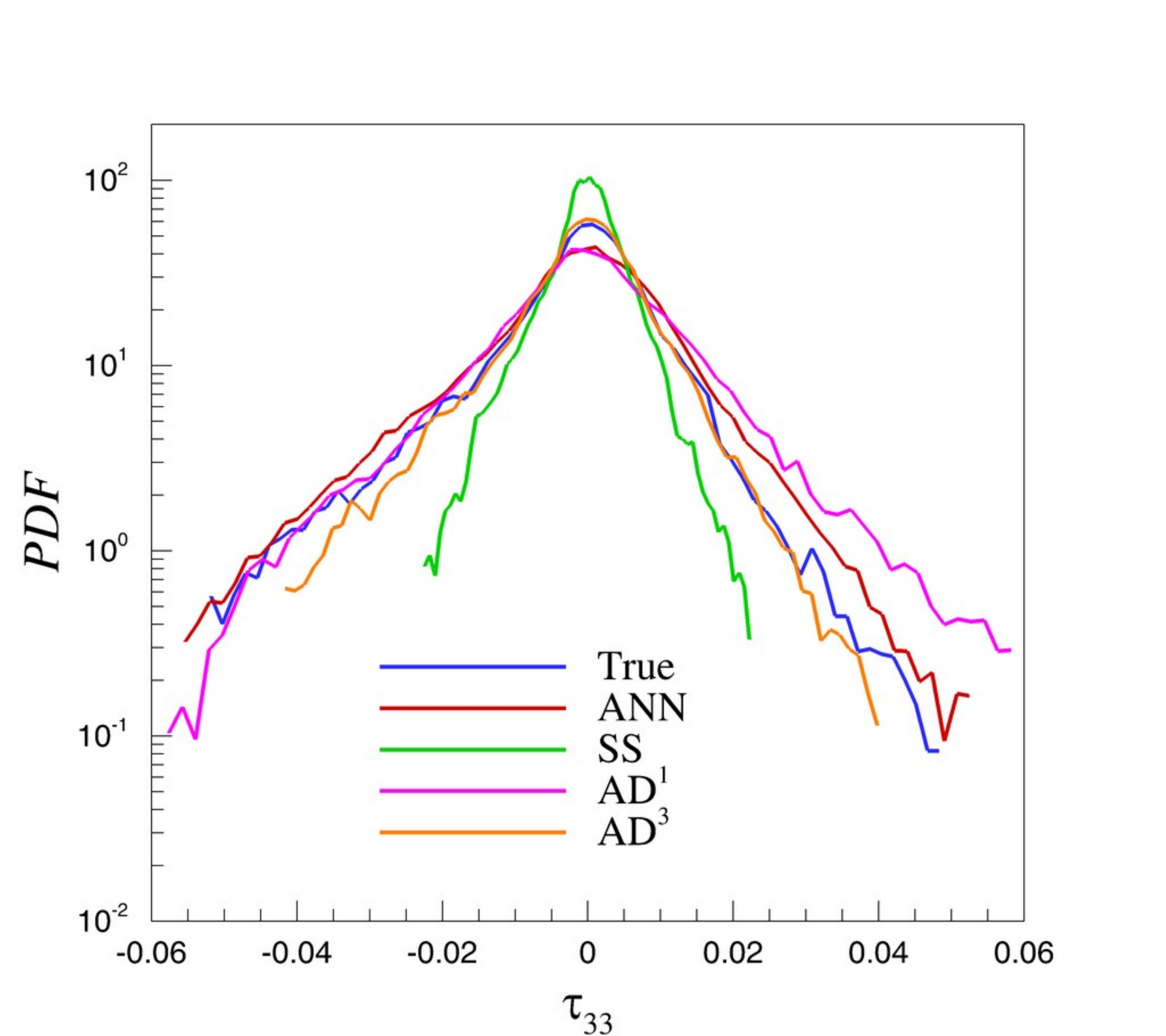}}
}
\caption{A-priori results for Kolmogorov turbulence subfilter stress predictions by the proposed architecture. Probability density functions for different subfilter stress components along with predictions by state of the art structural closures. Our data-driven architecture performs in a manner similar to these well established closure strategies without any explicit definition of a low-pass spatial filter.}
\label{fig:Kolmogorov_Closure_Comparison}
\end{figure}

A comparison of the proposed closure with the state of the art structural closure models is shown in Figure \ref{fig:Kolmogorov_Closure_Comparison} by using probability density functions for an assessment of deviatoric subfilter stress recovery performance. For the purpose of benchmarking we utilize a set of closures including the scale-similarity approach proposed by \cite{bardina1980improved} (denoted SS), the approximate deconvolution methodology given by \cite{stolz1999approximate} with 3 iterative deconvolutions  (which forms the conceptual analog of our proposed architecture and is denoted as AD\textsuperscript{3}) as well as the scale-similarity approach proposed by \cite{layton2003simple} which may be interpreted to be an approximate deconvolution methodology which simply one iteration (and hence denoted AD\textsuperscript{1}). We clarify that while the decision to utilize three iterative deconvolutions for the approximate deconvolution approach (i.e., AD\textsuperscript{3}) is rather arbitrary, past studies \citep{maulik2018explicit} have shown that a choice of the number of iterations between 3 and 5 is usually sufficient for satisfactory subfilter recovery.

Our performance assessments indicate that the proposed architecture (denoted ANN) performs in a similar fashion to other widely utilized structural subfilter modeling strategies. The AD\textsuperscript{3} approach can be seen to perform better (on average) than our proposed framework due to the fact that the specified filter utilized for the iterative deconvolution is the same as the one used for convolving the field. In comparison, we must emphasize that the proposed data-driven blind deconvolution performance is quite exceptional since no spatial filter shape is assumed. In some cases, we can also observe a slight improvement over the AD\textsuperscript{1} and SS a-priori implementations as well (both of which require the specification of a low-pass spatial filter). A quantitative characterization of the subfilter stress recovery is shown in Table \ref{table:Kolmogorov_TauMSE} where it may be seen that the proposed method results in mean-squared-errors with respect to true subfilter stresses with values near those of the popular closure models examined in this study.

\begin{table}
  \centering
  \begin{tabular}{p{1cm} p{1.8cm} p{1.8cm} p{1.8cm} p{1.8cm} p{1.8cm} p{1.8cm}}
    Model & $\tau_{11}\times10^{-5}$ & $\tau_{12}\times10^{-5}$ & $\tau_{13}\times10^{-5}$ & $\tau_{22}\times10^{-5}$ & $\tau_{32}\times10^{-5}$ & $\tau_{33}\times10^{-5}$ \\
    \hline
    ANN &  8.00 &  3.60 &  3.51 &  7.77 & 3.64  & 6.82  \\
    SS  &  6.76 &  5.62 &  5.91 &  6.76 & 5.91  &  7.95    \\
    AD\textsuperscript{1} &  31.34 &  35.82 &  21.13 & 31.33 &  21.12  & 36.82 \\
    AD\textsuperscript{3} &  2.46 &  1.69 &  1.82 & 2.46 & 1.83  & 2.91 \\
    \hline
  \end{tabular}
  \caption{Mean-squared-error values for deviatoric subfilter scale components with respect to the true subfilter scale stresses for Kolmogorov turbulence.}\label{table:Kolmogorov_TauMSE}
\end{table}

\subsection{Stratified Turbulence}

Another three-dimensional test case is given by a stratified turbulence data set obtained using $512^3$ degrees of freedom. The difference between this test case and the Kolmogorov turbulence test case (TGV) is that this particular simulation is carried out using an implicit large eddy simulation of the inviscid Euler equations (see \cite{maulik2017resolution} for details on numerics). It is well known that this framework provides a good estimation for the Navier-Stokes equations in the limit of infinite Reynolds numbers \citep{bos2006dynamics,zhou2014estimating,sytine2000convergence}. Therefore this test case may be assumed to be representative of a different fluid physics in comparison to the Kolmogorov turbulence problem.

\begin{figure}
\centering
\mbox{
\subfigure[$t=0$]{\includegraphics[width=0.44\textwidth]{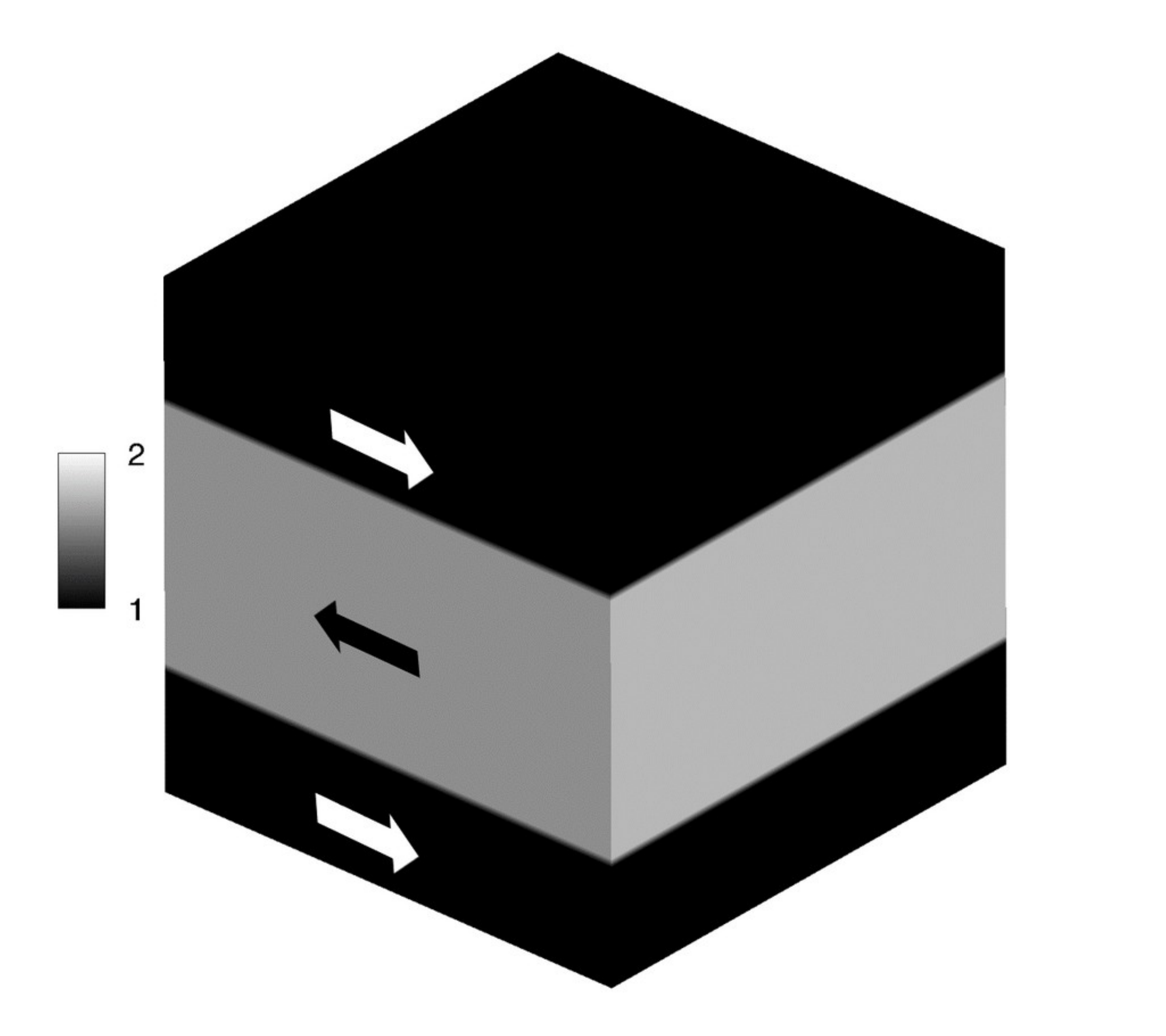}}
\subfigure[$t=1$]{\includegraphics[width=0.44\textwidth]{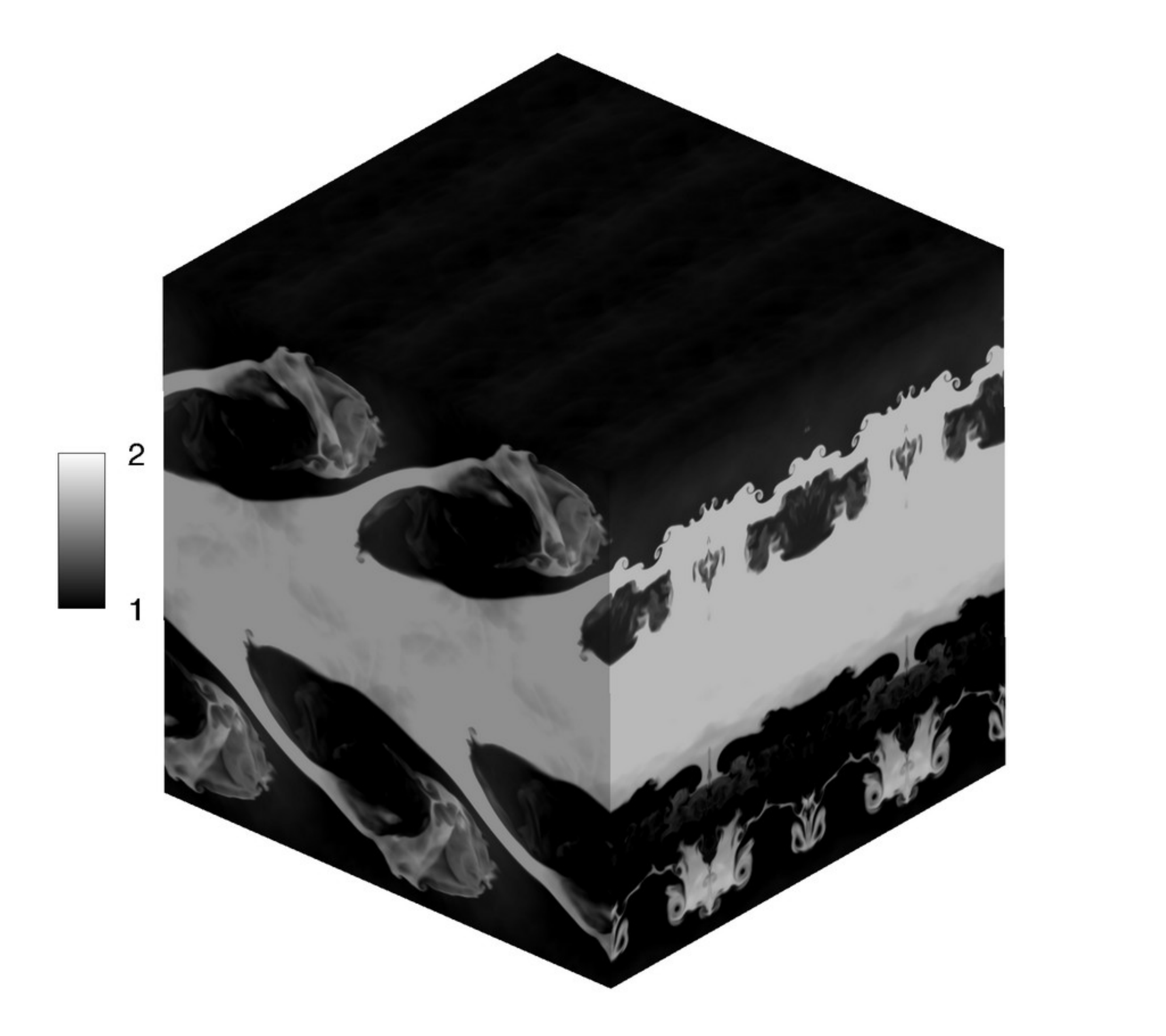}}
}\\
\mbox{
\subfigure[$t=3$]{\includegraphics[width=0.44\textwidth]{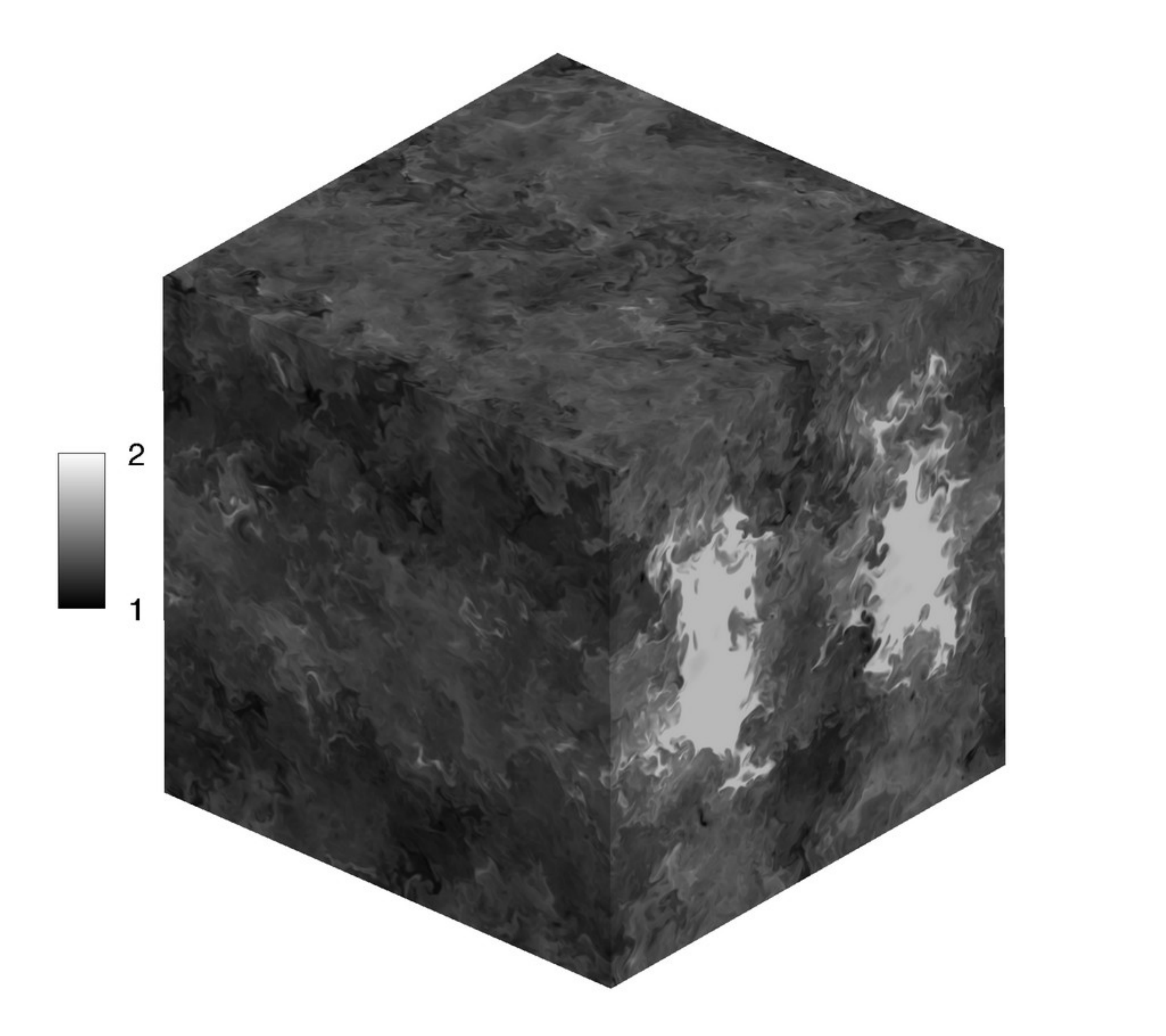}}
\subfigure[$t=5$]{\includegraphics[width=0.44\textwidth]{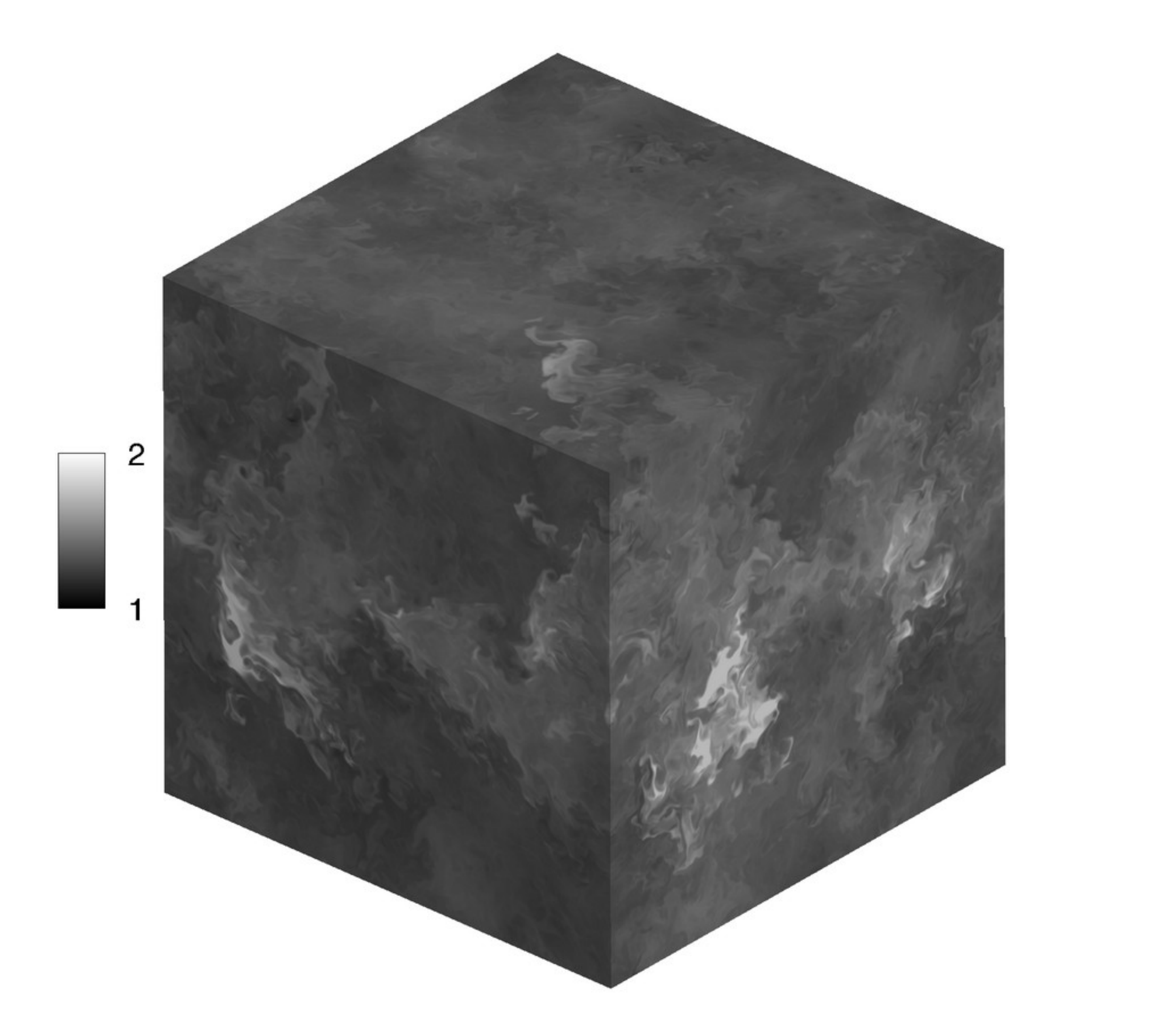}}
}
\caption{Evolution of density contours for the stratified turbulence problem through time.}
\label{KHI_Schem}
\end{figure}

Our stratified turbulence test case is obtained through the simulation of a 3D analog of the study presented in \cite{maulik2017resolution} to generate a Kelvin-Helmholtz instability (KHI) which eventually transitions through nonlinear interactions to fully developed compressible turbulence. Figure~\ref{KHI_Schem} displays density contours of the evolution of the system from its initial condition to a completely turbulent field. The system is evolved from a stratified initial condition with a denser fluid layer sandwiched between two lighter layers. The middle layer is given an initial velocity (in the negative $x$ direction) and the upper and lower layers are given an equal velocity magnitude in the opposite direction. The shearing velocities in both layers are specified to ensure the characteristics of a moderately compressible turbulent field (i.e., initial aggregate Mach number values of 0.54 and 0.75 in double shear layers).

\begin{figure}
\centering
\mbox{
\subfigure[Test data 1]{
\includegraphics[width=0.44\textwidth]{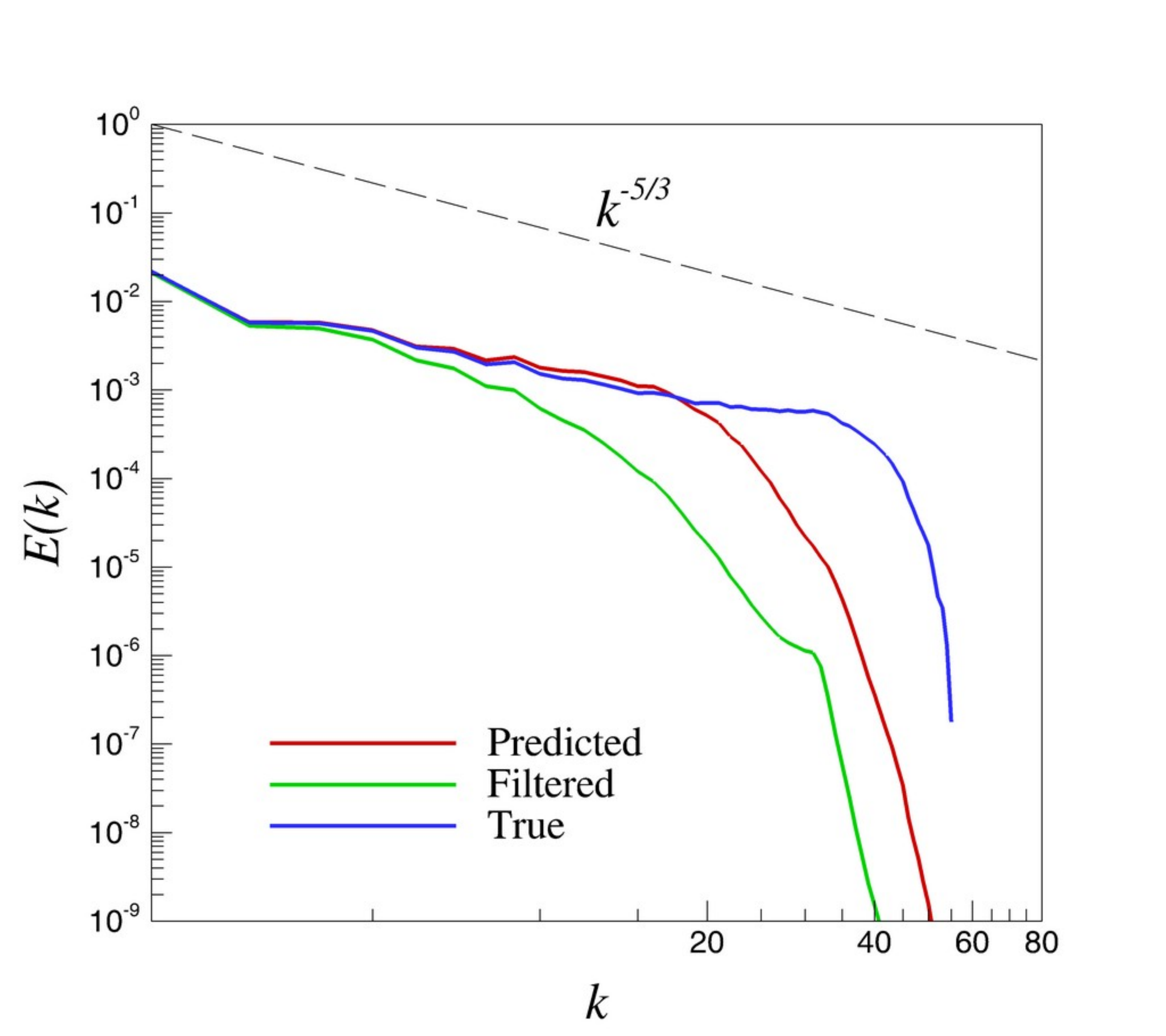}
\includegraphics[width=0.44\textwidth]{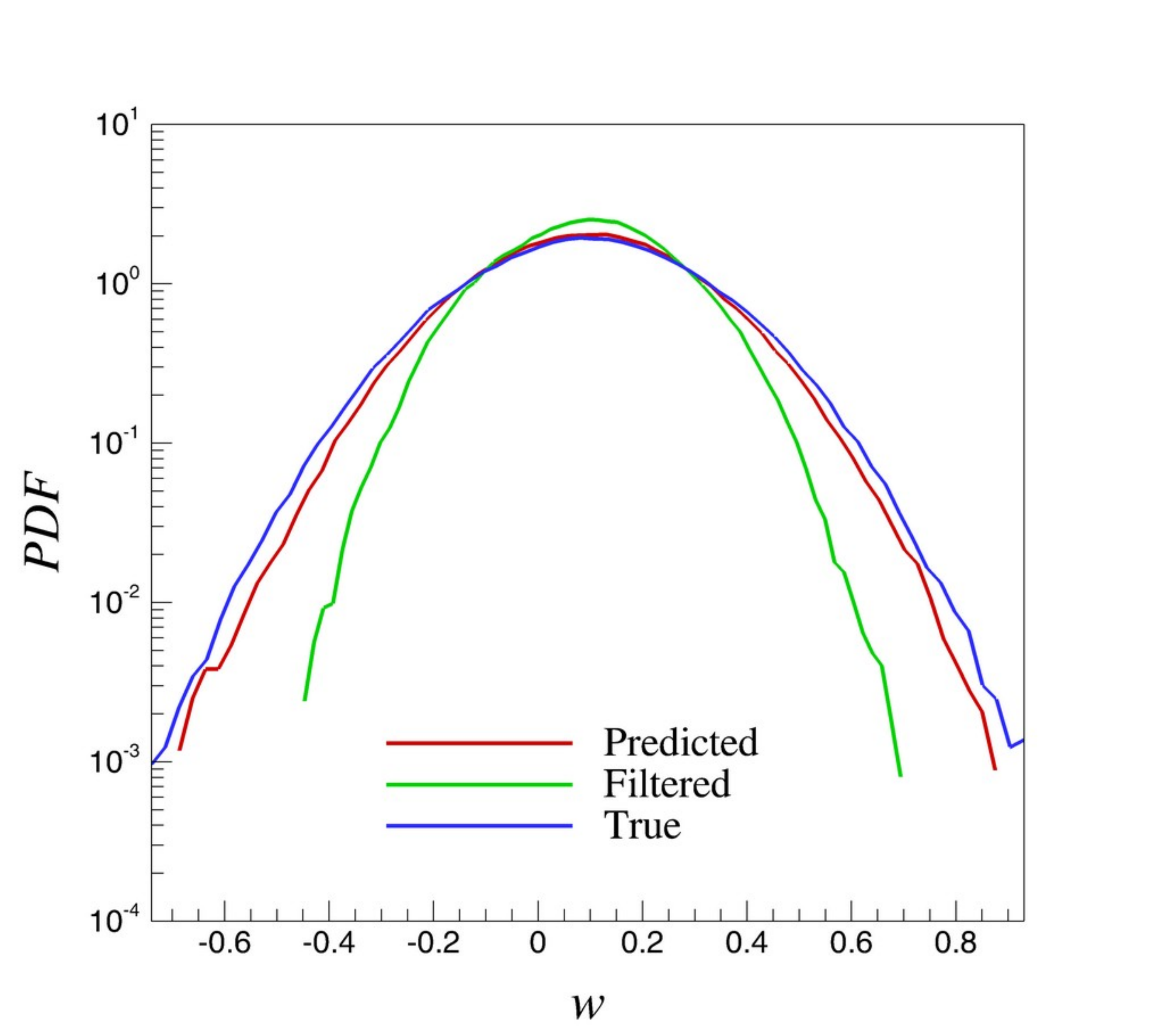}
}
}\\
\mbox{
\subfigure[Test data 2]{
\includegraphics[width=0.44\textwidth]{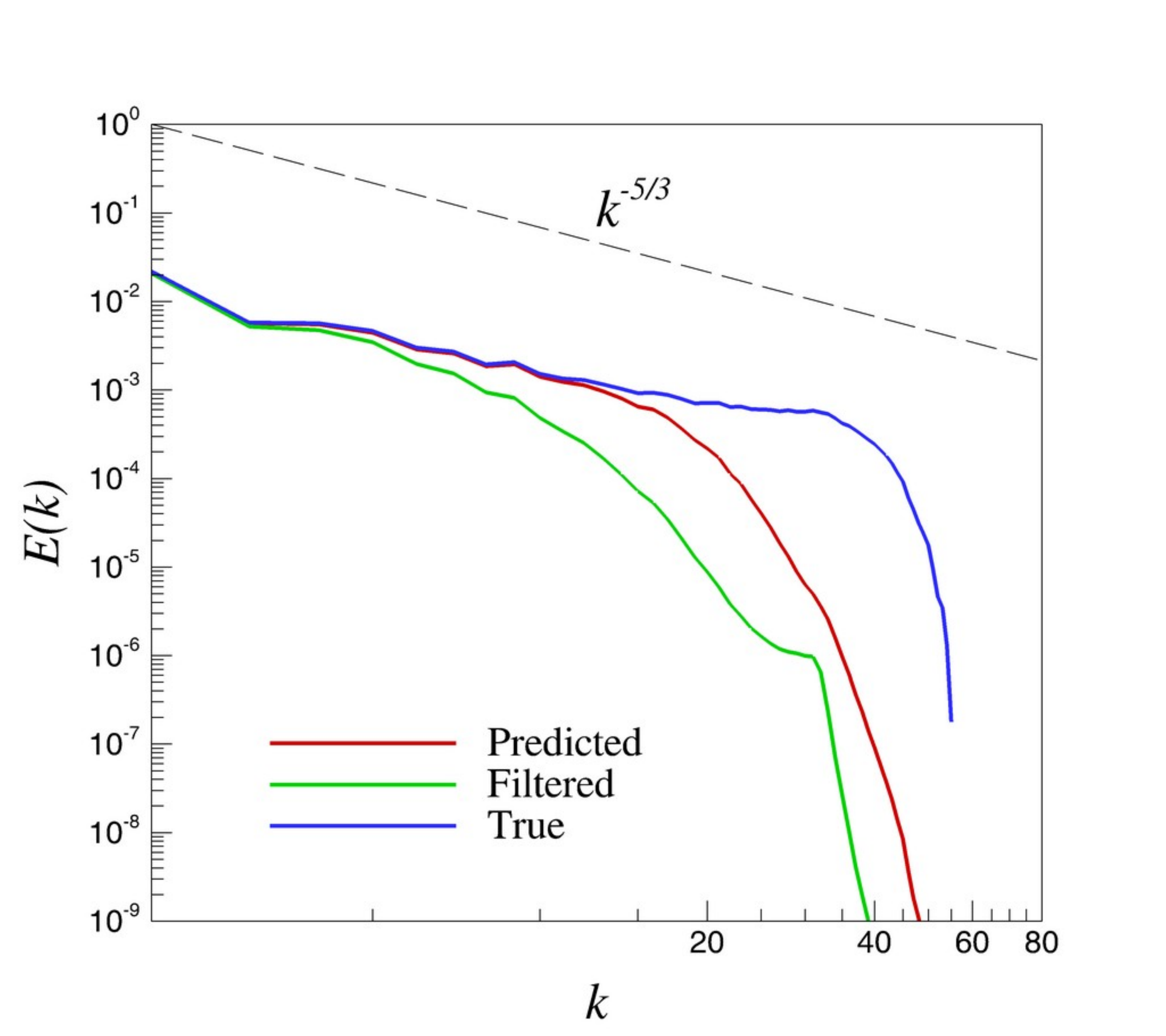}
\includegraphics[width=0.44\textwidth]{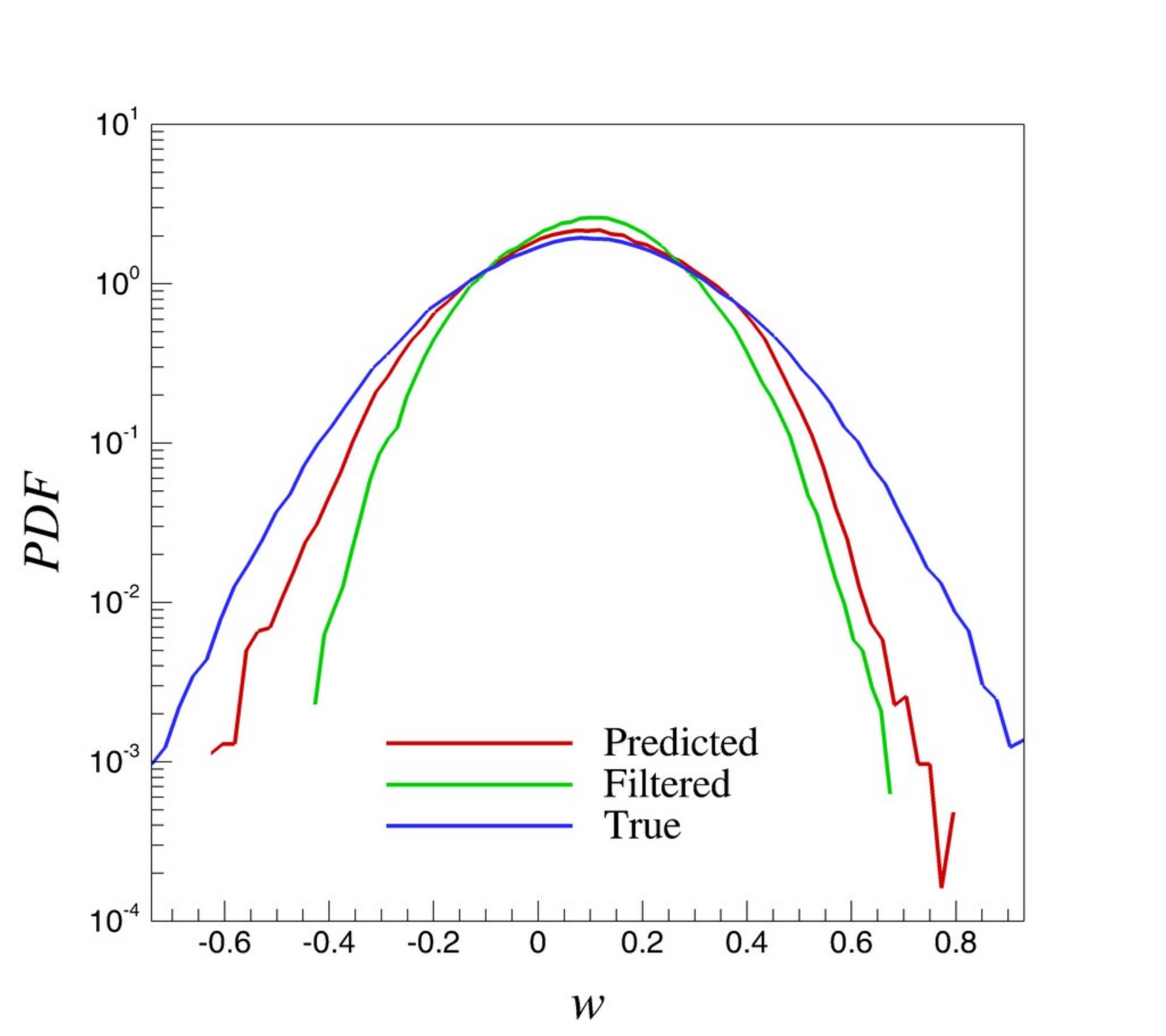}
}
}\\
\mbox{
\subfigure[Test data 3]{
\includegraphics[width=0.44\textwidth]{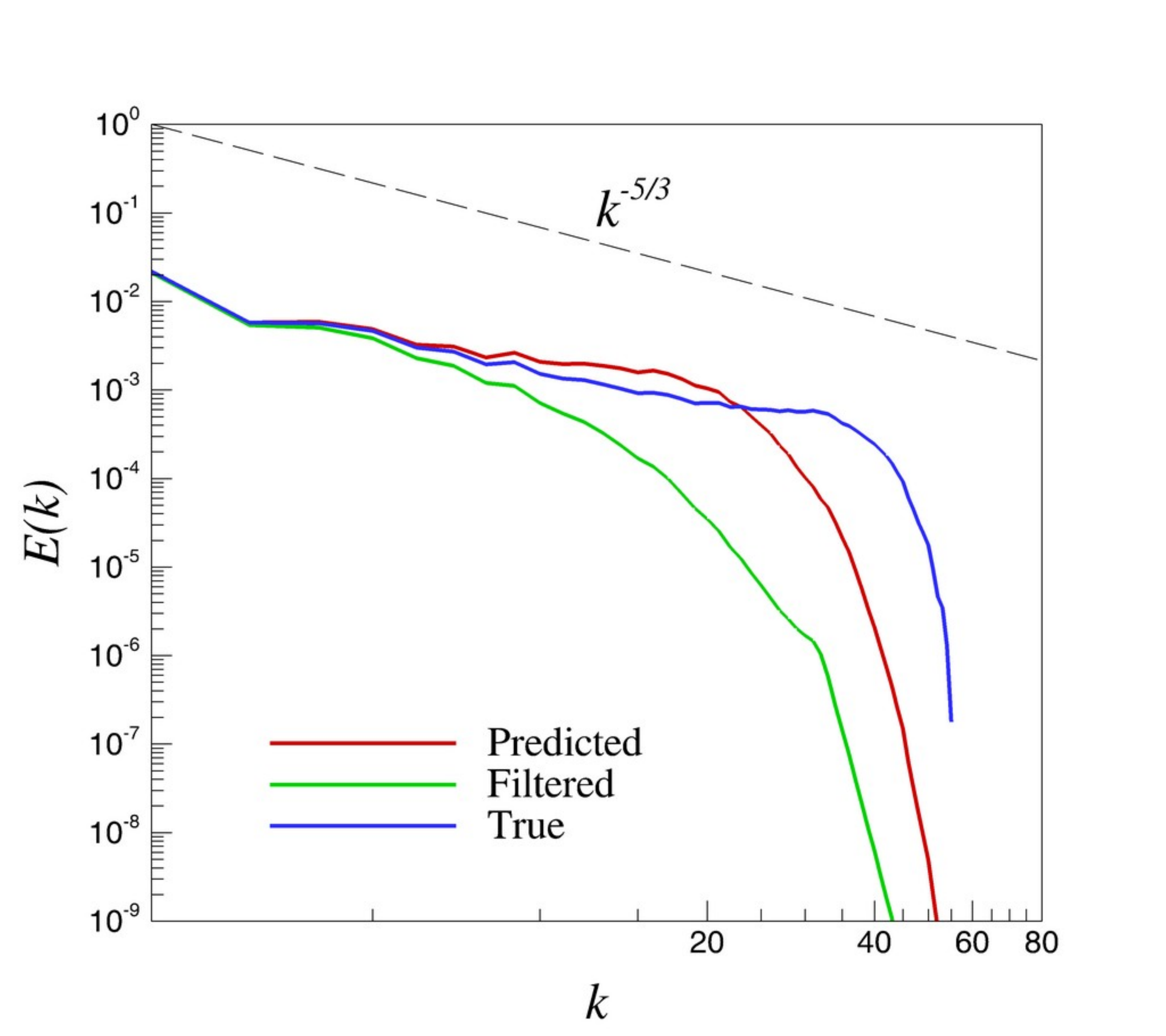}
\includegraphics[width=0.44\textwidth]{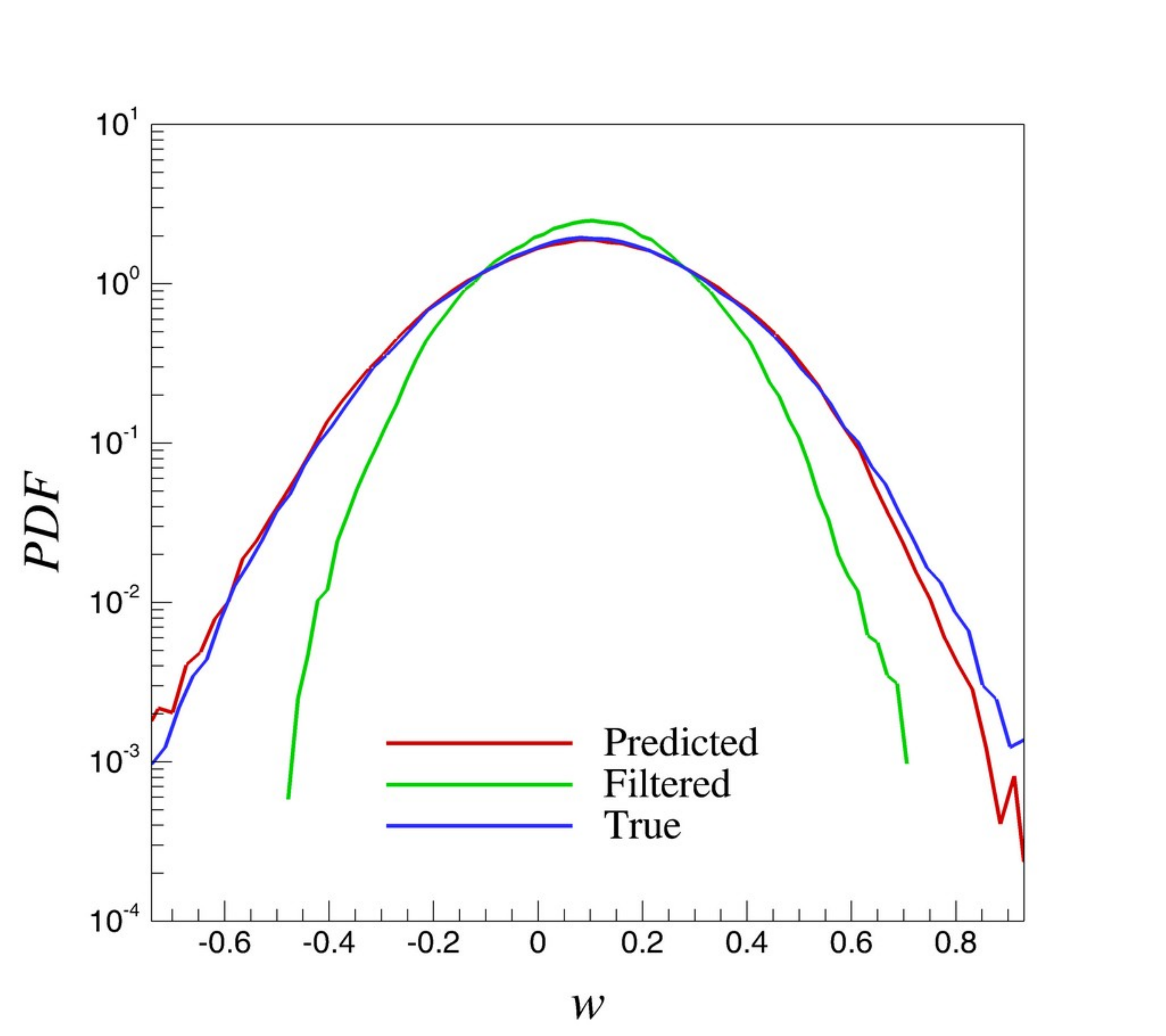}
}
}
\caption{A-priori results of the kinetic energy spectra (left) and PDF of the $z$ component of velocity (right) for stratified turbulence. Results for three different deconvolution test data sets shown.}
\label{fig:Spec3D_KHI_Filtered}
\end{figure}

We perform assessments similar to our previous test cases with three sets of testing data each for the examination of both deconvolution and noise reduction performance. Figure~\ref{fig:Spec3D_KHI_Filtered} shows the deconvolution performance of the proposed framework where we once again witness a notable recovery of the inertial range from the low-pass spatially filtered test data. This is also consistently observed for the differently perturbed test cases. PDF trends for the $z$ component of the velocity are examined as previously and reveal expected trends in high frequency recovery. We once again observe that the proposed framework performs better for test data 1 and 3 as compared to test data 2 where capture of the tails of the PDF is marginal. We may attribute this to the same reasons detailed in the previous subsection. We remark, however, that even for test data 2 our closure captures the peak of the PDF and the bulk of its distribution about the mean rather well (which leads to excellent inertial range capture).

\begin{figure}
\centering
\mbox{
\subfigure[Test data 1]{
\includegraphics[width=0.44\textwidth]{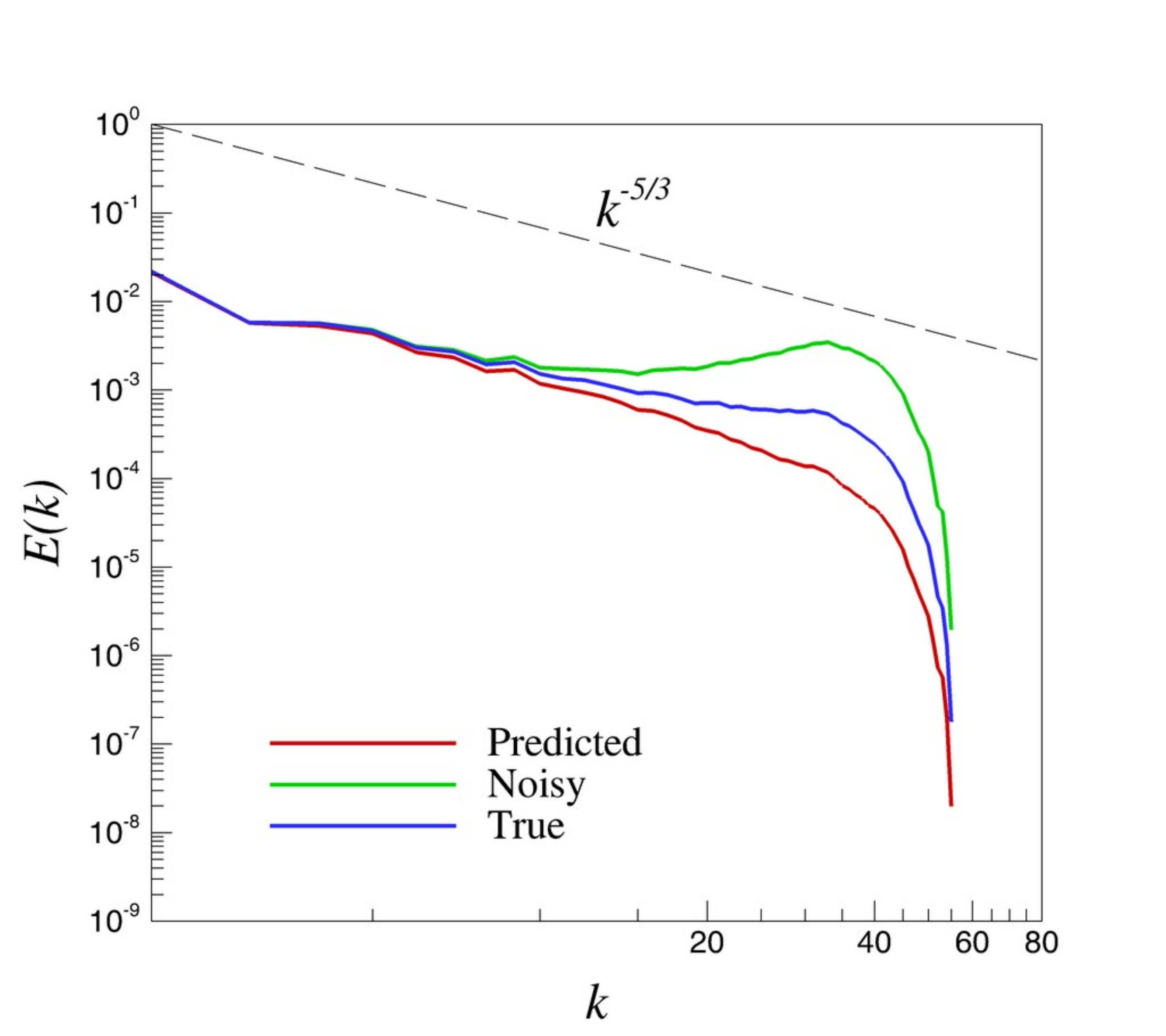}
\includegraphics[width=0.44\textwidth]{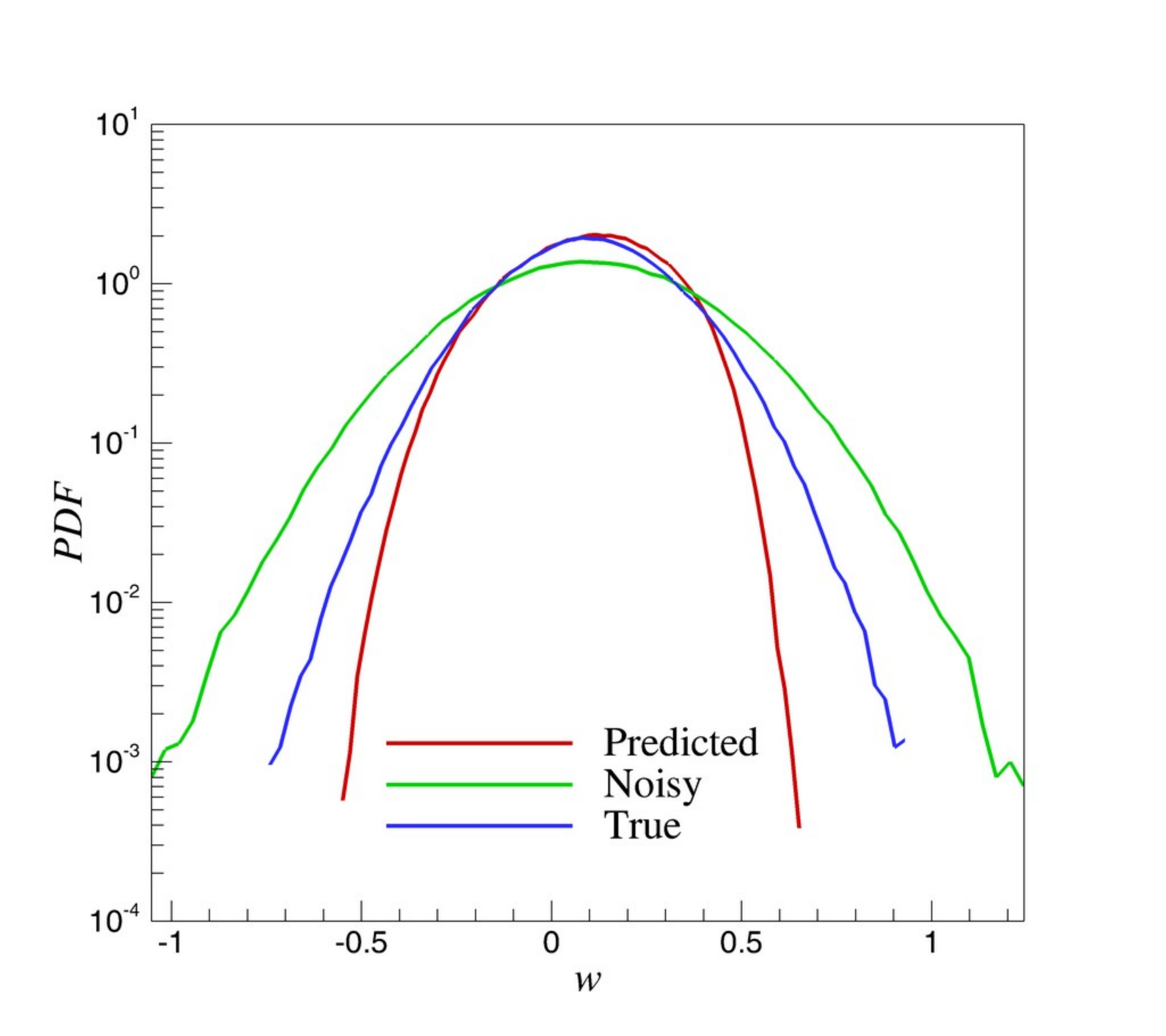}
}
}\\
\mbox{
\subfigure[Test data 2]{
\includegraphics[width=0.44\textwidth]{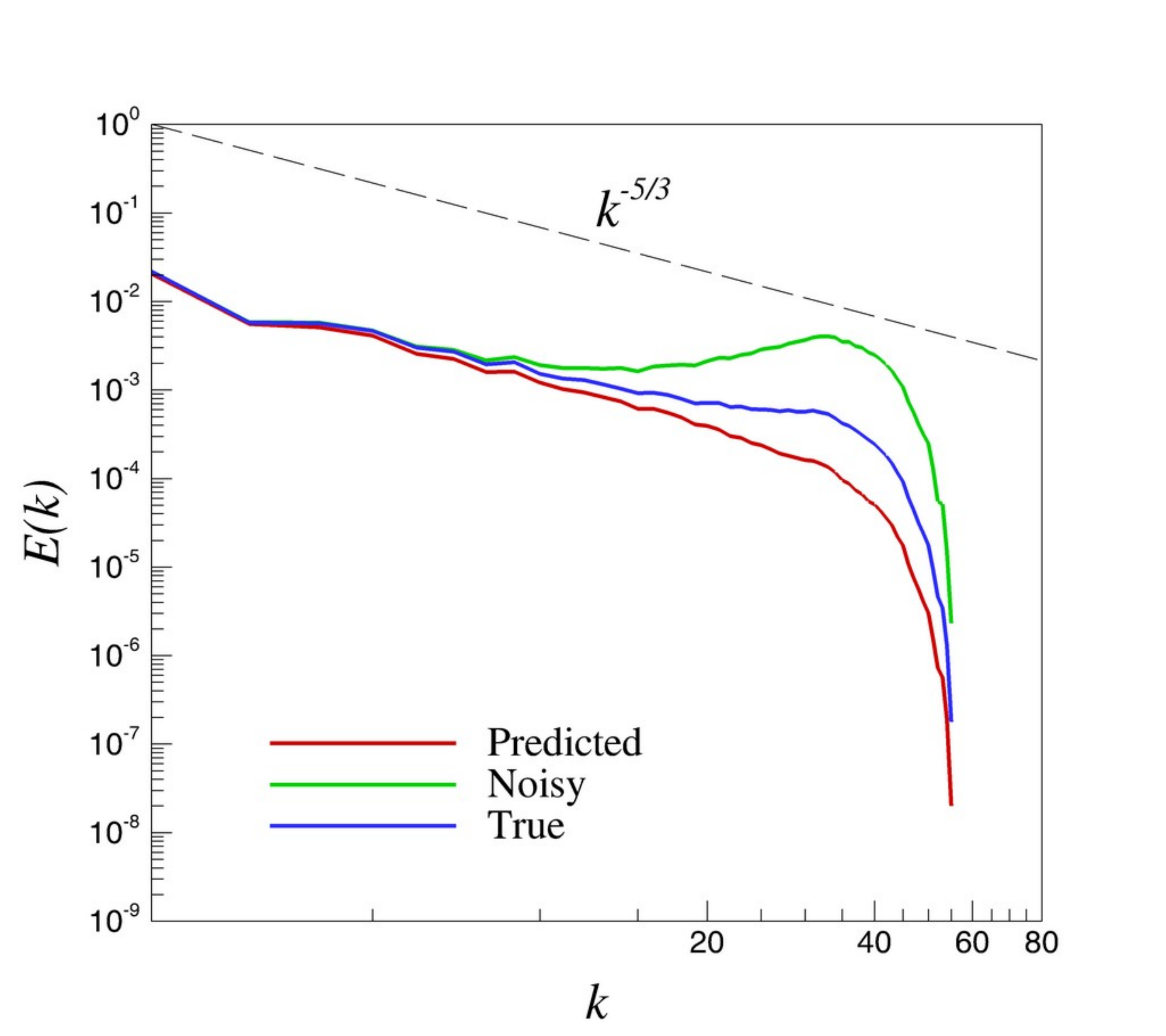}
\includegraphics[width=0.44\textwidth]{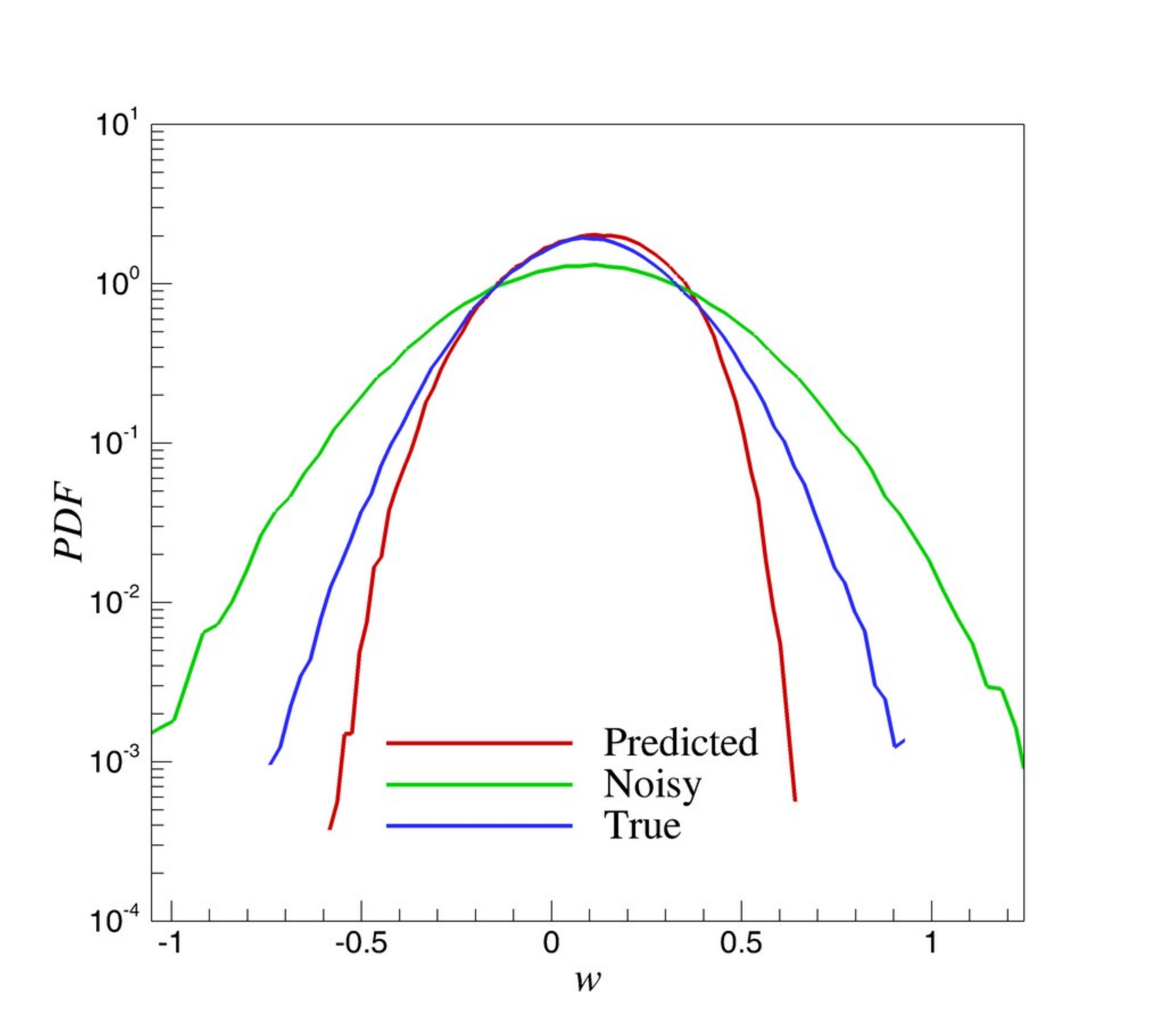}
}
}\\
\mbox{
\subfigure[Test data 3]{
\includegraphics[width=0.44\textwidth]{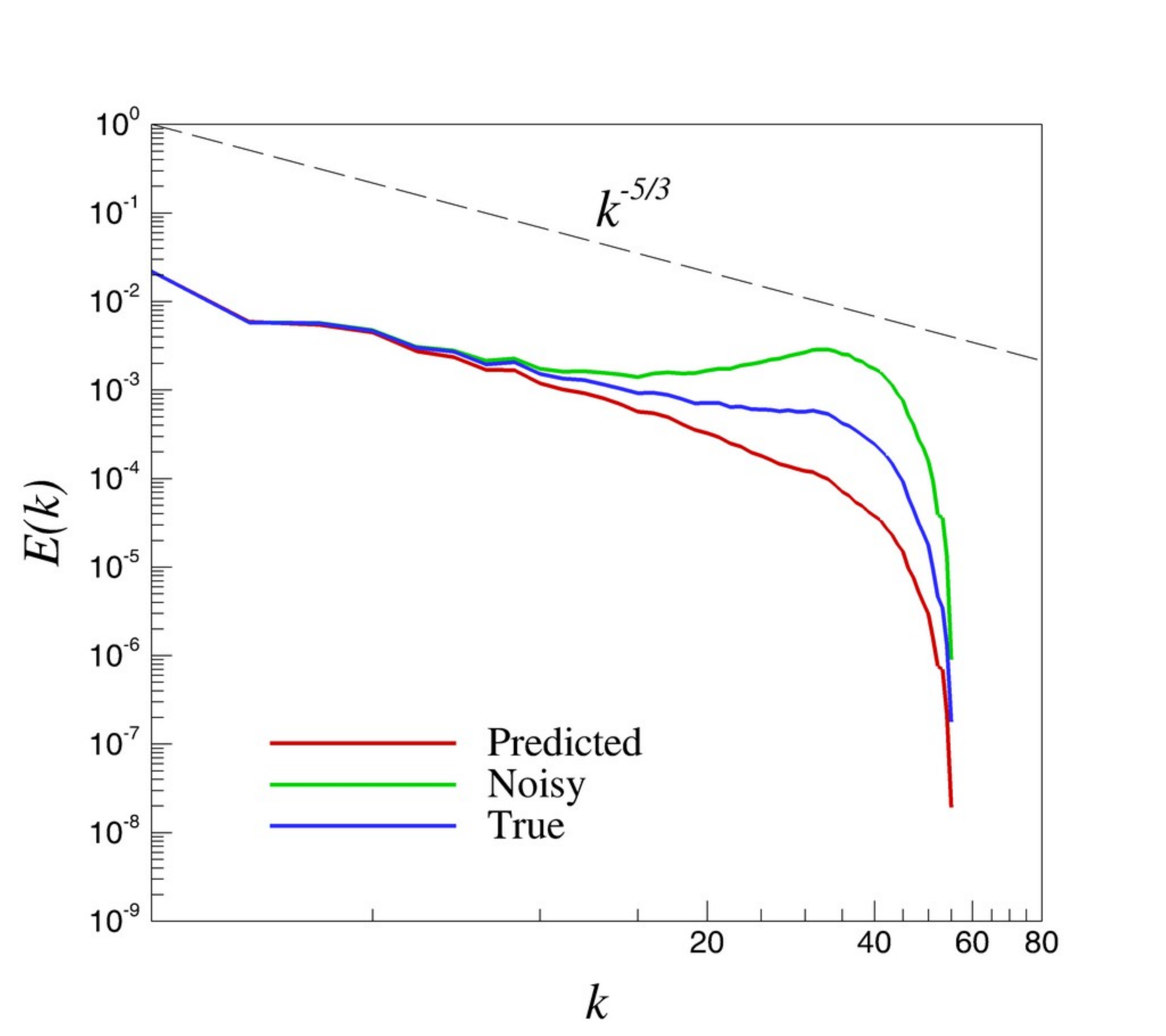}
\includegraphics[width=0.44\textwidth]{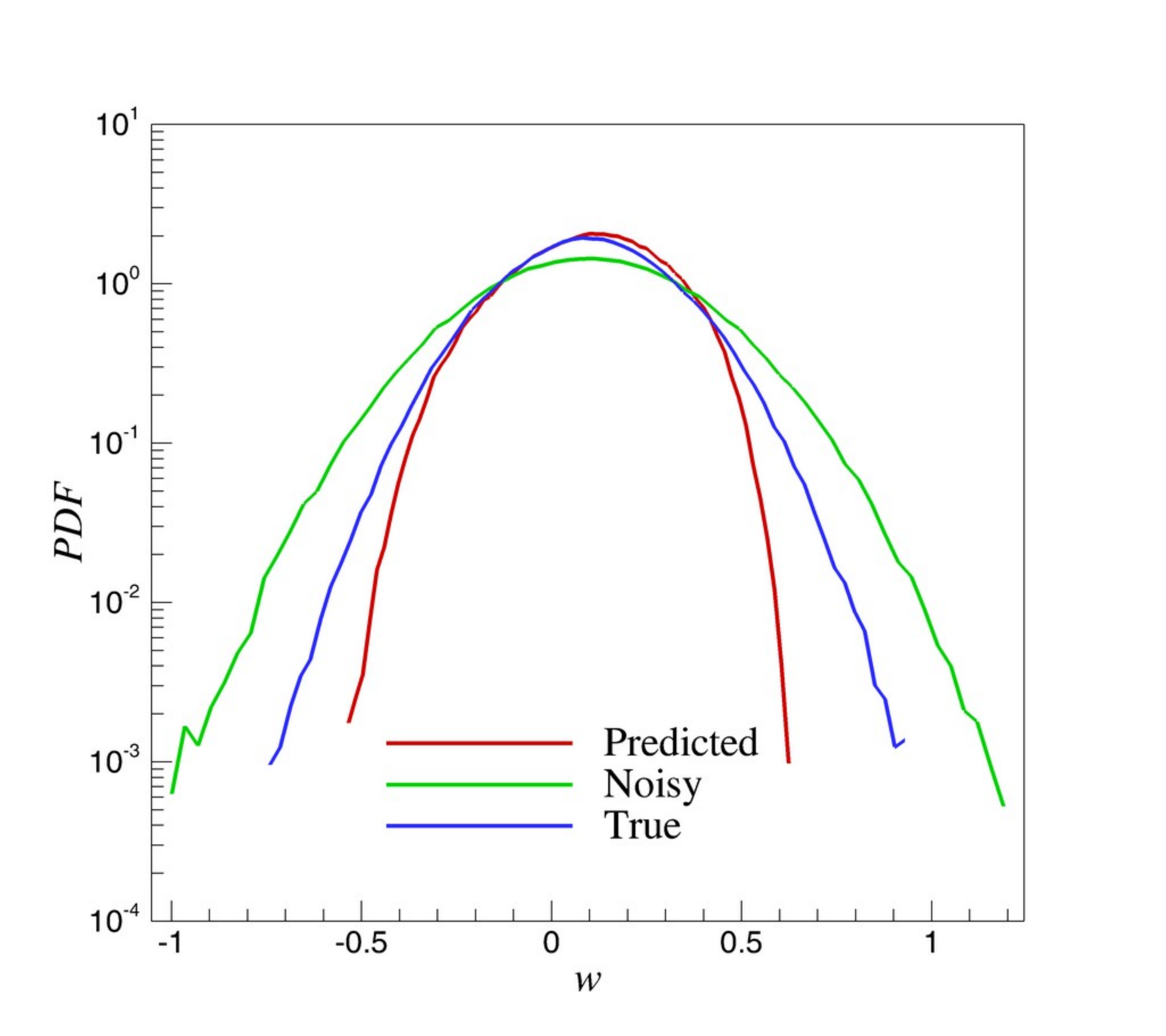}
}
}
\caption{A-priori results of the kinetic energy spectra (left) and PDF of the $z$ component of velocity (right) for stratified turbulence. Results for three different regularization test data sets shown.}
\label{fig:Spec3D_KHI_Noised}
\end{figure}

\begin{figure}
\centering
\mbox{
\subfigure[True]{\includegraphics[width=0.3\textwidth]{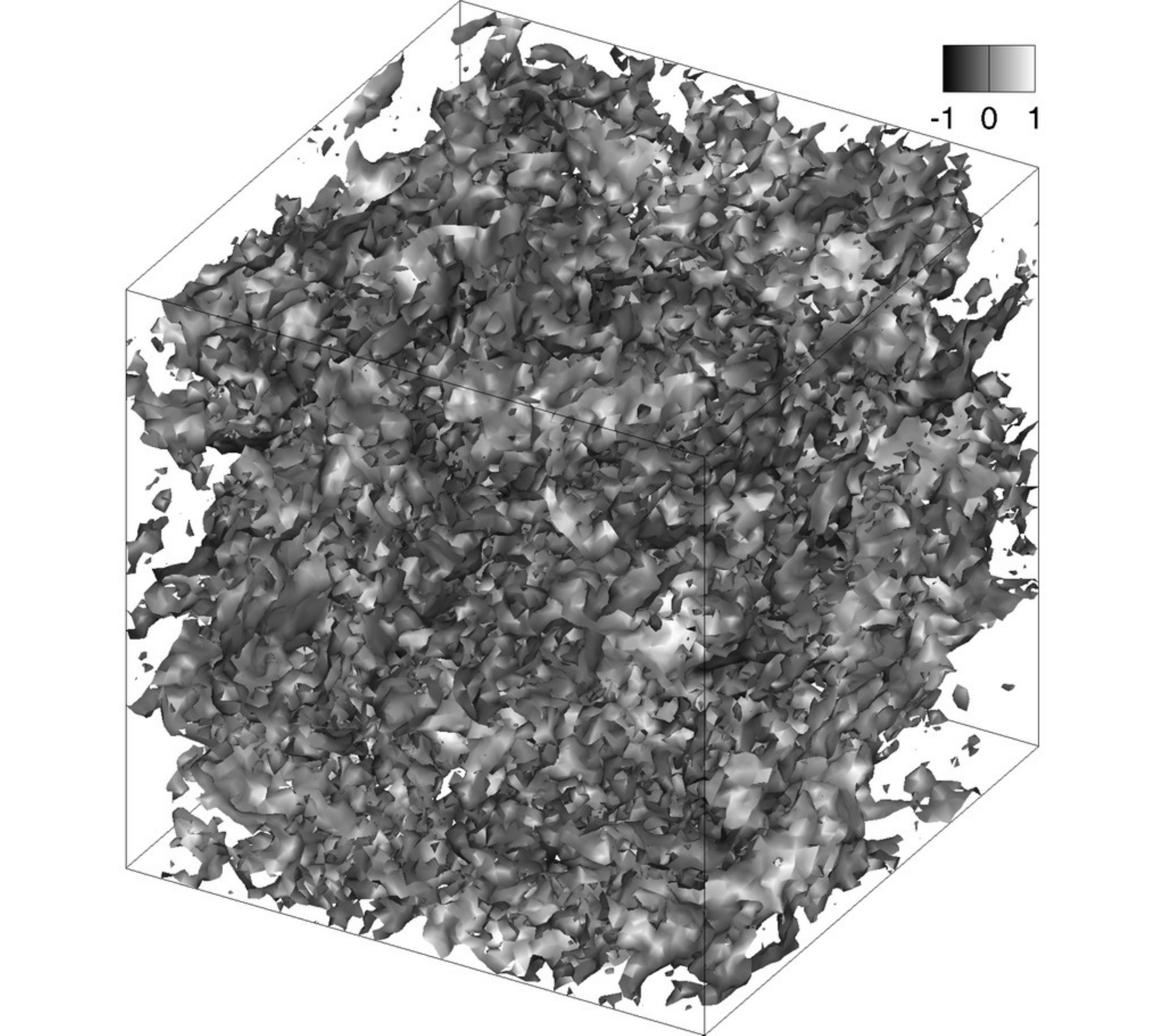}}
\subfigure[Filtered]{\includegraphics[width=0.3\textwidth]{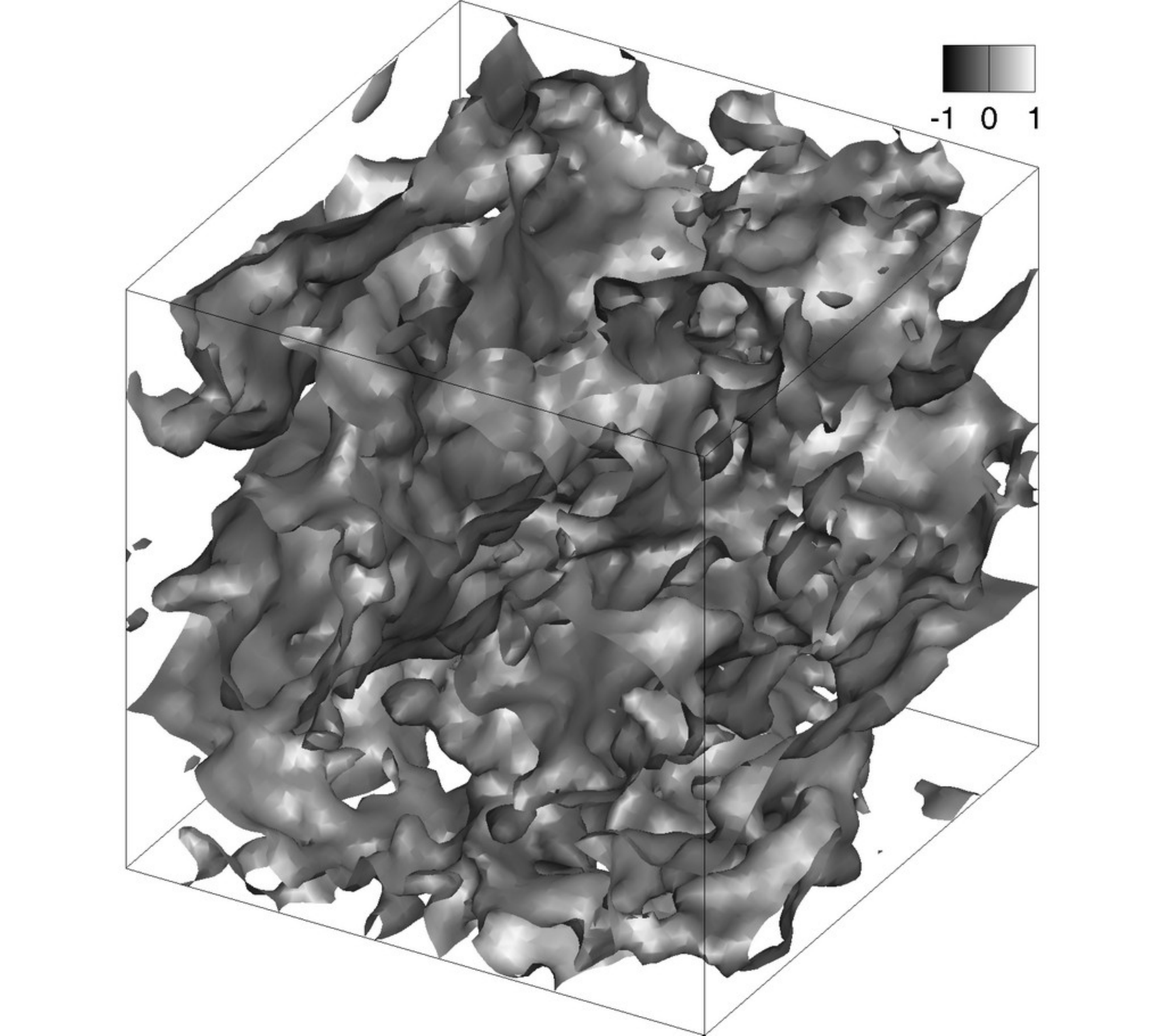}}
\subfigure[Recovered]{\includegraphics[width=0.3\textwidth]{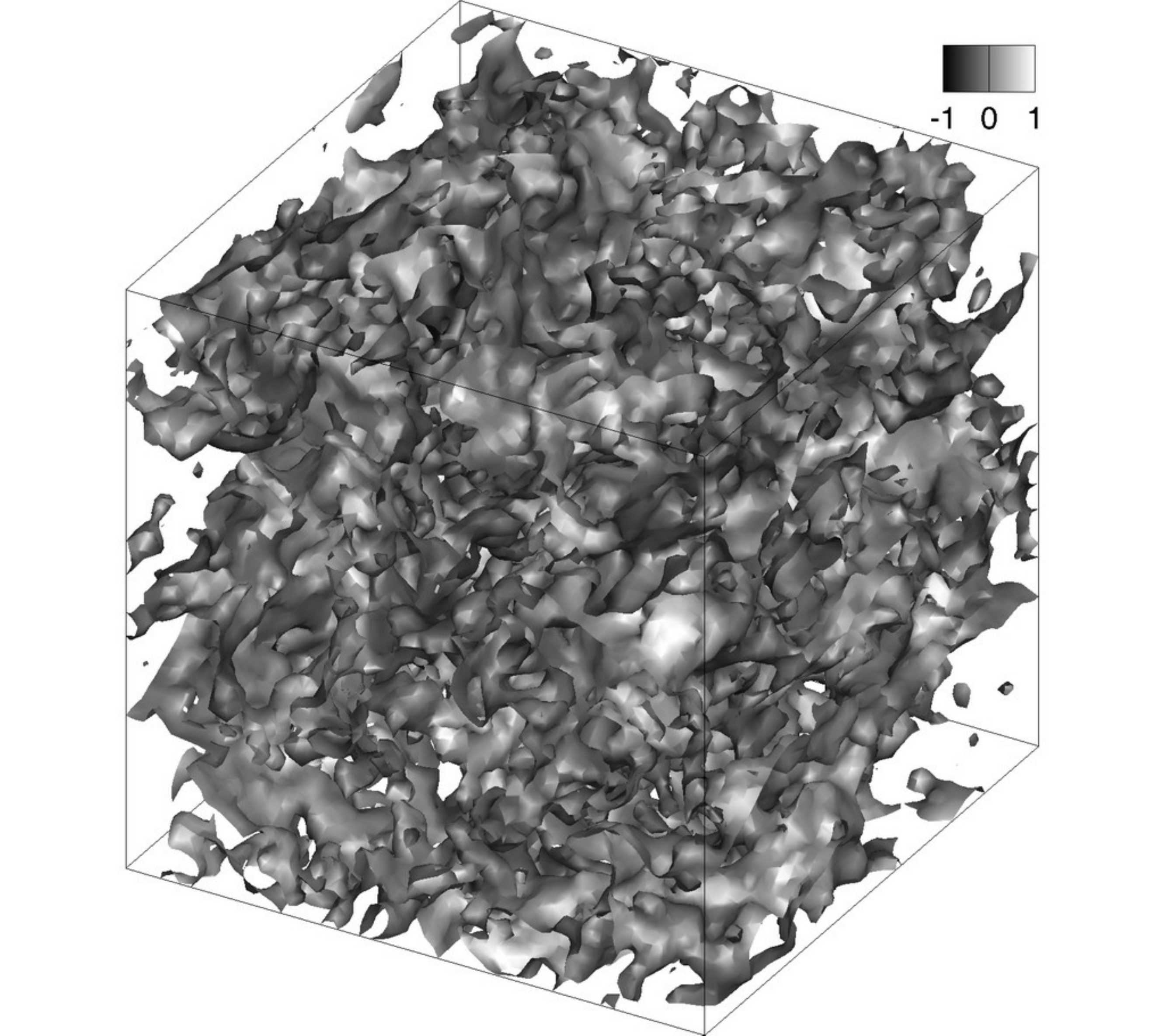}}
}\\
\caption{A-priori results for velocity field recovery from low-pass spatially filtered perturbations for stratified turbulence. Data shown for deconvolution test data 1. Isosurfaces for $x$ component of the velocity colored by $z$ component are shown: (a) true coarse-grained fields, (b) coarse-grained fields with Gaussian smoothing, and (c) coarse-grained fields reconstructed using proposed framework.}
\label{fig:field3D_KHI_1}
\end{figure}

\begin{figure}
\centering
\mbox{
\subfigure[True]{\includegraphics[width=0.3\textwidth]{3D_KHI_True.pdf}}
\subfigure[Noisy]{\includegraphics[width=0.3\textwidth]{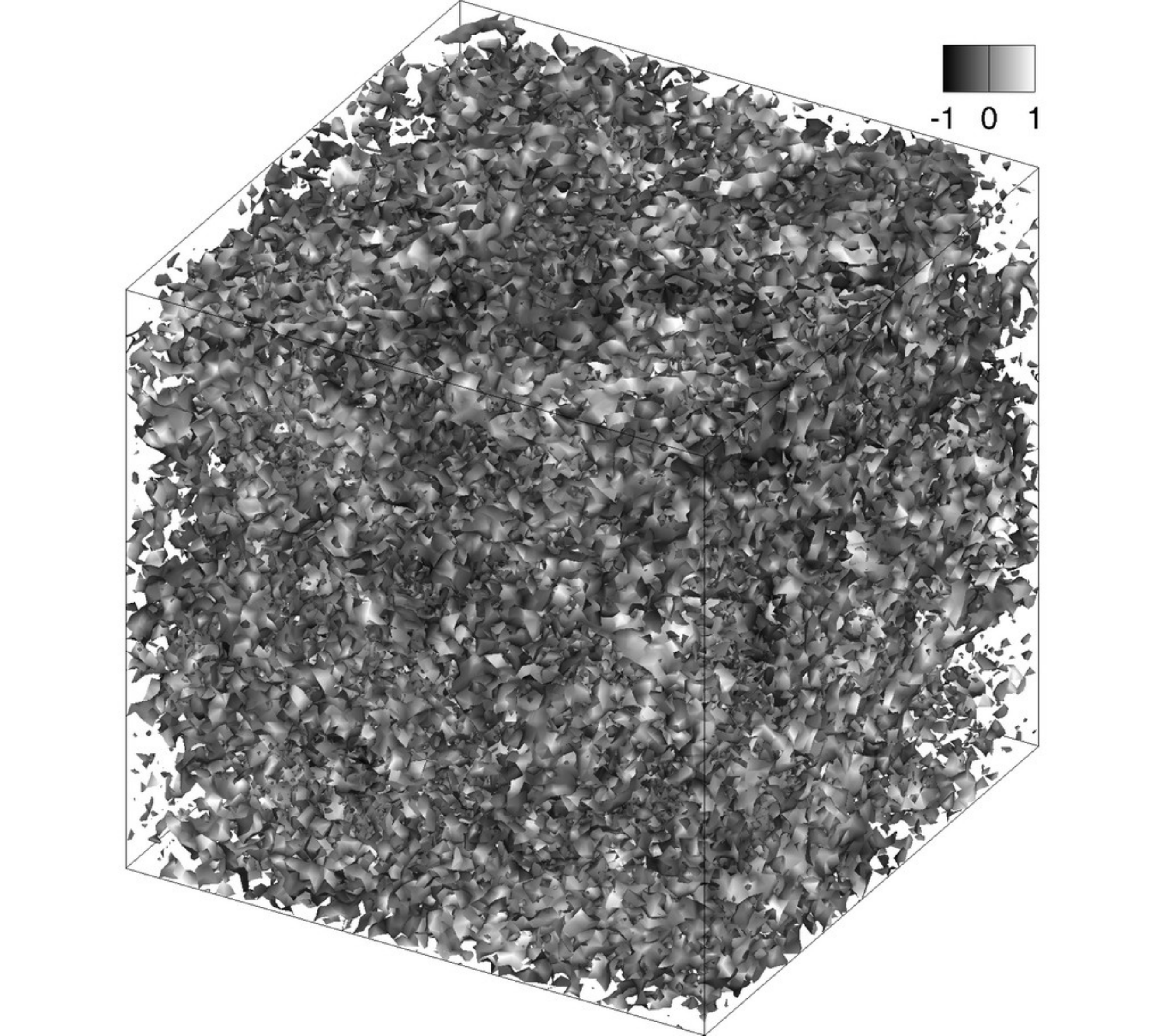}}
\subfigure[Recovered]{\includegraphics[width=0.3\textwidth]{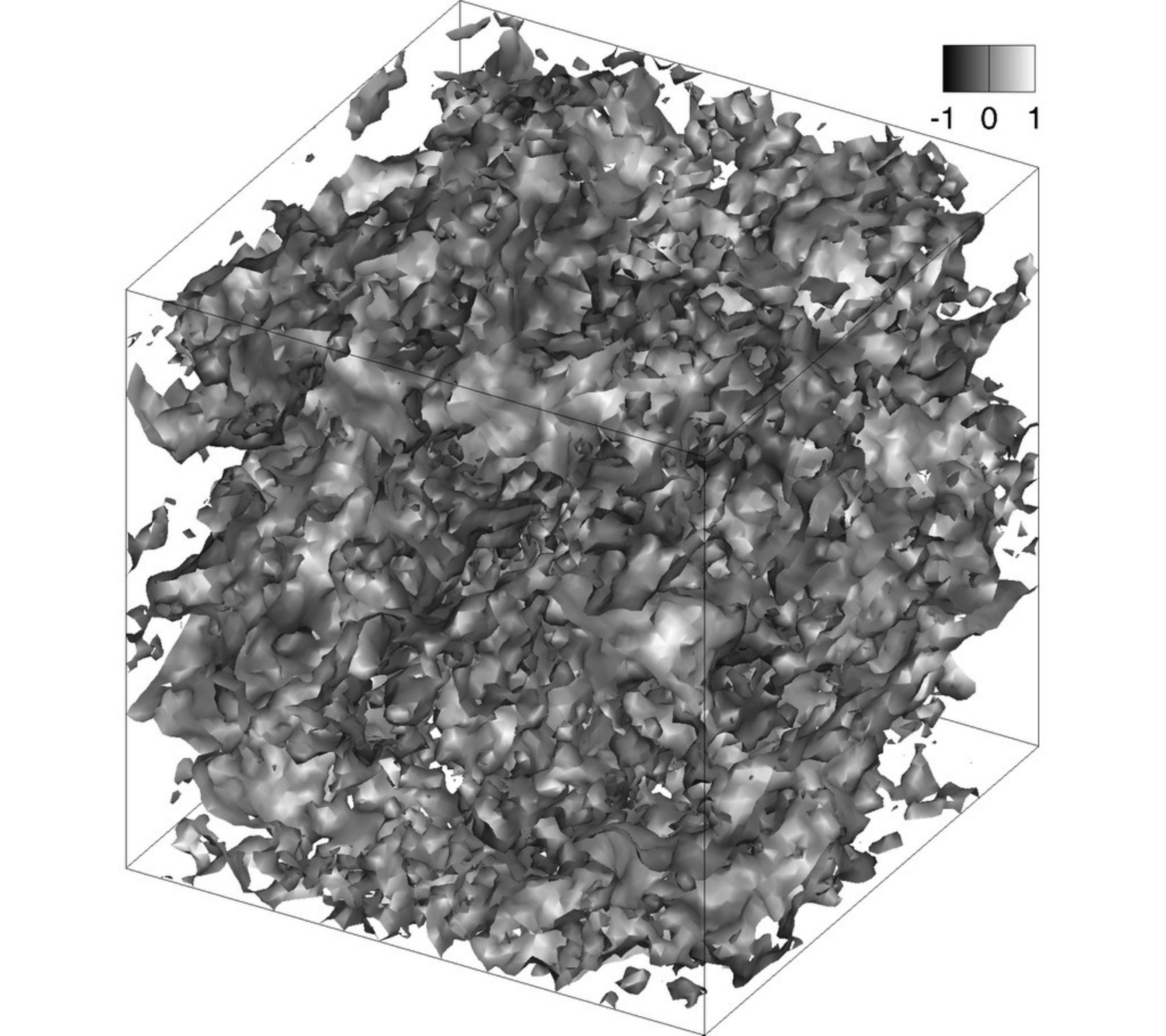}}
}
\caption{A-priori results for velocity field recovery from noisy perturbation inputs for stratified turbulence. Data shown for regularization test data 1. Isosurfaces for $x$ component of the velocity colored by $z$ component are shown: (a) true coarse-grained fields, (b) coarse-grained fields with Gaussian noise, and (c) coarse-grained fields reconstructed using proposed framework.}
\label{fig:field3D_KHI_2}
\end{figure}

The regularization performance of the proposed framework is tested for this particular test case with results described in Figure~\ref{fig:Spec3D_KHI_Noised}. Once again, we observe similar trends to those observed before with marginal improvement in inertial range recovery but effective capture of the peaks of the probability distribution function. The regularization ability of the ELM training procedure however leads to the now familiar reduction in tail capture accuracy. Aliasing errors are effectively removed as expected. Figures \ref{fig:field3D_KHI_1} and \ref{fig:field3D_KHI_2} qualitatively describe the deconvolution and regularization performance of the proposed framework where it can be seen that good recoveries of the true 3D field are obtained in both cases. A quantitative assessment of the performance of the closure for this framework may be observed in Table \ref{table:CV4} where trends similar to those observed for the Kraichnan and Kolmogorov test cases are recovered.

\begin{figure}
\centering
\mbox{
\subfigure[$\tau_{11}$]{\includegraphics[width=0.44\textwidth]{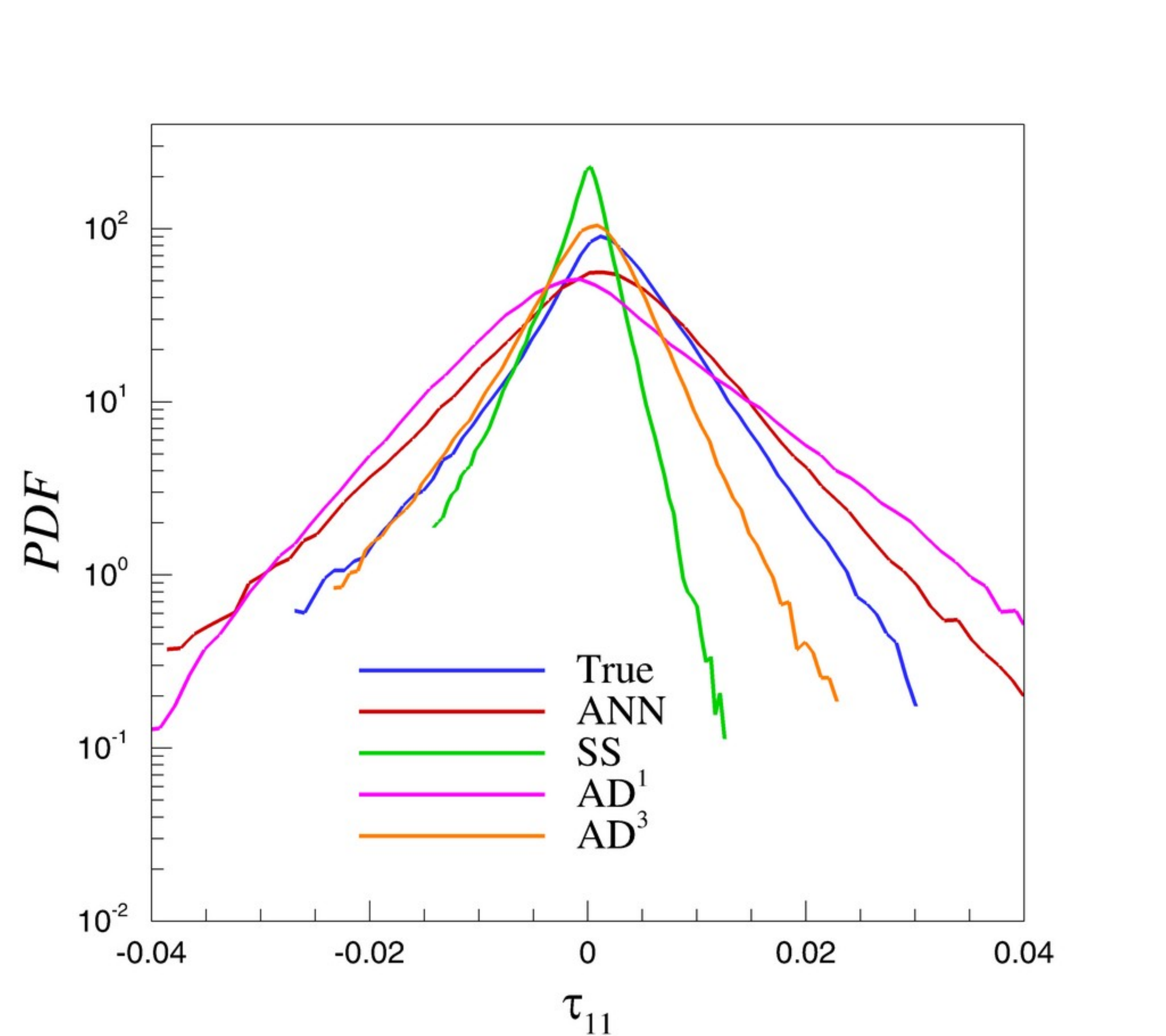}}
\subfigure[$\tau_{12}$]{\includegraphics[width=0.44\textwidth]{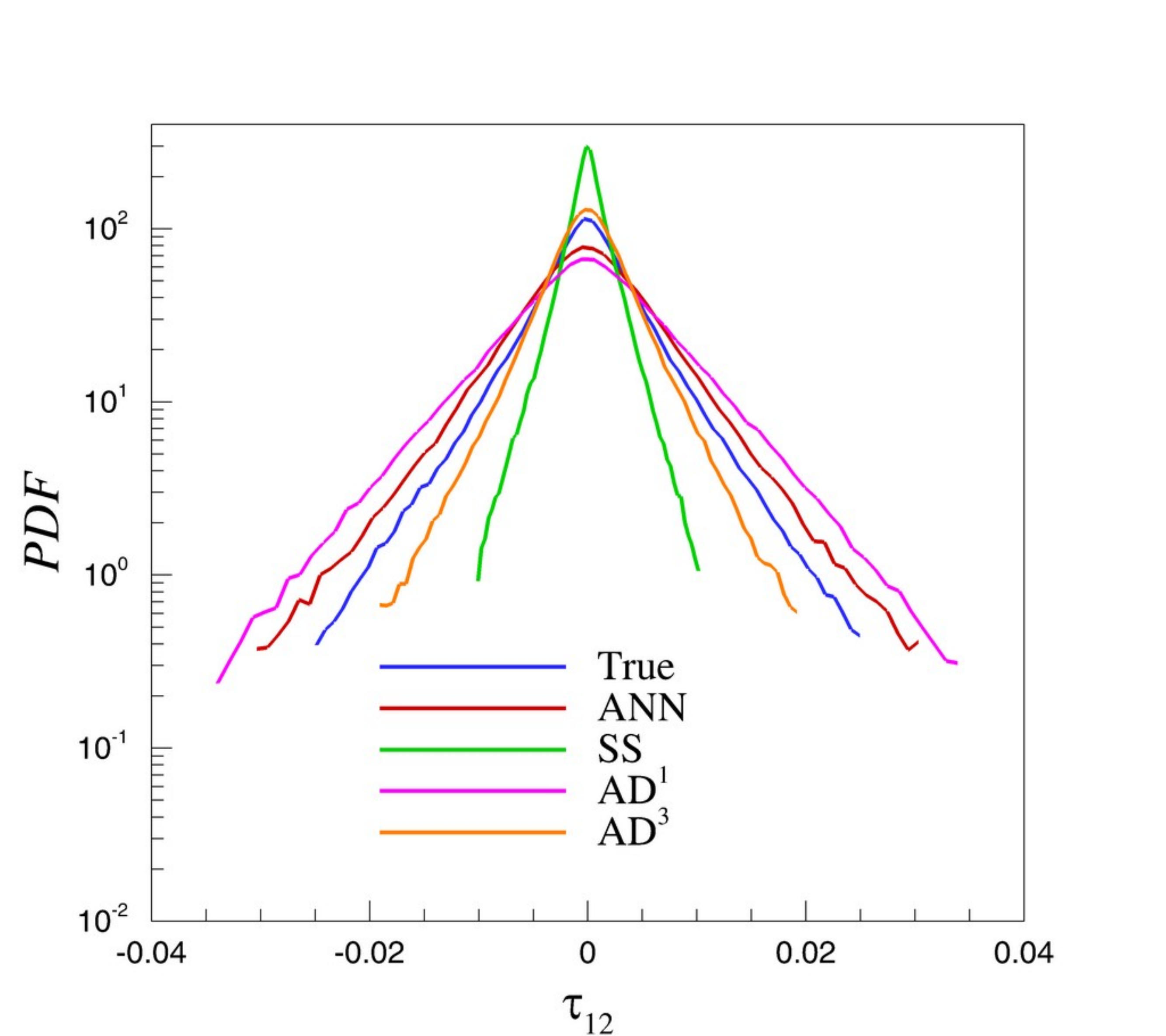}}
}\\
\mbox{
\subfigure[$\tau_{13}$]{\includegraphics[width=0.44\textwidth]{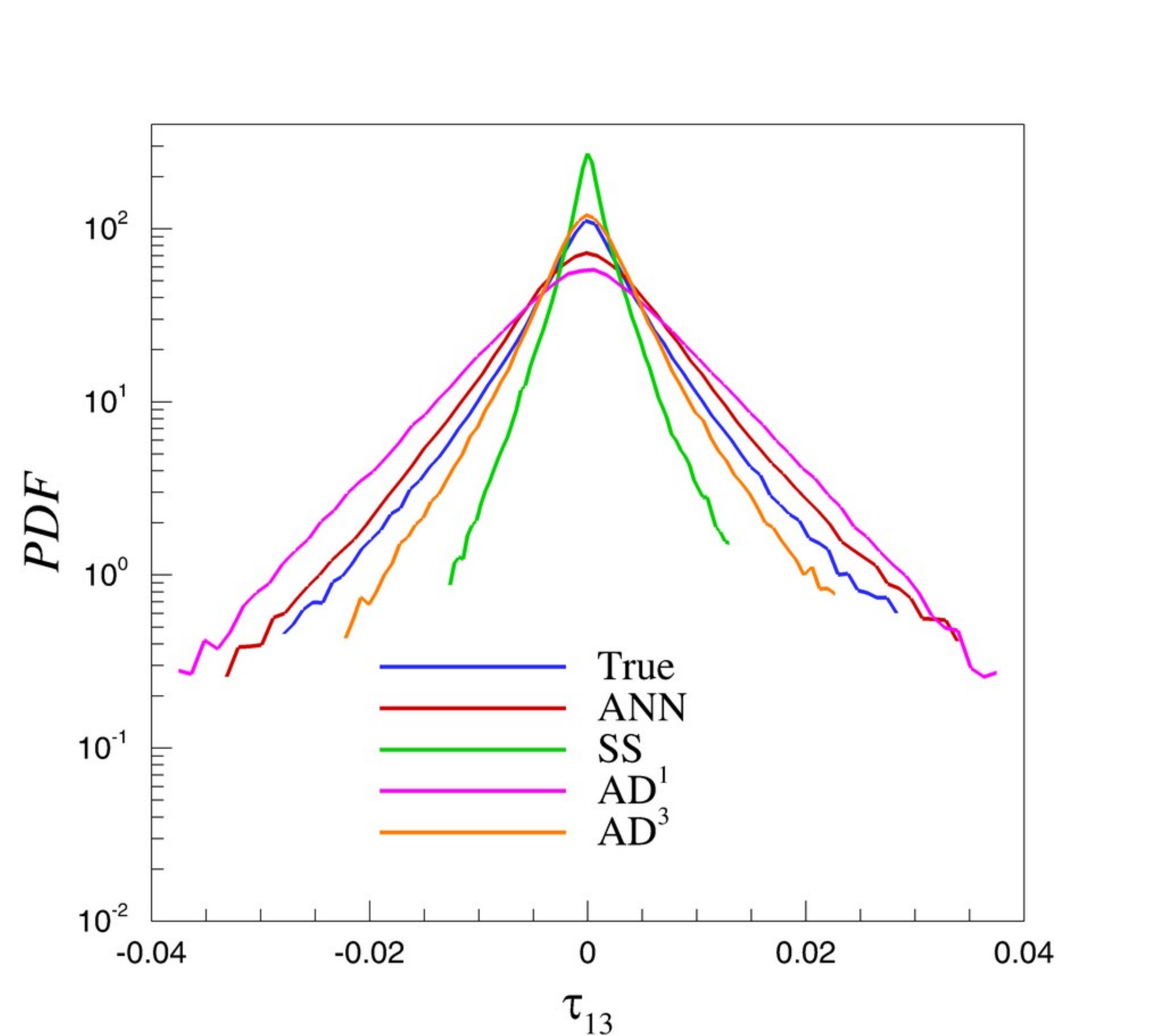}}
\subfigure[$\tau_{22}$]{\includegraphics[width=0.44\textwidth]{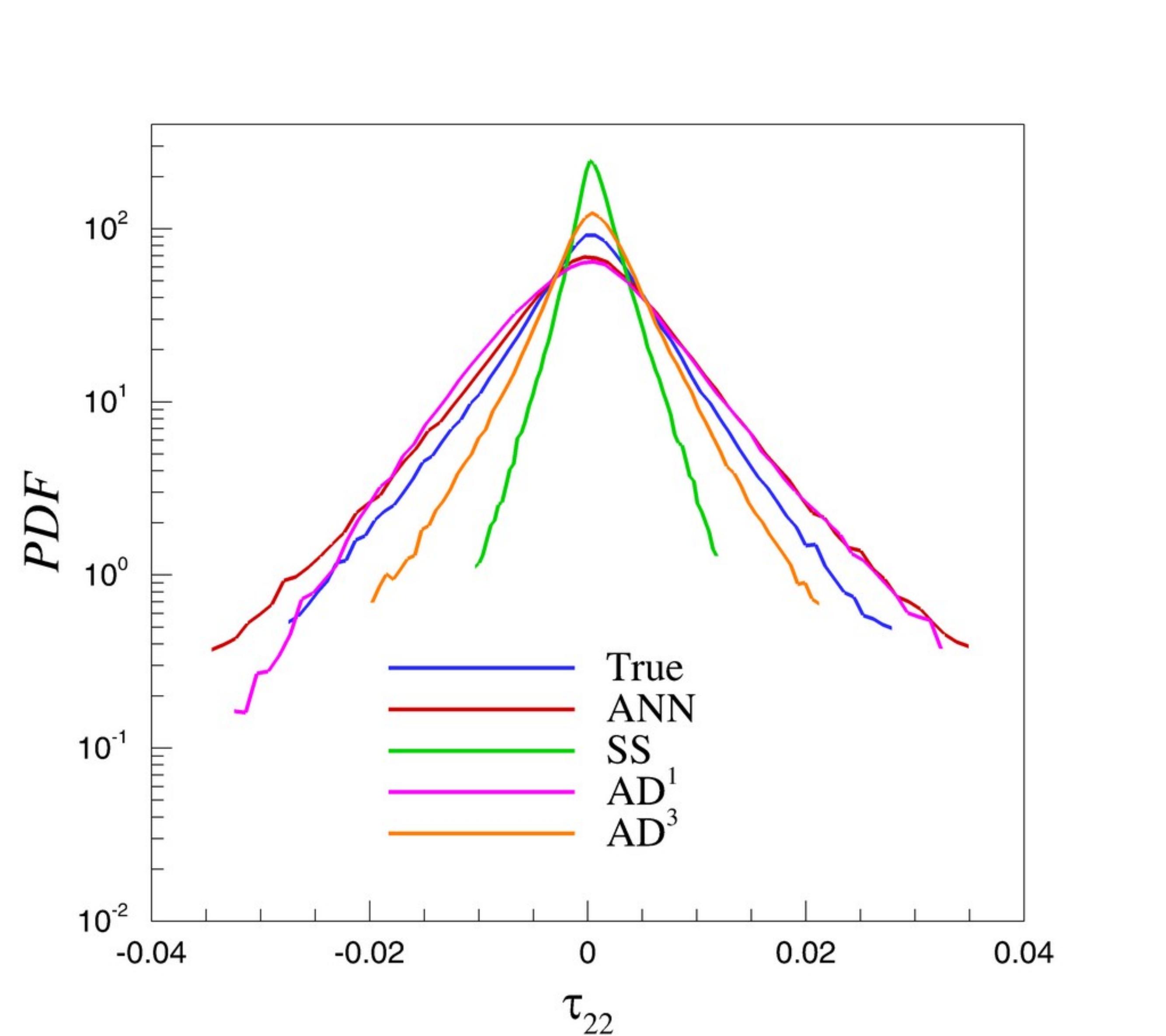}}
}
\\
\mbox{
\subfigure[$\tau_{32}$]{\includegraphics[width=0.44\textwidth]{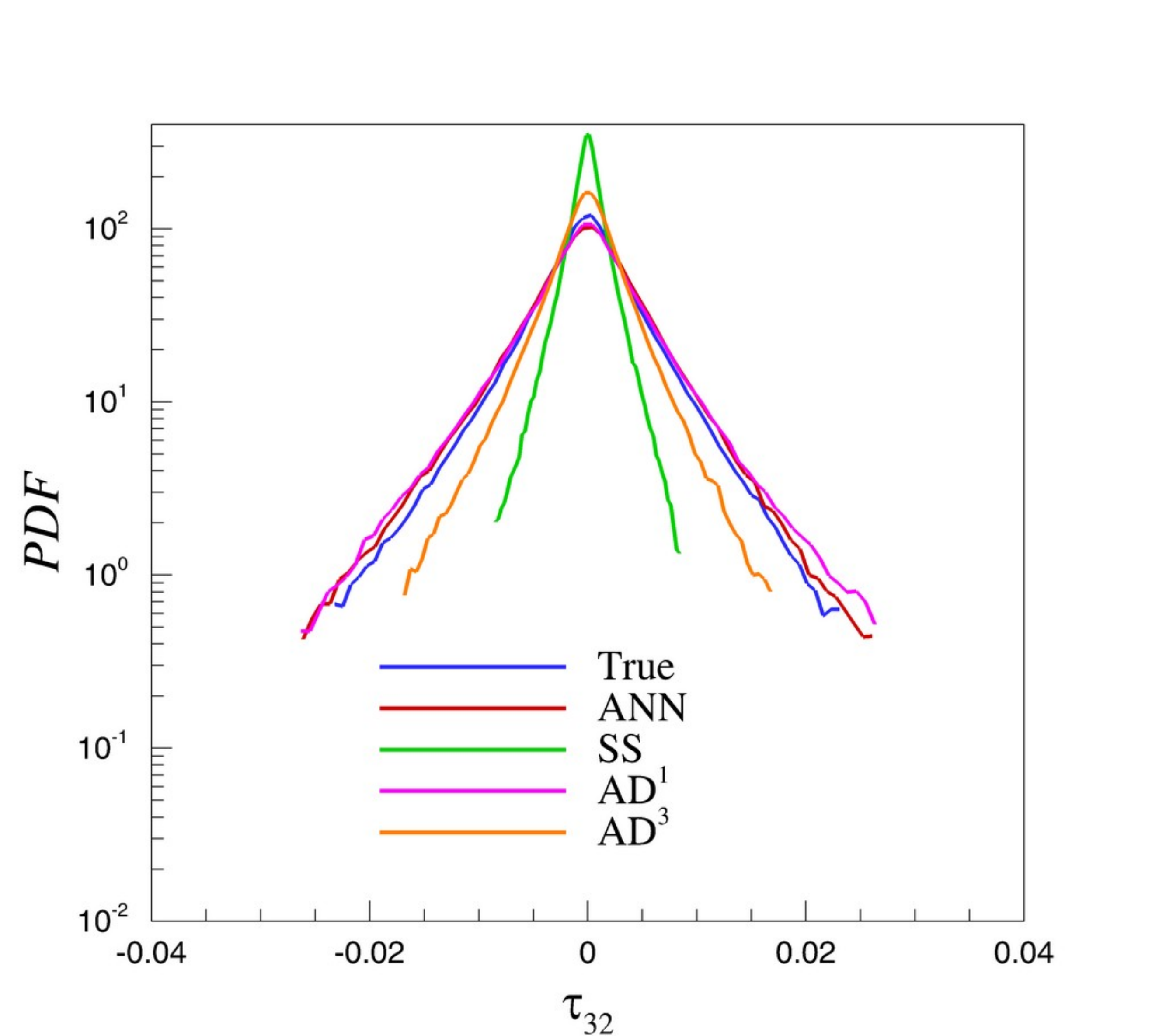}}
\subfigure[$\tau_{33}$]{\includegraphics[width=0.44\textwidth]{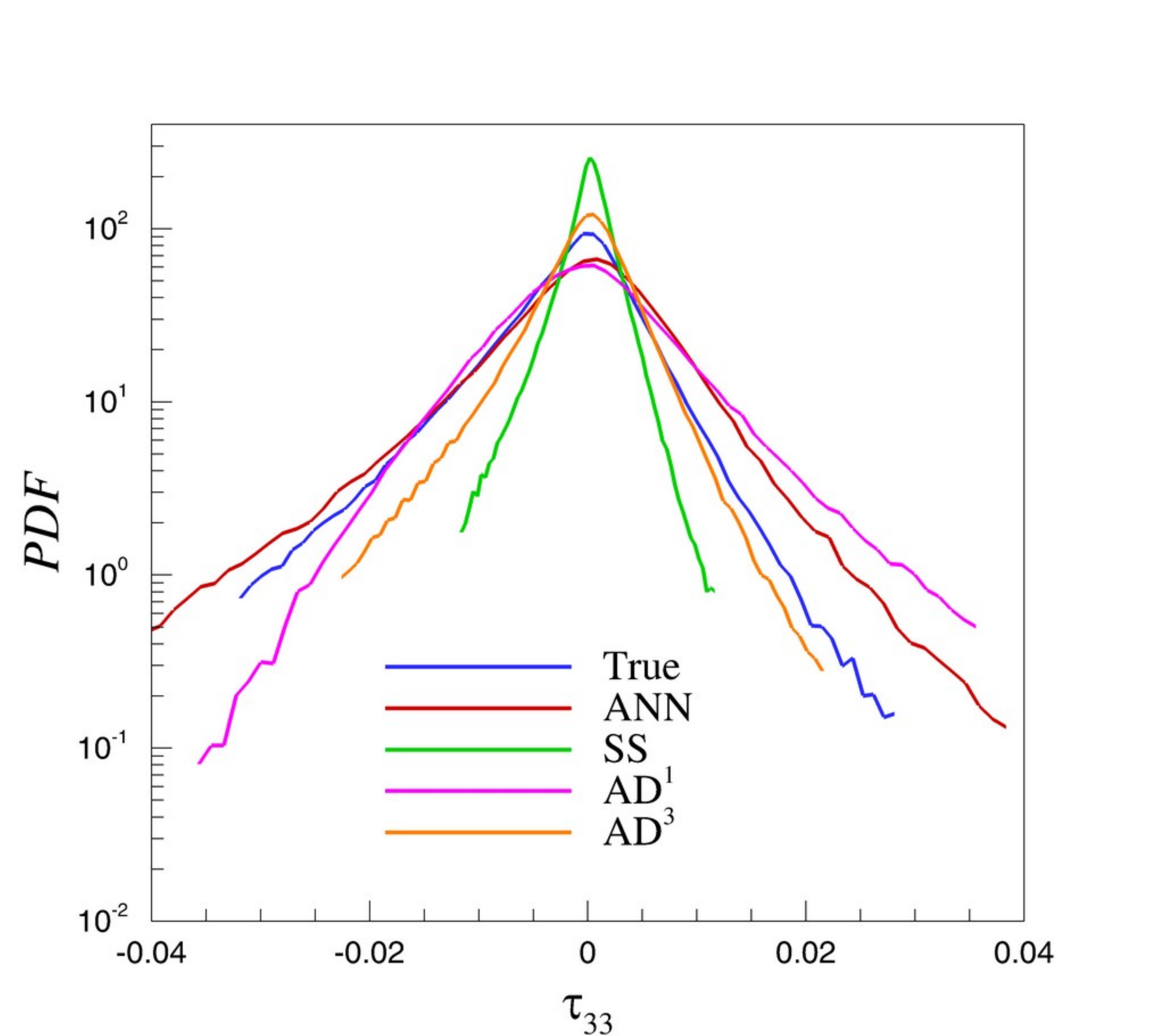}}
}
\caption{A-priori results for stratified turbulence subfilter stress predictions by the proposed architecture. Probability density functions for different subfilter stress components along with predictions by state of the art structural closures. Our data-driven architecture performs in a manner similar to these well established closure strategies without any explicit definition of a low-pass spatial filter.}
\label{fig:Stratified_Closure_Comparison}
\end{figure}

We outline a comparison with other structural closures in Figure \ref{fig:Stratified_Closure_Comparison} where trends similar to those observed in the Kolmogorov test case are witnessed. The AD\textsuperscript{3} approach performs in a superior manner as expected due to the prior specification of the Gaussian filter for both smoothing and iterative deconvolution. The data-driven closure, however, performs quite well in comparison to the chosen closure strategies and validates its application for the purpose of adequate data-driven deconvolution. A quantitative characterization of the subfilter stress recovery is shown in Table \ref{table:Stratified_TauMSE} where it is once again observed that the proposed method results in mean-squared-errors with values near those of the popular structural closure models examined in this study as observed in the Kolmogorov turbulence test case. We remark, once again, that the observed results for our approach are significant due to no prior specification of a filter kernel.

\begin{table}
  \centering
  \begin{tabular}{p{3cm} p{3cm} p{3cm}}
    \multicolumn{3}{c}{\textbf{Stratified turbulence}} \\
    \hline
    \multicolumn{3}{c}{\underline{\textbf{Deconvolution}}} \\
    \addlinespace[0.1cm]
    Dataset & Noised & Regularized \\
    Test data 1 & $4.03 \times 10^{-2}$ & $1.26 \times 10^{-2}$\\
    Test data 2 & $4.88 \times 10^{-2}$ & $1.38 \times 10^{-2}$\\
    Test data 3 & $3.25 \times 10^{-2}$ & $1.14 \times 10^{-2}$\\
    \hline
  \end{tabular}
  \begin{tabular}{p{3cm} p{3cm} p{3cm}}
    \multicolumn{3}{c}{\underline{\textbf{Regularization}}} \\
    \addlinespace[0.1cm]
    Dataset & Noised & Regularized \\
    Test data 1 & $4.19 \times 10^{-2}$ & $1.09 \times 10^{-2}$\\
    Test data 2 & $5.07 \times 10^{-2}$ & $1.18 \times 10^{-2}$\\
    Test data 3 & $3.38 \times 10^{-2}$ & $1.00 \times 10^{-2}$\\
    \hline
  \end{tabular}
  \caption{Mean-squared-error values for deconvolved and regularized fields obtained from the proposed architecture. Data shown from the three-dimensional stratified turbulence test case. Note that the mean-squared-error values are obtained from the $z$ component of the velocity field.}\label{table:CV4}
\end{table}

\begin{table}
  \centering
  \begin{tabular}{p{1cm} p{1.8cm} p{1.8cm} p{1.8cm} p{1.8cm} p{1.8cm} p{1.8cm}}
    Model & $\tau_{11}\times10^{-5}$ & $\tau_{12}\times10^{-5}$ & $\tau_{13}\times10^{-5}$ & $\tau_{22}\times10^{-5}$ & $\tau_{32}\times10^{-5}$ & $\tau_{33}\times10^{-5}$ \\
    \hline
    ANN &  6.02  &  2.60  &  3.12  &  3.82  & 1.82   &  5.15   \\
    SS  &  3.15  &  2.04  &  2.27  &  2.76  & 2.02   &  3.62   \\
    AD\textsuperscript{1} &  15.91 &  10.15 & 11.91 & 10.22 &  7.16  & 12.7\\
    AD\textsuperscript{3} &  2.01  &  1.18  &  1.29 & 1.60 &  1.12  & 2.05\\
    \hline
  \end{tabular}
  \caption{Mean-squared-error values for deviatoric subfilter scale components with respect to the true subfilter scale stresses for stratified turbulence.}\label{table:Stratified_TauMSE}
\end{table}

\section{Universality}
\label{sec:Univ}

The true value of the implementation of a data-driven framework emerges from an intelligent combination of our knowledge of physics from first principles with the robust techniques inherent to data-driven modeling. One of the motivations of studying homogeneous isotropic turbulence test cases in this investigation was to link the universal nature of the physics (as evidenced in the $k^{-5/3}$ scaling observed in the averaged kinetic energy spectra) through our proposed framework across flows exhibiting distinctly different physics. For instance, a great utility of any data-driven modeling framework would be to utilize training data from lower Reynolds numbers or shorter (high-fidelity) simulations to devise data-driven closures for runs with higher Reynolds numbers or longer simulations which are computationally intractable.

\begin{figure}
\centering
\mbox{
\subfigure[Filtered Inputs]{
\includegraphics[width=0.44\textwidth]{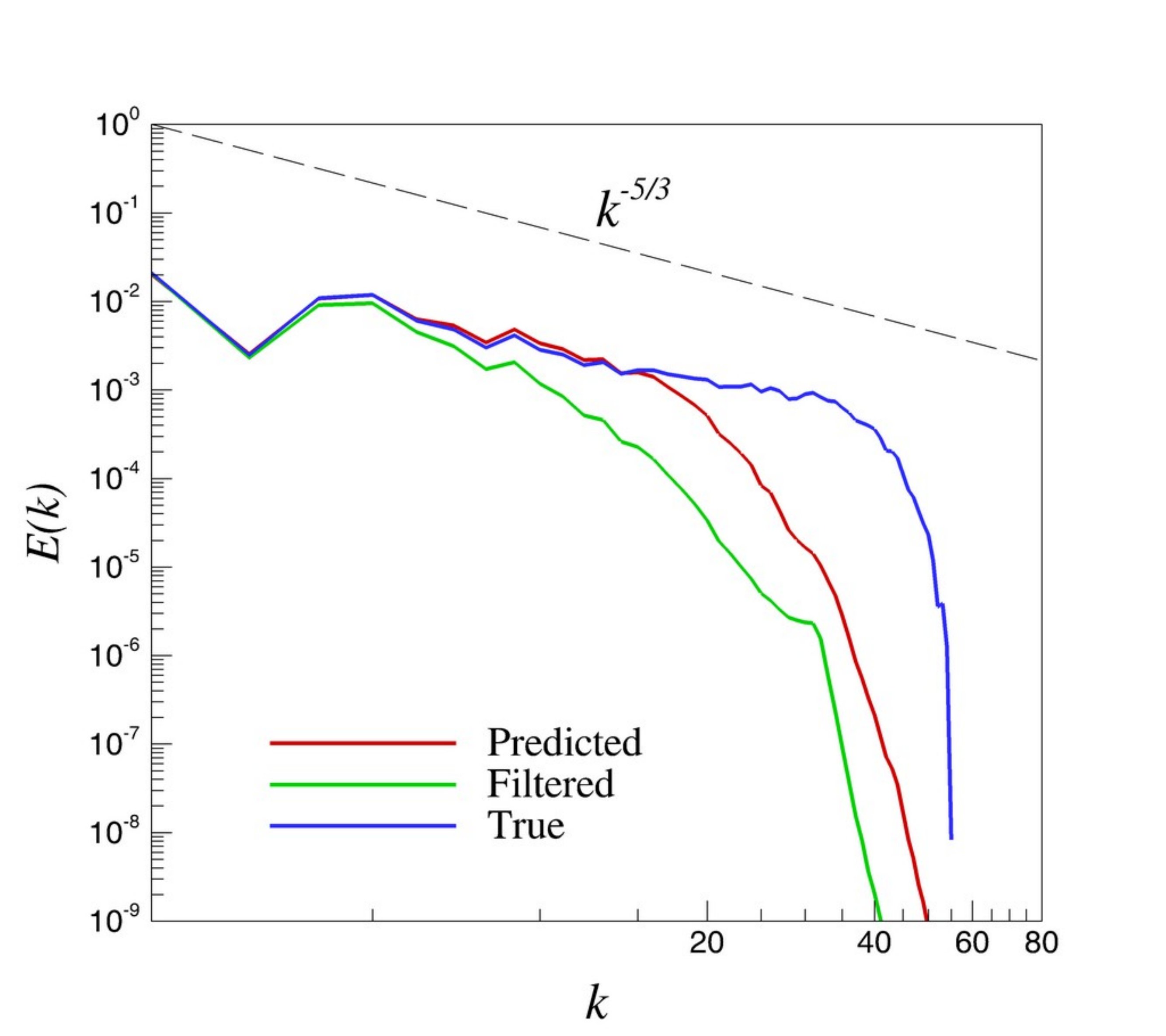}
\includegraphics[width=0.44\textwidth]{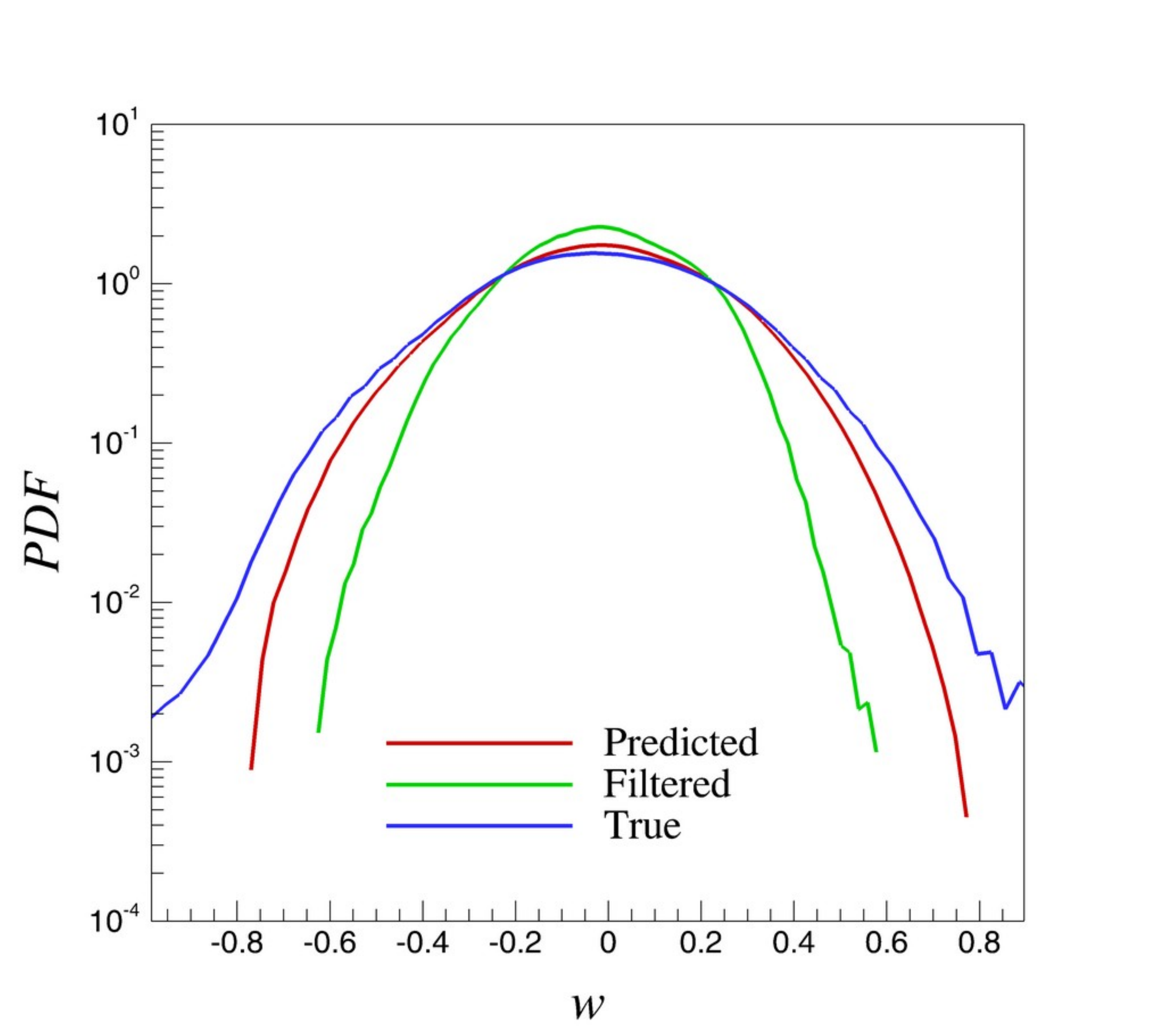}
}
}\\
\mbox{
\subfigure[Noised Inputs]{
\includegraphics[width=0.44\textwidth]{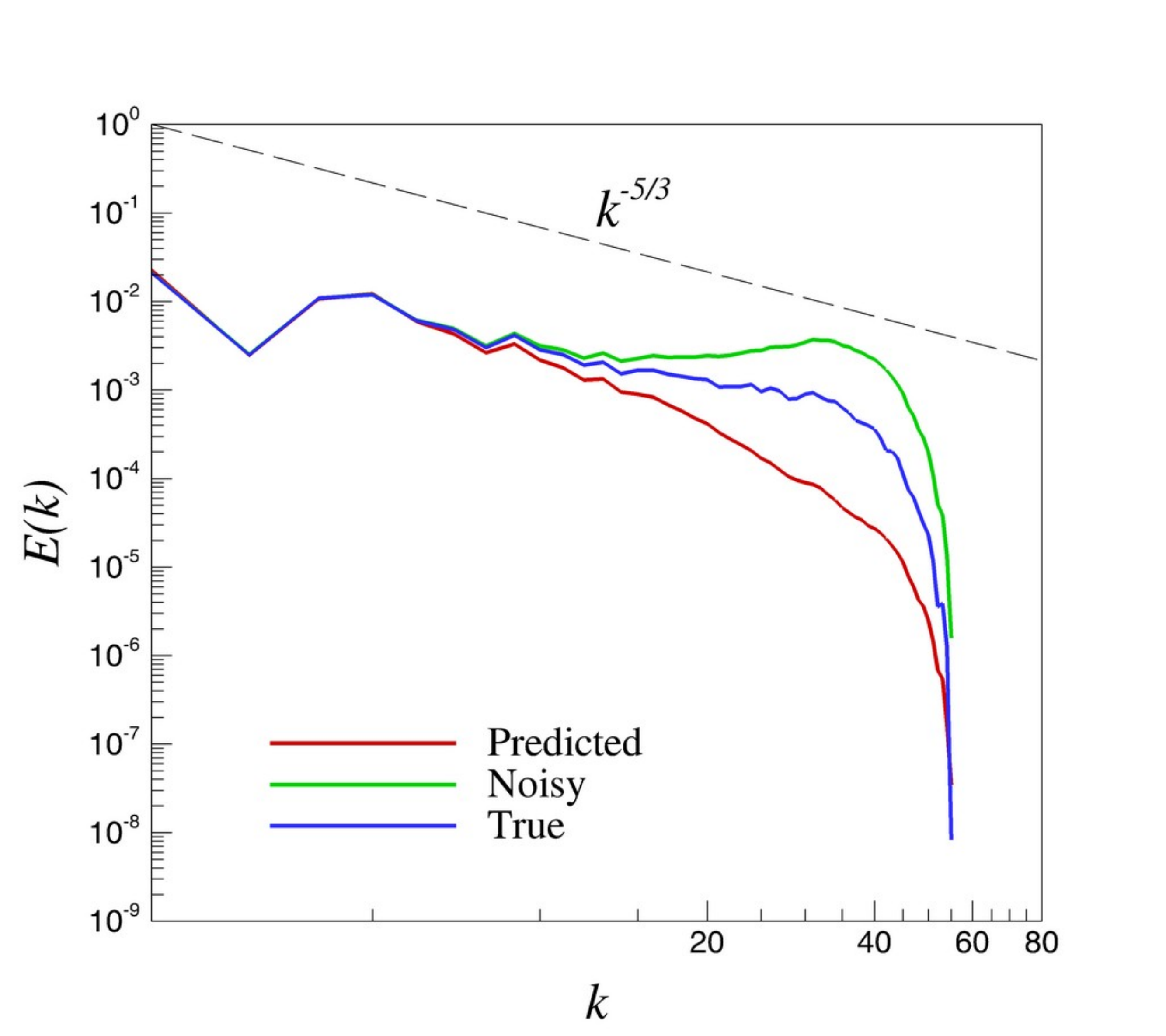}
\includegraphics[width=0.44\textwidth]{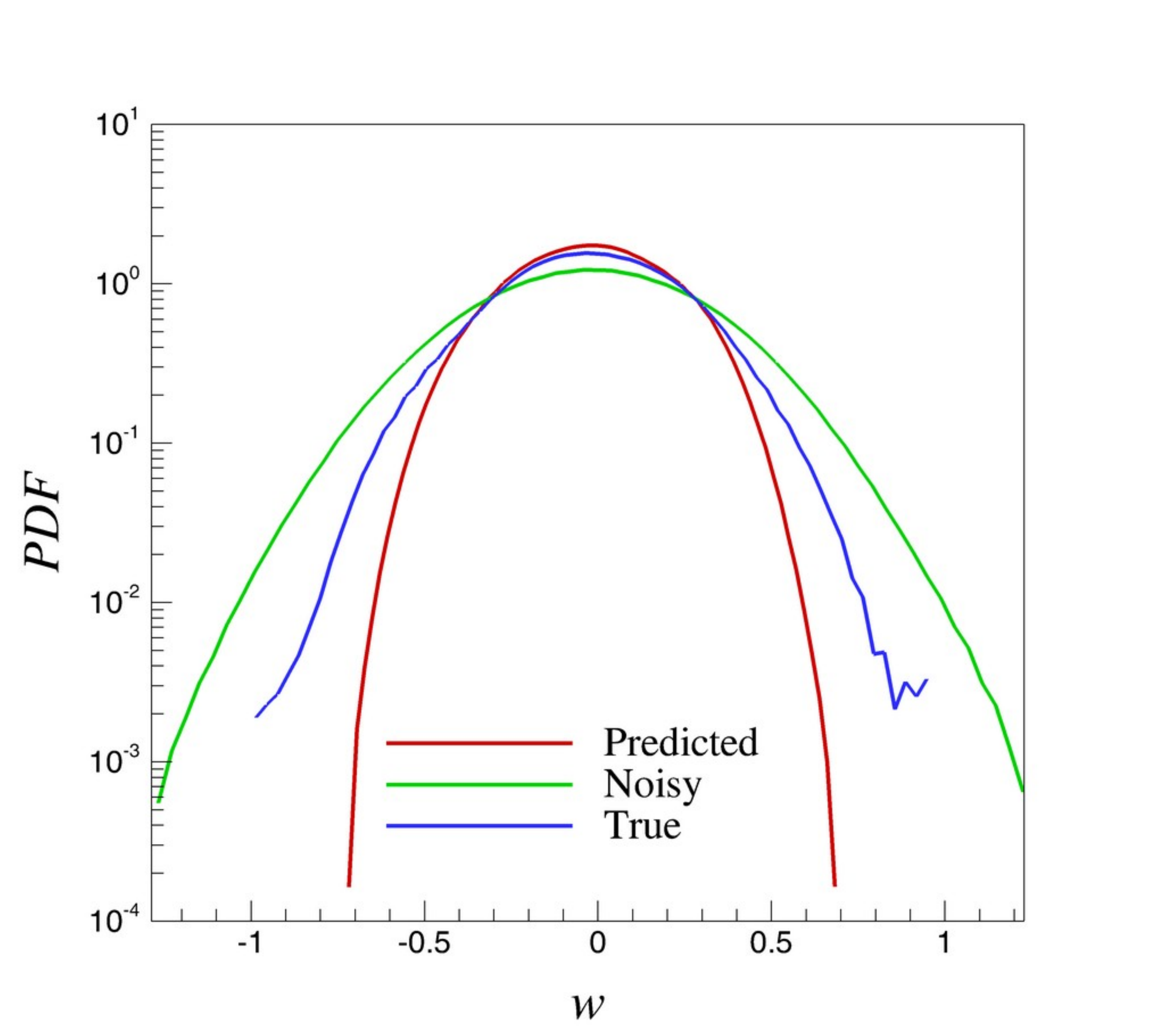}
}
}
\caption{A-priori results of the kinetic energy spectra (left) and PDF of the vorticity (right) for Kolmogorov turbulence. Here we utilize DNS data for the Taylor-Green vortex at $Re=1600$ to reconstruct an approximation to the true field for $Re=5000$.}
\label{fig:Reynolds_Numbers}
\end{figure}
\begin{figure}
\centering
\mbox{
\subfigure[Filtered Inputs]{
\includegraphics[width=0.44\textwidth]{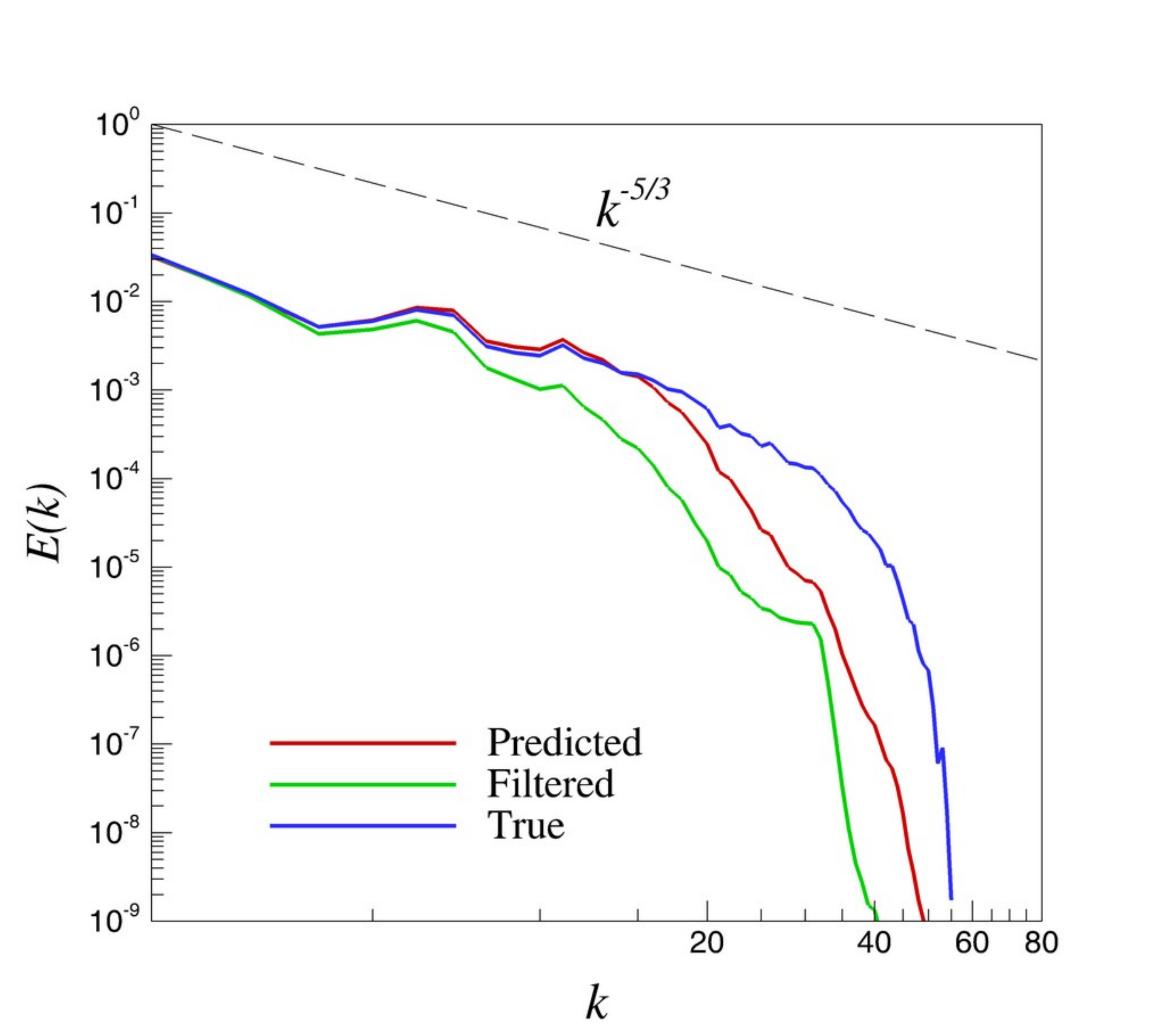}
\includegraphics[width=0.44\textwidth]{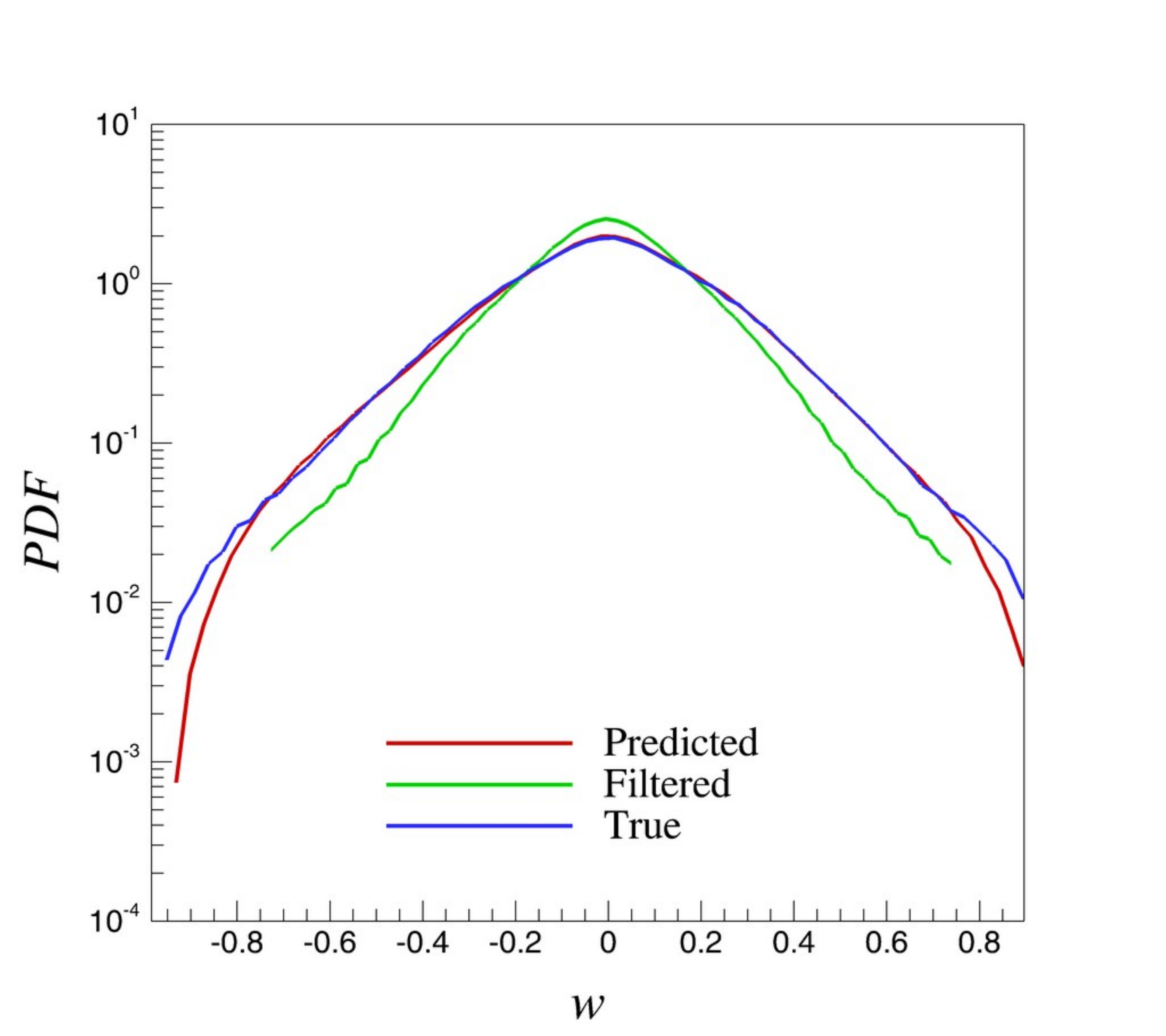}
}
}\\
\mbox{
\subfigure[Noised Inputs]{
\includegraphics[width=0.44\textwidth]{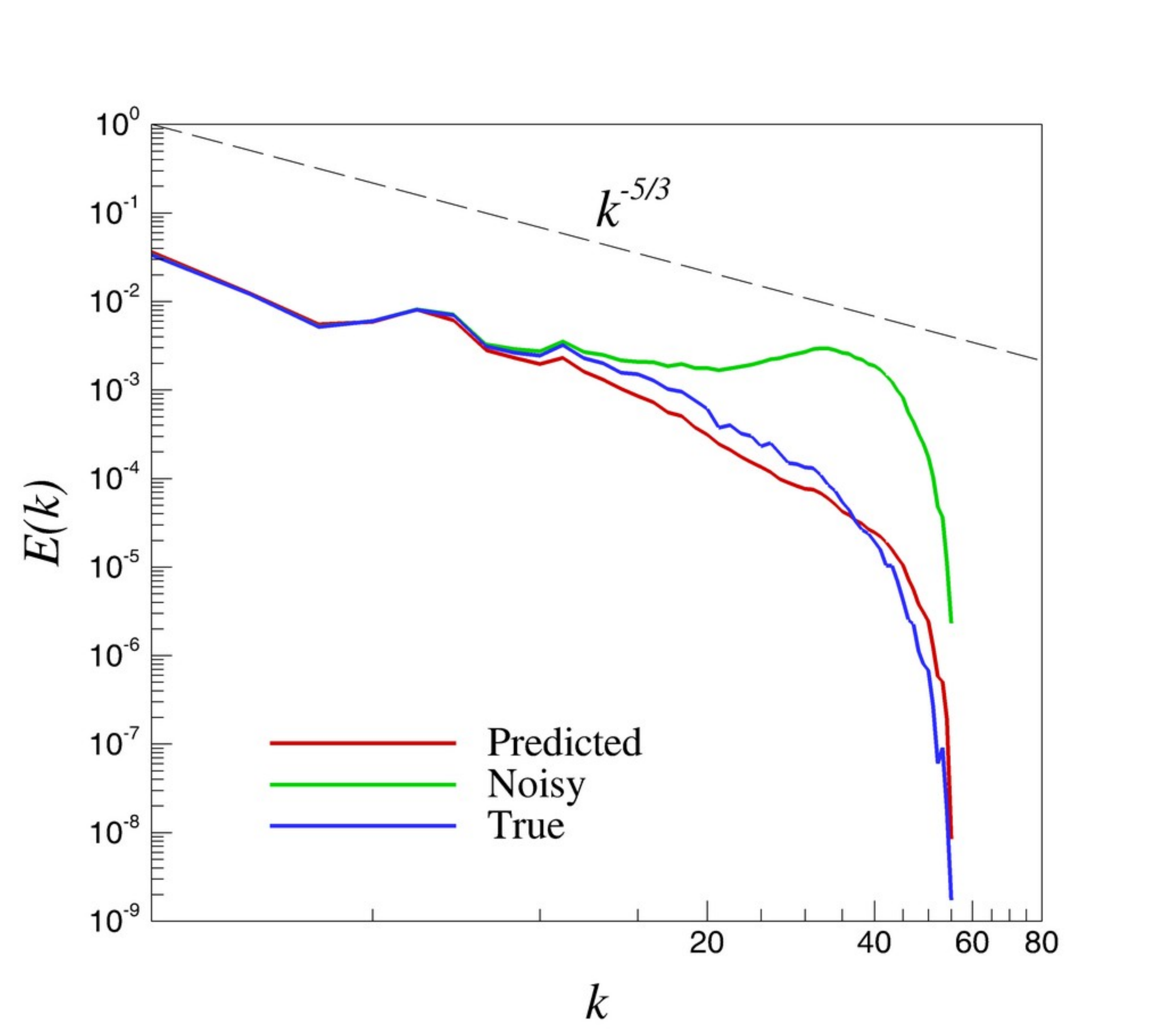}
\includegraphics[width=0.44\textwidth]{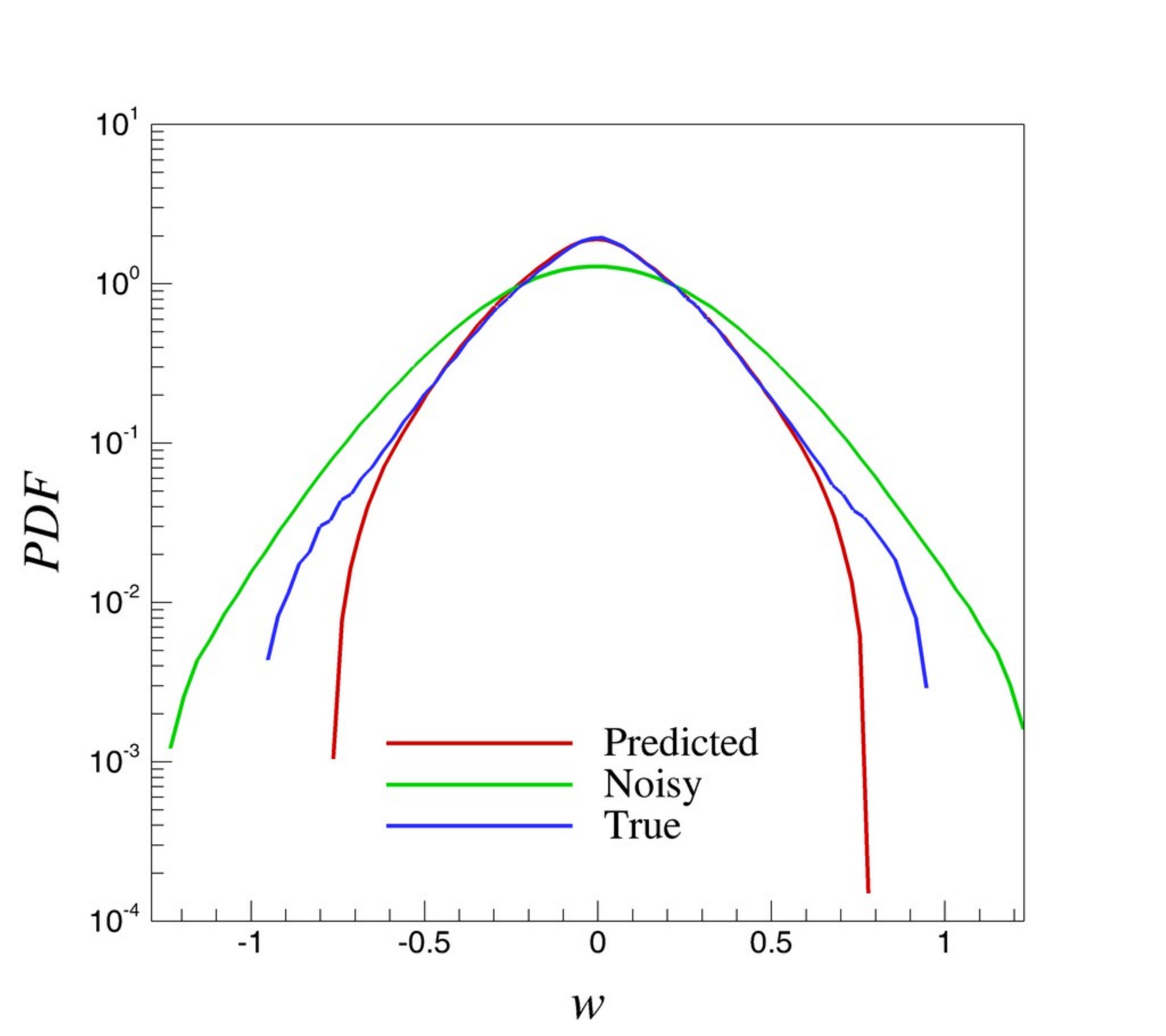}
}
}
\caption{A-priori results for the kinetic energy spectra (left) and PDF of the vorticity (right) for Kolmogorov turbulence. Here we utilize high fidelity data from the Taylor-Green vortex at time $t=15$ to obtain a reconstruction for the same test case at time $t=20$ for a $Re=1600$.}
\label{fig:NS_Time}
\end{figure}

We first examine the ability of the closure modeling strategy by using training data from a high-fidelity Taylor-Green vortex simulation at $Re=1600$ to recover true fields at $Re=5000$ as shown in Figure~\ref{fig:Reynolds_Numbers}. We observe that the framework is successfully able to replicate closure performance as shown in previous test cases. It tells us that the deconvolution (or regularization) ability has been learned without violating (as it appears) the underlying physics of homogeneous isotropic turbulence. Our next assessment shown in Figure~\ref{fig:NS_Time} utilizes the closure modeling performance of the framework when training data is utilized at time $t=15$ for a Taylor-Green vortex test case to obtain true field reconstructions at $t=20$ (for the same simulation). It can once again be observed that the closure recovery is exceptional with trends similar to those exhibited by previous test cases.

\begin{figure}
\centering
\mbox{
\subfigure[Filtered Inputs]{
\includegraphics[width=0.44\textwidth]{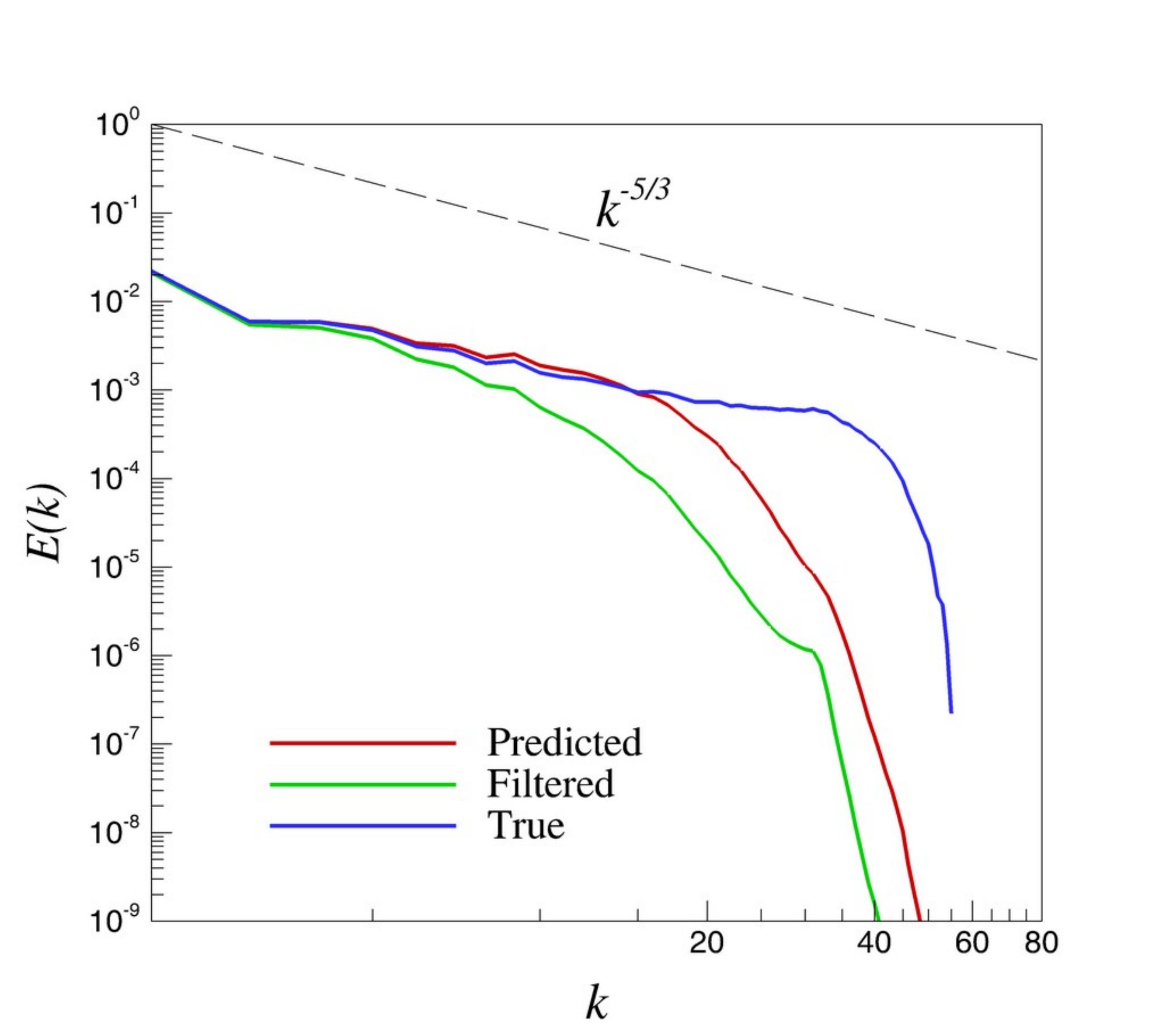}
\includegraphics[width=0.44\textwidth]{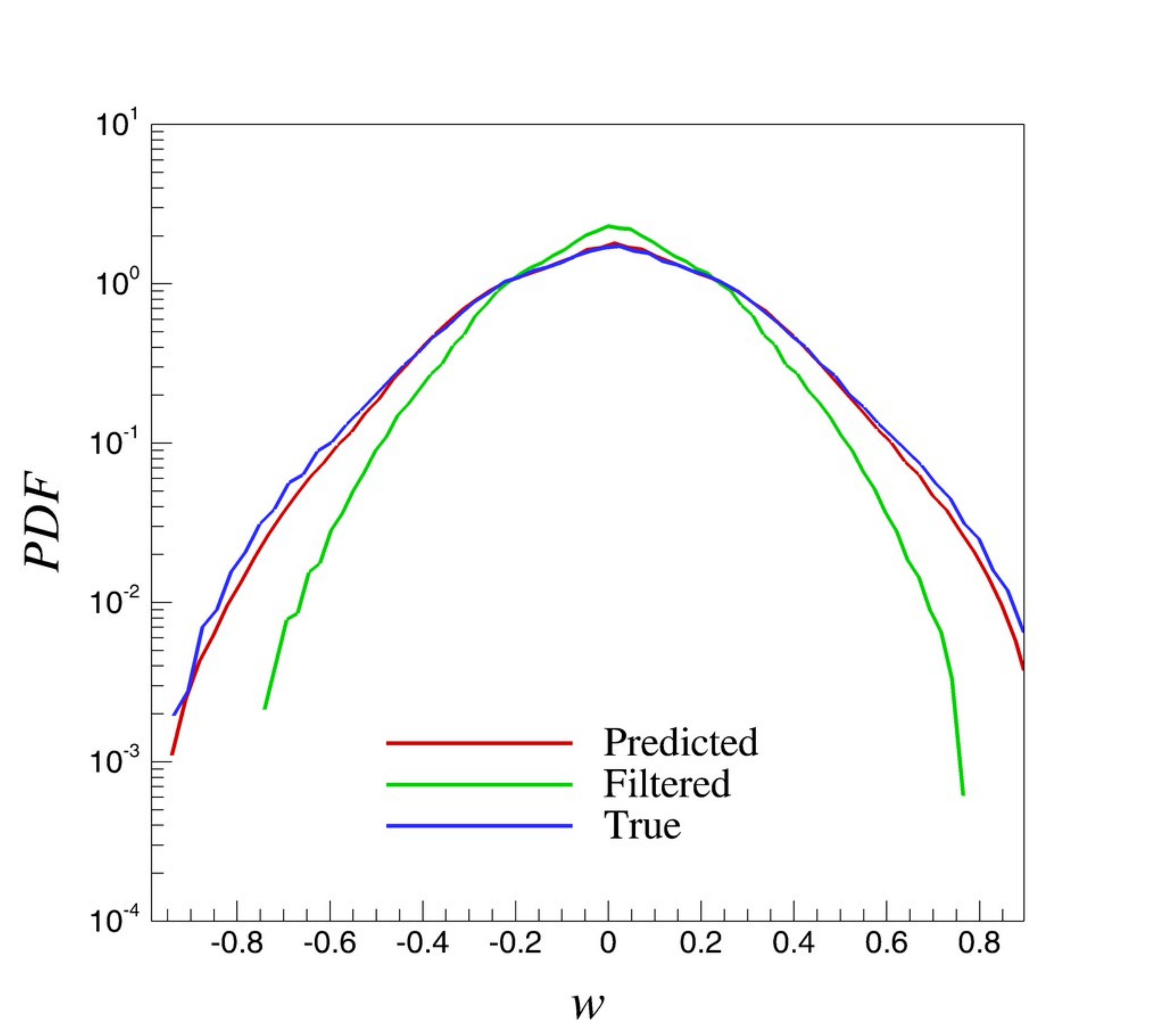}
}
}\\
\mbox{
\subfigure[Noised Inputs]{
\includegraphics[width=0.44\textwidth]{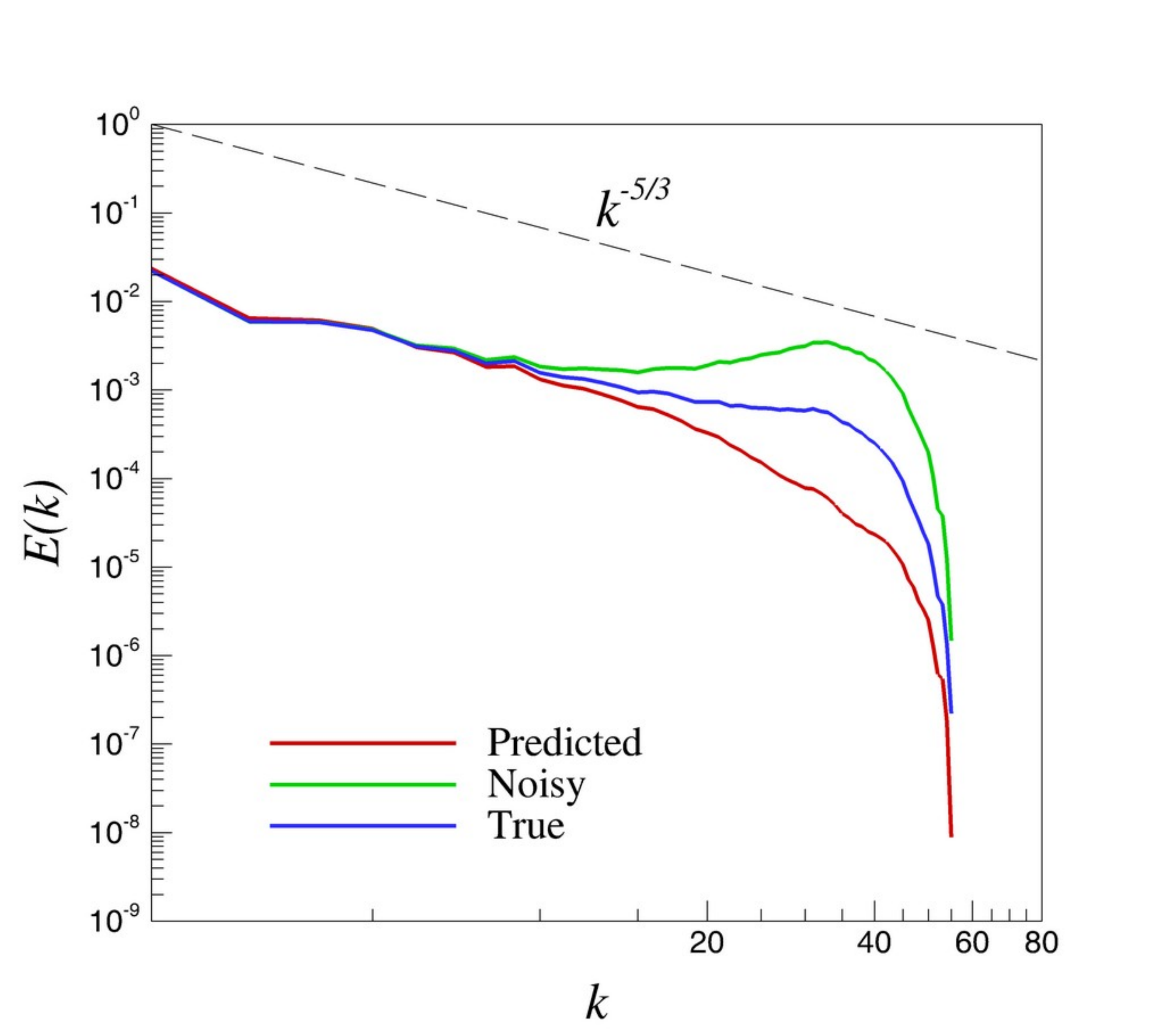}
\includegraphics[width=0.44\textwidth]{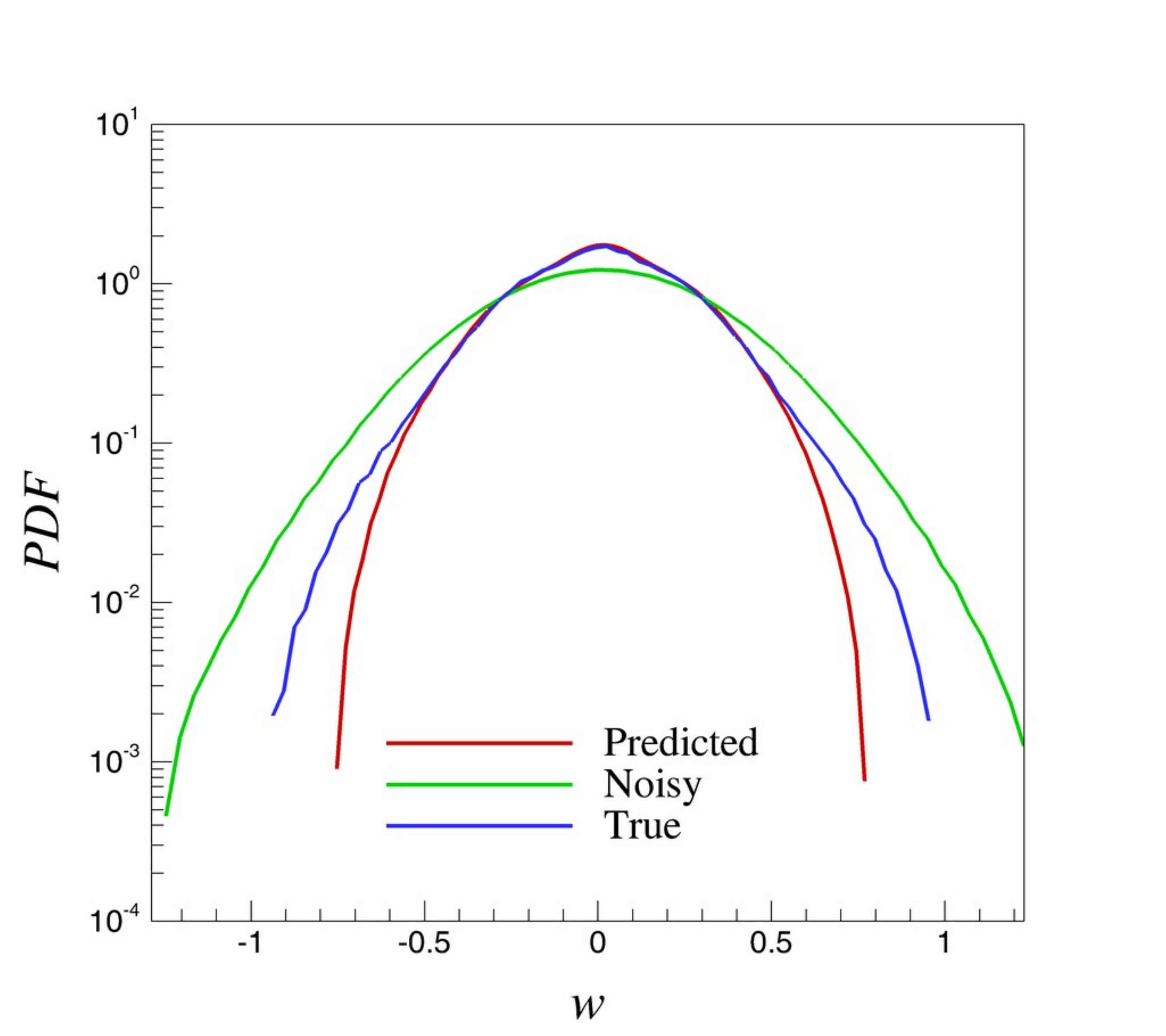}
}
}
\caption{A-priori results of the kinetic energy spectra (left) and PDF of the vorticity (right) stratified turbulence. Here we utilize DNS data for the Taylor-Green vortex at $Re=1600$ to reconstruct an approximation to the true field for the stratified turbulence test case generated from the inviscid Euler equations.}
\label{fig:NS_Euler}
\end{figure}

\begin{figure}
\centering
\mbox{
\subfigure[Filtered Inputs]{
\includegraphics[width=0.44\textwidth]{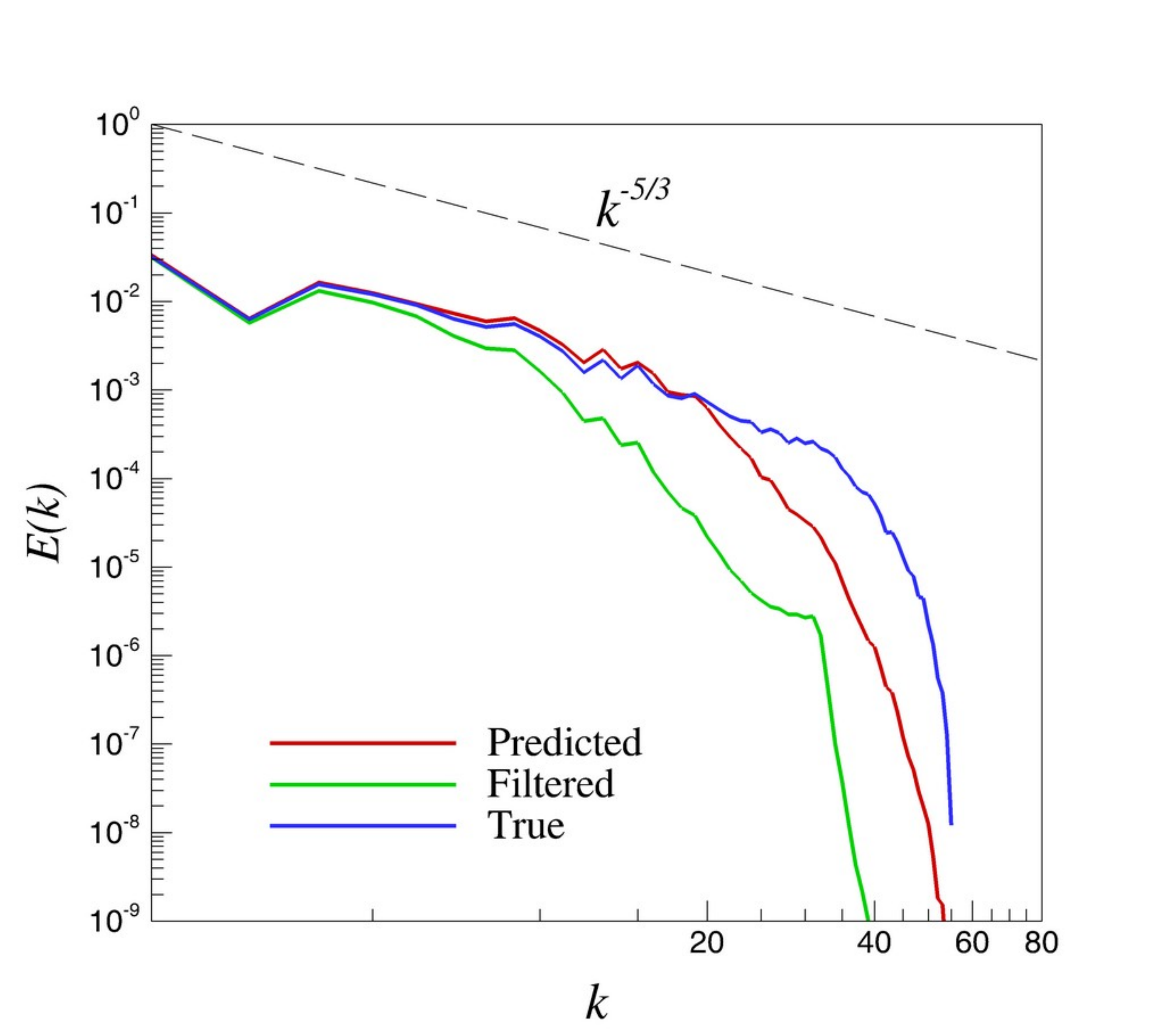}
\includegraphics[width=0.44\textwidth]{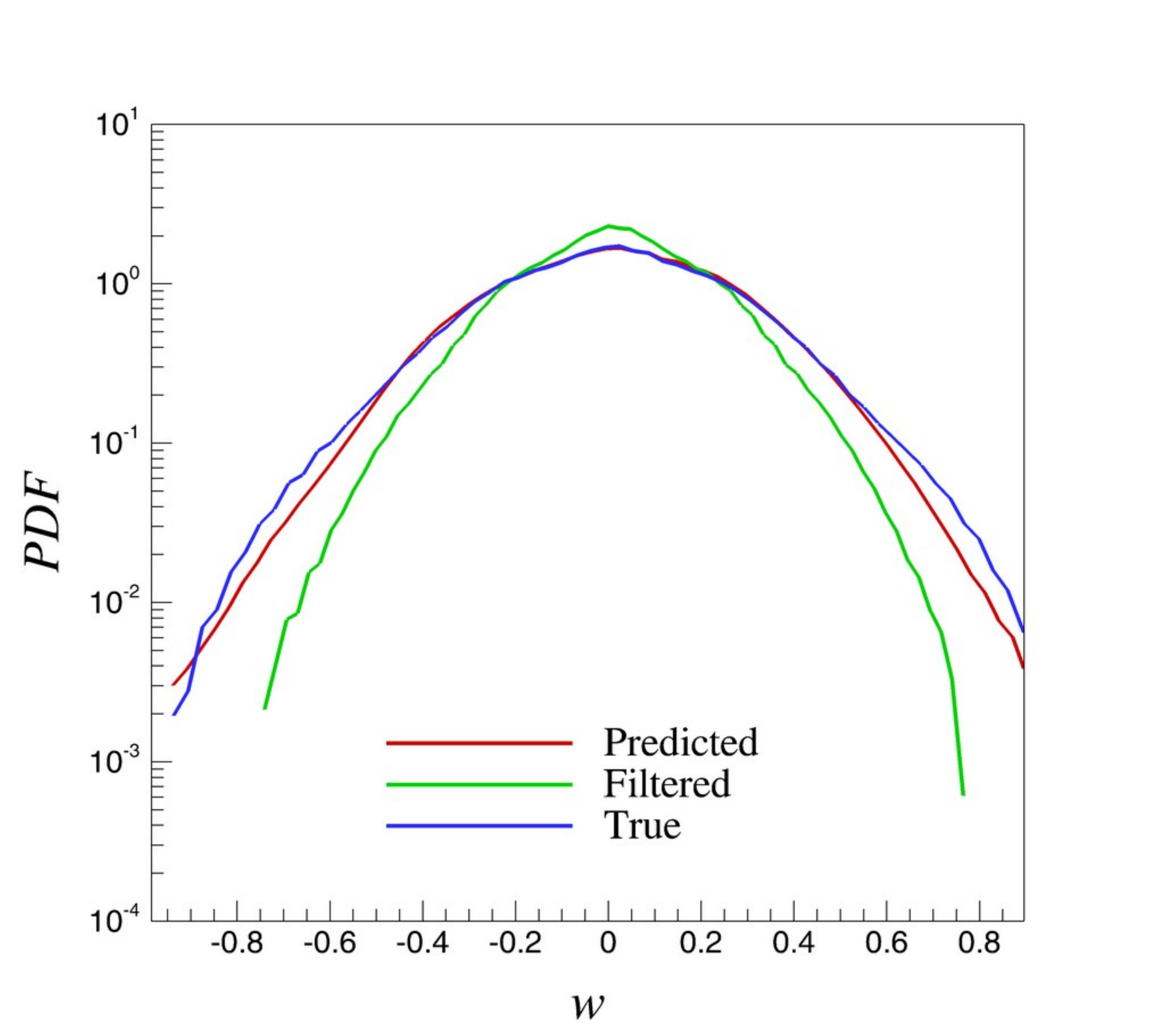}
}
}\\
\mbox{
\subfigure[Noised Inputs]{
\includegraphics[width=0.44\textwidth]{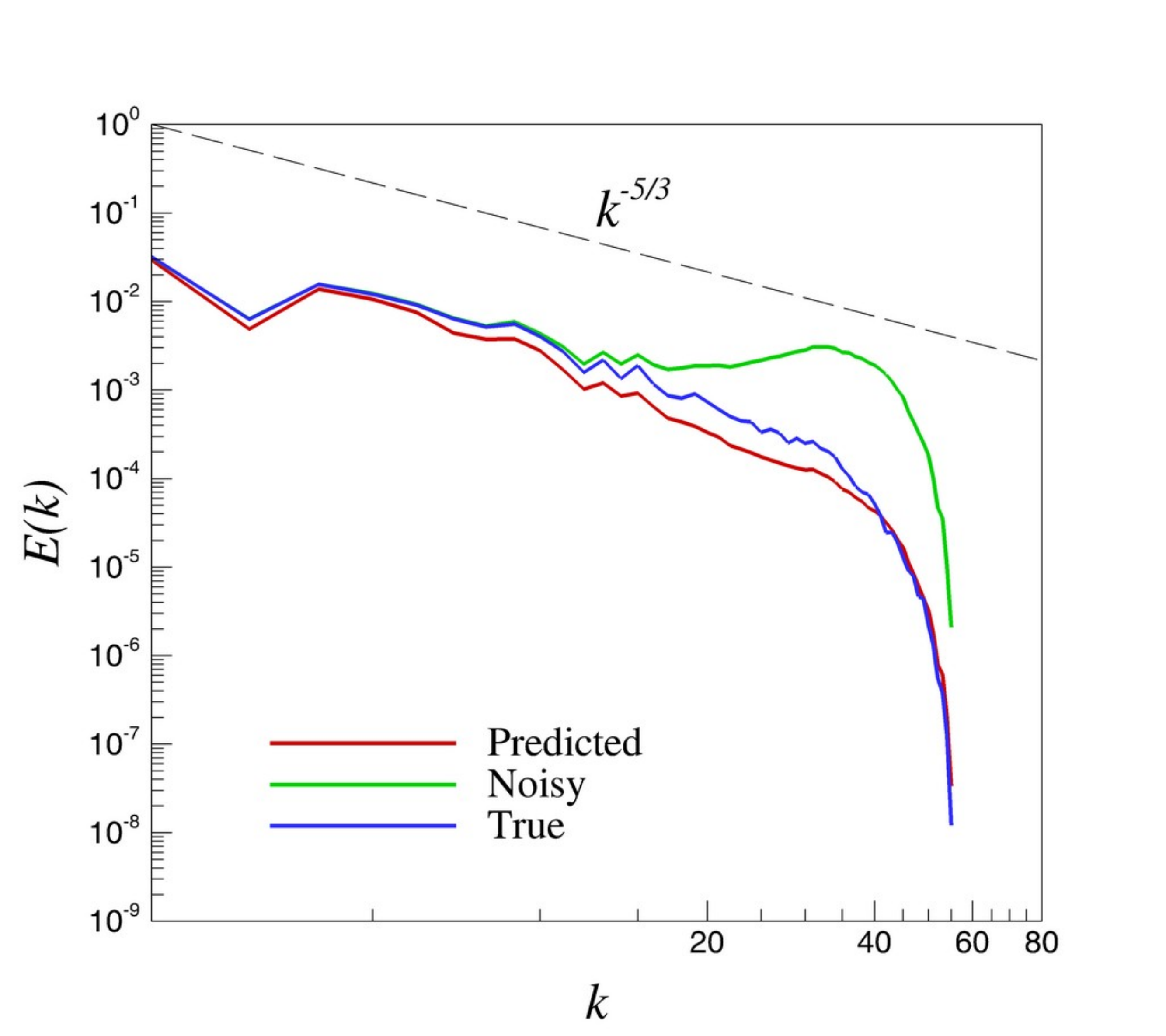}
\includegraphics[width=0.44\textwidth]{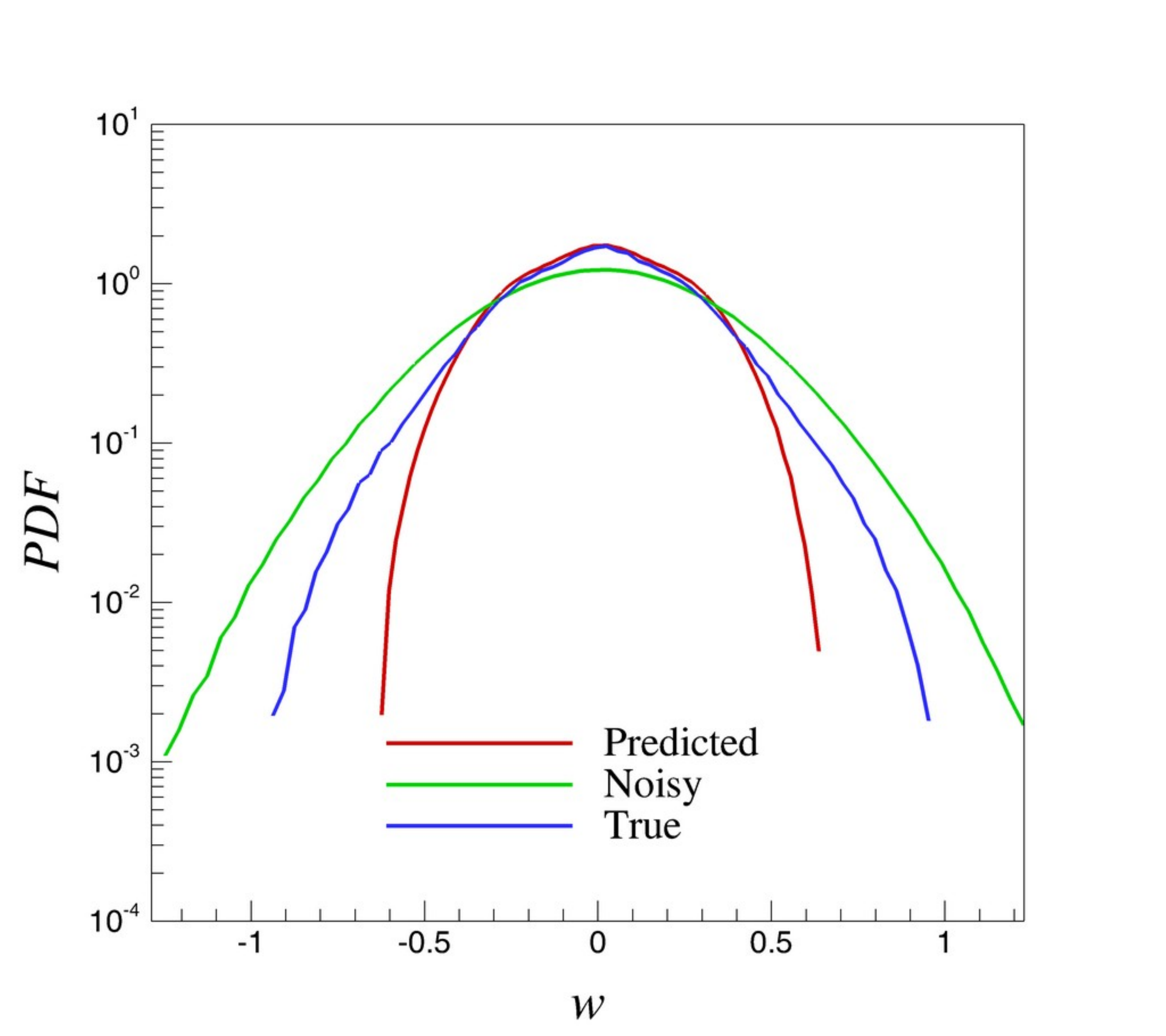}
}
}
\caption{A-priori results of the kinetic energy spectra (left) and PDF of the vorticity (right) for Kolmogorov turbulence. Here we utilize high fidelity data from the stratified turbulence problem to train a prediction for the Taylor-Green vortex at $Re=1600$.}
\label{fig:Euler_NS}
\end{figure}

Next, we present cross-validation results for different flows. Here, the proposed blind deconvolution ANN closure is trained from a completely different simulation data (i.e., testing with different flow configurations). Figure~\ref{fig:NS_Euler} demonstrates the ability of the data-driven closure to predict subfilter terms for the stratified KHI turbulence test case where TGV simulation data at $Re=1600$ is utilized to train our ANN. In this case we highlight that the only link between training and testing data is the $k^{-5/3}$ cascade and the magnitude of perturbations to the coarse-grained field. It can be seen that a remarkable inertial range recovery and regularization performance is still observed for both the low-pass spatially filtered and noised testing data. Analogously, in Figure~\ref{fig:Euler_NS}, we invert our training and testing data with stratified turbulence data being used to train the ELM which is further used for deconvolution and regularization of an appropriately perturbed solution field given by a Taylor-Green vortex simulation at $Re=1600$. Once again, previously observed trends in the output of the closure model are recovered. From the evidence provided above, we may tentatively conclude that for flows exhibiting similar cascade characteristics, subfilter recovery and high wavenumber regularization are reliant on the filter radius and magnitude of high frequency perturbation alone. This allows for exciting possibilities for 3D turbulence closures, for instance, to implicitly learn filter shapes from physical experiments and reproduce accurate trends through coarse-grained numerical simulations within a wide range of Reynolds numbers.

\section{Concluding Remarks}

An artificial neural network architecture is proposed for the data-driven deconvolution and regularization of low-pass spatially filtered turbulence fields. Both 2D and 3D test cases are examined with training data sets obtained through the coarsened and perturbed versions of high fidelity simulations for canonical homogeneous isotropic turbulence and stratified compressible turbulence problems. Two types of perturbations are tested, one in which the true data is filtered using a Gaussian kernel and the other which has a certain quantum of noise added to the field to represent high wavenumber aliasing errors. Testing data sets are generated through a spatial shifting procedure as well as through the utilization of slightly different magnitudes of filter radius and noise. This ensures that deconvolution and regularization performance of the proposed architecture remains localized to \emph{physics} and not numerical artifact. The proposed architecture is tested independently on Kraichnan, Kolmogorov and stratified compressible turbulence test cases and is able to provide an estimate of the deconvolved variable as examined through the energy spectra and probability density functions of the recovered fields. For noisy data, the innate regularization of the ELM training approach results in smoother predictions for the recovered variable. For our training mechanism, the ELM approach is chosen due to its exceptional speed of training in comparison with traditional gradient based methods as well as its excellent generalization ability (which avoids the problem of overfitting). The ability of the proposed approach for subfilter scale content recovery is benchmarked against several popular structural closure modeling strategies. It is observed that our the data-driven framework yields a similar deconvolution performance without the explicit specification of a filter kernel. In addition, we also test the universality of our data-driven closure by utilizing testing and training data sets from different simulations displaying the same classical Kolmogorov cascade. This leads to an exciting observation, as described in Section \ref{sec:Univ}, which tells us that the proposed framework may be utilized to leverage our understanding of the cascade of energy in three-dimensional turbulence to obtain reconstructions for flows across different physics. Indeed, the blind deconvolution procedure appears to be linked across a wide range of physics solely through the underlying filter radius that arises in LES flow computations due to coarse-graining and implicit (or explicit) numerical dissipation.

A natural follow-up to this investigation is to test our proposed approach in a fully a-posteriori analysis. One of our primary goals in subsequent investigations is also to address the issue of sampling for training data. Sampling strategies must be devised to ensure that the proposed framework is exposed to data from many physical regimes (as opposed to indiscriminate selections in parameter regimes). This is an important distinction that has significant implications on the performance of any proposed data-driven modeling framework. Basically, a data-driven model is \emph{only} as good as the data it has been trained on and can only reproduce physical behaviors similar to those it has seen in training. In addition to sampling strategies, it is also important to develop outlier identification systems for noisy data. Data preprocessing, an active area of research in the data science community, must therefore be integrated into our framework. From this point of view, the fast training times of the proposed architecture suggest the use of multiple networks trained to act in a `committee' for aggregate subfilter predictions of flow datasets with variable spatial and temporal characteristics. This ensures robustness towards outliers since aggregate predictions are likely to be more accurate than those from solely one trained network.

Since, apart from its training data, an ANN's performance is heavily dependant on its architecture, it is necessary to estimate the performance of the proposed framework for several different configurations in terms of network architectures (number of inputs, number of outputs, number of neurons, activation functions etc). Fortunately, the speed of training aids us in generating regularized network weights for extremely large datasets with considerable ease. In conclusion, this work features a preliminary glimpse at an exciting avenue for the next generation in data-driven turbulence closures.

\bibliography{jfm-references}
\bibliographystyle{jfm}

\end{document}